\documentclass[a4paper,11pt]{article}
\pdfoutput=1 
\usepackage{jheppub}
\usepackage[T1]{fontenc}
\usepackage{float}
\usepackage{neuralnetwork}
\usepackage{booktabs}
\usepackage{multirow}
\usepackage{xstring}
\usepackage{xpatch}
\usepackage{array}
\usepackage{ulem}

\newcommand{\la}{\left\langle}
\newcommand{\ra}{\right\rangle}

\newcommand{\lp}{\left(}
\newcommand{\rp}{\right)}

\newcolumntype{H}{>{\iffalse}c<{\fi}@{}}
\makeatletter
\xpatchcmd{\linklayers}{\nn@lastnode}{\lastnode}{}{}
\xpatchcmd{\linklayers}{\nn@thisnode}{\thisnode}{}{}
\makeatother

\title{\boldmath A new generation of simultaneous fits to LHC data using deep learning}

\author{Shayan Iranipour,}
\author{Maria Ubiali}

\affiliation{DAMTP, University of Cambridge, Wilberforce Road, Cambridge, CB3 0WA, United Kingdom}

\emailAdd{s.iranipour@damtp.cam.ac.uk}
\emailAdd{m.ubiali@damtp.cam.ac.uk}

\numberwithin{equation}{section}
\numberwithin{figure}{section}
\numberwithin{table}{section}

\keywords{Parton Distribution Functions, Effective Field Theories,
  Drell-Yan Processes, LHC Phenomenology, High-Luminosity LHC.}
\arxivnumber{2201.07240}

\abstract{We present a new methodology that is able to yield a simultaneous
determination of the Parton Distribution Functions (PDFs) of the
proton alongside any set of
parameters that determine the theory predictions;
whether within the Standard Model (SM) or beyond it.
The \texttt{SIMUnet} methodology is based on an extension of the {\tt NNPDF4.0} neural network
architecture, which allows the addition of an extra layer to simultaneously determine PDFs
alongside an arbitrary number of such parameters.  We illustrate its 
capabilities by simultaneously fitting PDFs
with a subset of Wilson coefficients within the Standard Model Effective Field Theory framework and show how the
methodology extends naturally to larger subsets of Wilson coefficients
and to other SM precision parameters, such as the strong coupling
constant or the heavy quark masses.}

\makeatletter
\newcommand\notsotiny{\@setfontsize\notsotiny\@vipt\@viipt}
\makeatother

\begin{document}
\maketitle
\flushbottom

\section{Introduction}
The successful operation of the Large Hadron Collider (LHC) at CERN is
enabling us to scrutinise the fundamental laws of nature to an
unprecedented degree and for a wide range of energy scales. Despite many
indications that the Standard Model (SM) of particle physics cannot be
a complete description of nature and despite theoretical
arguments pointing towards physics beyond the Standard Model (BSM) at
the TeV scale, no direct evidence for new physics at the TeV scale has been
gathered so far at colliders.

Far from being discouraging, it is an
exciting time for particle physics: the precision level reached by the
LHC experiments gives us the unique chance of investigating the
effects of new particles whose masses are far above the TeV scale, but
still produce observable effects at the scales within the direct kinematical
reach of the LHC.
Unlike for direct searches, which are
limited by the energy reach of the collider, indirect searches are limited
only by the theoretical and experimental control over the processes
under inspection. As such, they are becoming increasingly relevant at
the LHC as more data is collected and better theoretical predictions
are devised.

Despite the broad consensus on the need for precision, the paradigm of
indirect searches is often reduced to performing better measurements
and improving the accuracy of theoretical predictions. However, for an indirect
detection of new physics, it is pivotal to have a robust framework that is able to globally
interpret all subtle deviations from the SM predictions that might arise.  While
huge
progress has been made in determining key ingredients of theoretical
predictions from the data \cite{Britzger:2021ocj}, such as the Parton Distribution Functions
(PDFs) of the proton~\cite{nnpdf40,Bailey:2020ooq,Hou:2019efy,Alekhin:2017kpj},
the strong coupling constant~\cite{Zyla:2020zbs}, and the coefficients of a suitable
parametrisation of the effects of heavy new states via the addition of
higher dimensional operators to the SM Lagrangian, such as the SMEFT~\cite{Ethier:2021bye,Ellis:2020unq},
it is not yet evident how to combine all these partial fits into a global interpretation of
the LHC data.

These simultaneous determinations are sometimes essential to retrieve
a correct interpretation of the LHC data.
For example, in Ref.~\cite{Forte:2020pyp} it was shown that any
determination of the strong coupling constant $\alpha_s$ from a process which depends
on PDFs, such as hadronic processes or deep-inelastic
scattering (DIS), generally does not lead to a correct result
unless the PDFs are determined simultaneously along with $\alpha_s$.
The case of the determination of the strong coupling constant is
particularly relevant because of its strong correlation with the PDFs.
However, similar considerations apply to the simultaneous determination
of any physical parameter in PDF-dependent processes, such as for
example the
determination of the top quark mass~\cite{Alekhin:2016jjz}, or of the
electroweak (EW) parameters, such as the
$W$-mass~\cite{Bagnaschi:2019mzi}.
In the latter case, the correlation of PDFs
and the EW parameters is in principle weaker than in the case of the
strong coupling constant, but the very high accuracy
which is sought suggests that currently available results, specifically in $W$-mass determination, should be
reconsidered with care ~\cite{H1:2018mkk}.
Going beyond unpolarised PDFs, if more exclusive or spin-dependent
observables were included in a fit, then the simultaneous extraction of polarised and
unpolarised PDFs and of fragmentation functions and PDFs are the next
obvious frontiers. 
In the former case, the analysis of semi-inclusive measurements
involving the production of an identified hadron in the final state
employs observables that are often presented as ratios of
spin-dependent to spin-averaged cross sections, thus 
unpolarised and polarised PDFs should be treated simultaneously in an 
universal QCD analysis~\cite{Ethier:2020way,Khalek:2021ulf}.
In the second case,  the analysis of semi-inclusive measurements
involving the production of an identified hadron in the final state can be accurately performed only if the
final-state fragmentation functions are determined
alongside the initial-state PDFs~\cite{Ethier:2020way,Bertone:2018ecm}.

Recent findings indicate that caution should be taken not only in the
determination of SM precision parameters from hadronic processes, but
also for the Wilson coefficients that parametrise the effects of heavy
new physics via Effective Field Theory
(EFT) expansions. Such extensions of the SM Lagrangian determine the effect of
physics, that lives well above the energy scale reached by the LHC, by adding
higher dimensional operators to the SM Lagrangian, whose coefficients are
suppressed by powers of the new physics scale.
Although the proton structure parametrised by PDFs is intrinsically a
low-energy quantity and, as such, it should in principle be
separable from the high-energy new physics imprints, 
the complexity of the LHC environment might well intertwine them,
first and foremost because the same high-energy data is used to
constrain both the PDFs and the SMEFT parametrisation.
The effects of a simultaneous determination of the Wilson
coefficients of the SMEFT~\cite{Brivio:2017vri} and of the proton PDFs has been pioneered in several
recent studies~\cite{Greljo:2021kvv,Carrazza:2019sec,ZEUS:2019cou,CMS:2021yzl}. 
These studies reveal that, while with current DIS and Drell-Yan data the interplay can be kept under
control, once High Luminosity LHC (HL-LHC) data are considered,
neglecting the PDF interplay could potentially miss new physics manifestations
or misinterpret them.

In this paper we present a new methodology, dubbed
\texttt{SIMUnet}, which allows for a truly simultaneous determination of the
PDFs alongside any physical parameter that enters theoretical
predictions, whether a precision SM parameter, or the Wilson
coefficients of some EFT expansion. 
%
\texttt{SIMUnet} is based on an extension of the {\tt n3fit}
methodology and the {\tt NNPDF4.0} neural
network architecture~\cite{nnpdf40,nnpdfcode}, which treats both the PDFs and
the parameters fitted alongside PDFs on a completely
equal footing. We employ analogous deep learning techniques to
{\tt NNPDF4.0}, however, here we introduce an additional layer to
the network that captures the sensitivity of the data to
the parameters considered, leaving the first two layers of the architecture
to capture the data dependence upon the underlying PDFs.
This work was made possible thanks to the open-source
availability of the full {\tt NNPDF4.0} framework~\cite{nnpdfcode}.

To display the potential of {\tt SIMUnet}, in this work we
apply it on the simultaneous fit of the PDFs and of the SMEFT Wilson
coefficients considered in Ref.~\cite{Greljo:2021kvv}.
The rationale behind this choice stems from the fact that the
parameter dependence for the linear and quadratic contributions of the
Wilson coefficients of the dim-6 SMEFT expansion can be easily
incorporated within our framework. Furthermore, the
results of this study can be benchmarked against those obtained earlier, thus also 
assessing the effect of using a methodology that overcomes the
limitations of the methodology employed in Ref.~\cite{Greljo:2021kvv}.

In what follows, in Sect.~\ref{sec:theory} we outline a general framework that
allows for the dependence of the theoretical predictions on the parameters
considered to be obtained in an efficient way.  We
also describe how a simple method using $K$-factors can be employed in specific
cases, such as EFT expansions. In Sect.~\ref{sec:methodology}
we outline how our methodology works and how the neural network
architecture can exploit the fast interface discussed in Sect.~2 so as to
constrain external parameters alongside PDFs.
In Sect.~\ref{sec: results} we turn to discuss the results we obtain,
showcasing the methodology's ability to use current high-mass Drell-Yan (DY)
tails from the LHC to constrain not only the PDF, but also to simultaneously
determine the best fit Wilson coefficients to the data.  We demonstrate how
our methodology can recover degenerate (flat) directions in EFT space and how
the introduction of HL-LHC projected data is able to eliminate this
flat direction.  In Sect.~\ref{sec:closure} we stress test our proposed methodology using the closure
testing strategy, whereby we artificially contaminate the input data with a
particular choice of Wilson coefficients and demonstrate how our
approach is not only able to retrieve this choice, but is sufficiently robust to also simultaneously
replicate a known underlying PDF law. Finally, in Sect.~\ref{sec: conclusions} we
conclude this study and discuss possible future avenues using the
methodology presented in this paper.

\section{Theoretical predictions in a simultaneous fit\label{sec:theory}}
In this section we outline a general formalism that allows for the
fast interface of theory predictions and to isolate their dependence on the physical parameters that we may want
to fit simultaneously with the PDFs.
We then focus on simplifying this approach through the use of multiplicative
$K$-factors before discussing in the next section how we extend the
{\tt NNPDF4.0} framework to fit PDFs along with other physical parameters.

\subsection{Fast interface for theory predictions \label{sec: fktable}}

A formalism that is widely adopted in order to obtain a fast interface
of theory predictions in a PDF fit is the so-called Fast Kernel (FK)
table approach. The latter was formulated in
Refs.~\cite{Ball:2010de,Ball:2012cx,Bertone:2016lga} and it allows for the convolution of
PDFs with the partonic cross section in an efficient way by reducing the
non-trivial integral convolution to a tensor product using pre-computed
integrals on interpolation grids. Moreover, encoded within the FK table, is the
evolution kernel operator which performs the Dokshitzer–Gribov–Lipatov–Altarelli–Parisi
(DGLAP) evolution of the PDFs,
allowing for the Bjorken-\(x\) dependence of the PDF to be fitted at fixed
scale, \(Q_0\), before evolving the factorization scale to the relevant kinematic
scale, \(Q\), in a fast way.
Using the notation employed in \cite{Ball:2012cx},
we can write the theoretical prediction for any hadronic cross section as
\begin{equation}
T^{\rm hh}_I = \sum_{i,j=1}^{N_{\rm pdf}}\,\sum_{\alpha,\beta=1}^{N_x} \Sigma^I_{\alpha\beta i j}\,N_{\alpha i}^0\,N_{\beta j}^0\equiv N^0\cdot \Sigma_I\cdot N^0,
\end{equation}
where $I$ indicates a specific hadronic observable included in a PDF fit; $\alpha,\beta$ are the indices of
the interpolation grids in $x$-space for the first and second parton
respectively; $i,j$ are the indices of the PDFs of the initial-state partons
that contribute to the observable $I$ and $N^0$ are the (neural-network)
parametrisation of the independent PDFs at the initial scale $Q_0$. The
computation of the hadronic observables is reduced to a bilinear product over an
interpolation grid in the $x_{1,2}$ space and the basis of the input PDFs for a
given process. The quantity $\Sigma$ is the FK-table, which incorporates both the evolution of
the PDFs from the initial scale to the scale of the measured
observable and the partonic cross sections associated to each of the partonic channels
that enter the computation of the hadronic cross section. For
processes involving only one hadron and one lepton, like DIS, the expression
is even simpler and reads:
\begin{equation}
 T^{\rm hl}_I = \sum_{i=1}^{N_{\rm pdf}}\,\sum_{\alpha=1}^{N_x} \Sigma^I_{\alpha i}\,N_{\alpha i}^0\equiv \Sigma_I\cdot N^0.
\end{equation}
We can collectively refer to the theory prediction for a generic
observable - whether it involves one or two hadrons in the initial
states - as
\begin{equation}
  \label{eq:genlumi}
 T_I =\Sigma_I\cdot L^0,
\end{equation}
where $L^0$ indicates either the parametrisation of one independent
PDF at the initial scale or the product of two of them.

If we now consider how the theoretical predictions $T_I$ for each 
observable $I$ explicitly depend upon a single
parameter $c$, that could be for example the strong coupling constant, $\alpha_s$,
or the top mass, $m_t$, the entire dependence upon this parameter is contained in
the FK-tables $\Sigma$, since $L^0$ only captures the initial
scale parametrisation of PDFs, which is fitted from the data. 
For clarity of notation we focus
here on the case where only one single parameter is fitted alongside PDFs, but
the generalisation to more parameters is a straightforward extension of the
following argument, simply requiring the use of multi-variate Taylor expansions.
Schematically we can write the FK-tables as
\begin{equation}
    \Sigma_I(c) = \left[\hat{\sigma}(c)\otimes\Gamma(c)\right]_I
\end{equation}
where \(\hat\sigma\) is the partonic cross section and \(\Gamma\) are
the evolution kernels that evolve the PDFs from the initial scale to the scale of $T_I$.
The shorthand \(\otimes\) denotes the usual convolution, where for sufficiently well behaved functions \(f, g\) we have
\begin{equation}
    (f\otimes g)(x) = \int_x^1\frac{dy}{y}f(y)g\left(\frac{x}{y}\right).
\end{equation}
For a standard PDF-only determination, the parameter \(c\) is typically fixed to
a certain value during the computation of the FK table.  In general both
$\hat\sigma$ and $\Gamma$ depend on the given parameter (the strong coupling
being one such example), whilst in some other cases only the partonic cross sections,
$\hat\sigma$, depends on the parameter under consideration (the Wilson
coefficients of the SMEFT expansion being one such example).  

If we now want to fit the parameter $c$ alongside the PDFs, we need a fast interface to the dependence of each $\Sigma_I(c)$
upon the parameter \(c\)\footnote{ In Sect.~3 we will see that the
  reason for this is that gradient descent will assess the optimal
  value of \(c\) at every step during learning by repeatedly evaluating theory predictions with different \(c\) values.}.
One possible way to achieve such a fast interface is to assume that
both the evolution kernel and the partonic cross section are suitably
analytic such that they can be accurately described by their Taylor expansion about some point
\(c^*\). The closer the point \(c^*\) is to the actual parameter, the greater the validity of truncating
the Taylor series. Dropping from now on the observable index $I$, we can write
\begin{align}
    \Sigma(c) =& \sum_{p,q} \frac{(c-c^*)^{p+q}}{p!q!}\,\frac{\partial^p\hat\sigma(c^*)}{\partial c^p}\otimes
        \frac{\partial^q\Gamma(c^*)}{\partial c^q}\notag\\
    =& \sum_k (c-c^*)^k
        \sum_{\substack{p, q:\\p+q=k}} \frac{1}{p!q!}\,\frac{\partial^p\hat\sigma(c^*)}{\partial c^p}\otimes
        \frac{\partial^q\Gamma(c^*)}{\partial c^q}\notag\\
  =& \sum_k(c-c^*)^k\Sigma_{k}(c^*),
     \label{eq:expansion}
\end{align}
whereby for each power of \(c\) we have an order-by-order FK table,
\(\Sigma_k(c^*)\), that can be pre-computed before the fit, with
the dependence on the parameter \(c\) being isolated from the FK table. In
this way the task of querying the FK table for various values of \(c\) has been
reduced to computing individual FK tables for each order and taking a weighted
sum of these tensors which is a computationally trivial and fast operation\footnote{Importantly, such an operation can be implemented using the purely \texttt{TensorFlow}
functionality already present in the {\tt n3fit} methodology, thus allowing for
the gradients to be computed using automatic differentiation techniques
\cite{baydin2018automatic}.}.

To make the above discussion more concrete, in App.~\ref{sec: example_alphas} we explicitly consider the case whereby we wish to
isolate the FK table dependence on $\alpha_s$. 
The strong coupling constant $\alpha_s(\mu_r)$ and its evolution with
the scale from the reference value $\alpha_s(m_Z)$ to the
renormalization scale associated with each observable $\mu_r\sim Q$,
with $Q$ being the scale of the hard process, appears both in DGLAP
evolution kernel $\Gamma$ and in the partonic cross sections
$\hat{\sigma}$. In App.~\ref{sec: example_alphas} we show explicitly
how the dependence on  $\alpha_s(m_Z)$ and its evolution from $m_Z$ to
$\mu_r\sim Q$ are combined in the FK tables. Then, following our above
notation, we identify \(c=\alpha_s(m_Z)\) and
\(c^*=\alpha_s^\text{PDG}(m_Z) = 0.1179(10)\)  is the PDG value for
the strong coupling evaluated at the \(Z\)-boson
mass~\cite{Zyla:2020zbs} and explicitly derive the first two terms in the
Taylor expansion of Eq.~\eqref{eq:expansion} in the case of
$\alpha_s$. In the same appendix we also mention the approach one would have to use to Taylor expand about the central values of the electroweak parameters that enter a PDF fit.

\subsection{Observable dependence on the SMEFT Wilson coefficients  \label{sec: example_wc}}

We now focus on the parameters that we are specifically considering in this work,
to display the potential of our approach. Specifically, 
we are interested in fitting \(N\) parameters \(\{c_n\}\) with $n=1,\dots, N$, each of these
parameters associated to the Wilson coefficient of a given operator in the SMEFT expansion,
\(\mathcal{O}_n\).  We adopt an operator normalisation such that
\begin{equation}
\label{eq:lagrangian}
\mathcal{L}_{\rm SMEFT} = \mathcal{L}_{\rm SM} + \sum_{n=1}^{N} \frac{c_n}{v^2} \,\mathcal{O}_n \, ,
\end{equation}
with $N$ indicating in this specific case the number of operators that contribute to a given
observable and $c_n$ being the (dimensionless) Wilson coefficient associated to $\mathcal{O}_n$.
The quantity \(v\) refers to the energy scale of new physics, making the Wilson
coefficients dimensionless.

To include the effects of the corrections coming from the dim-6 operators
included in the SMEFT expansion in Eq.~\eqref{eq:lagrangian} in the
theoretical prediction $T_I$ for any observable included in the fit, 
one has to augment the SM partonic cross sections with the effects of
the relevant operators with the linear and quadratic modifications of
the SM cross section that the operators induce. 
Note that here we do not include the RGE running of the Wilson
coefficients, rather we treat them as fixed numbers. This is justified
by the fact that the anomalous dimensions of the operators that we
consider in this work are either Yukawa-suppressed or suppressed by
NLO EW corrections~\cite{Greljo:2021kvv}. To include the scale
dependence of the Wilson
coefficients~\cite{Gonzalez-Alonso:2017iyc,Jenkins:2013wua,Jenkins:2013zja},
we would have to fit the scale dependent Wilson coefficients at some fixed scale and use the relevant $\beta$-functions to
evolve the operator to the relevant scale using some pre-computed evolution kernel, which
would be factored into the FK-tables, exactly as is done for the PDF evolution.
Furthermore, given that the SMEFT operators that  we consider here do not modify the PDF DGLAP
evolution, the corrections will only appear in the partonic cross
sections $\hat\sigma$ and, unlike in the case of the Taylor expansion around the PDG value of $\alpha_s$
discussed in App.~\ref{sec: example_alphas}, here the sum is exact (and
not approximated) when the Taylor expansion is truncated at order $k=2$ (linear and quadratic corrections). 
The linear SMEFT corrections for each observable $I$ (from now
omitting the index $I$) can be parameterised as
\begin{equation}
  \label{eq:mult_k_fac_app1}
   R_{\rm SMEFT}^{(n)} \equiv \displaystyle \left( {\cal L}_{ ij}^{\rm NNLO} \otimes d\widehat{\sigma}_{ij,{\rm SMEFT}}^{(n)}\right)
 \big/ \left( {\cal L}_{ ij}^{\rm NNLO} \otimes d\widehat{\sigma}_{ij,{\rm SM}} \right) \, , \quad
 n=1\,\ldots, N \, ,
\end{equation}
with ${\cal L}_{ ij}^{\rm NNLO}$ being the partonic luminosity
evaluated at NNLO QCD
\begin{equation}
  \label{eq:lumidef}
\mathcal{L}_{ij}(\tau,M_X) = \int_{\tau}^1 \frac{d x}{x}~f_i (x,M_X) f_j (\tau/x,M_X) ~,
\end{equation}
$d\widehat{\sigma}_{ij,{\rm SM}}$
the bin-by-bin partonic SM cross section, and $d\widehat{\sigma}_{ij,{\rm SMEFT}}^{(n)}$
the corresponding partonic cross section associated to the interference between 
$\mathcal{O}_n$ and the SM amplitude $\mathcal{A}_{\rm SM}$ when setting $c_n = 1$.
Likewise, the ratio encapsulating the quadratic effects is defined as
\begin{equation}
  \label{eq:mult_k_fac_app2}
  R_{\rm SMEFT}^{(n,m)} \equiv \displaystyle \left( {\cal L}_{ ij}^{\rm NNLO} \otimes d\widehat{\sigma}_{ij,{\rm SMEFT}}^{(n,m)}\right)
  \big/ \left( {\cal L}_{ ij}^{\rm NNLO} \otimes d\widehat{\sigma}_{ij,{\rm SM}} \right) \, , \quad
 n,m=1\,\ldots, N \, ,
\end{equation}
with the bin-by-bin partonic cross section
$d\widehat{\sigma}_{ij,{\rm SMEFT}}^{(n,m)}$ now being evaluated from the squared amplitude $\mathcal{A}_n\mathcal{A}_m$
associated to the operators $\mathcal{O}_n$ and $\mathcal{O}_m$ when $c_n = c_m = 1$.
The partonic cross sections in these ratios are computed at LO in
QCD\footnote{Notice that, given that the SMEFT corrections are then
  multiplied by the SM predictions in Eq.~\eqref{eq:theory_k_fac},
  which do include NNLO QCD and NLO EW corrections, the SMEFT
  K-factors inherit factorisable higher-order radiative
  corrections~\cite{Greljo:2017vvb,Torre:2020aiz}. To account for the
  full NLO QCD effects in the SMEFT one should include the relevant diagrams in the partonic cross section computation.}.
In terms of  Eqs.~(\ref{eq:mult_k_fac_app1}) and~(\ref{eq:mult_k_fac_app2}), we can define the EFT $K$-factors as
\begin{eqnarray}
\label{eq:theory_k_fac}
  K(\{c_n\}) &=& 1+\sum_{n=1}^{N} c_n R_{\rm SMEFT}^{(n)}
  +\sum_{n,m=1}^{N} c_n c_m R_{\rm SMEFT}^{(n,m)} \, ,
  \end{eqnarray}
  which allow us to express a general cross section accounting for
  the dim-6 operators in Eq.~(\ref{eq:lagrangian}) as
\begin{eqnarray}
  \label{eq:theory_k_fac_app3}
 T &=& T^{\rm SM} \times  K(\{c_n\})\, 
\end{eqnarray}
where the $T$ is the SMEFT-modified theoretical prediction, $T^{\rm SM}$ is the state-of-the-art SM theoretical prediction
including NNLO QCD and NLO EW corrections and $K(\{c_n\})$ are the
SMEFT $K$-factors defined in Eq.~\eqref{eq:theory_k_fac}. 
%
%
%
The coefficients associated with the linear (quadratic)
corrections $ R_{\rm SMEFT}^{(n)}$ ( $R_{\rm SMEFT}^{(n,m)}$) in Eq.~(\ref{eq:theory_k_fac}) can be 
precomputed before the fit using a reference PDF set and then kept
fixed\footnote{The effect of varying the input NNLO PDF in
  Eqs.~(\ref{eq:mult_k_fac_app1}) and~(\ref{eq:mult_k_fac_app2}) was 
  quantitatively assessed in Ref.~\cite{Greljo:2021kvv} and it was
  found to be at most at the percent level.}
and the coefficients $\{c_n\}$ can be fitted alongside
the PDF parameters by multiplying the FK-tables, $\Sigma$, for each measurement included in
the fit by the $K$-factors encapsulating the whole dependence on the
parameters that we want to fit alongside the PDFs. Schematically this reads
\begin{equation}
  \label{eq:factored}
  \Sigma(\{c_n\})=\left[\hat\sigma \otimes \Gamma\right]\times K(\{c_n\})=\Sigma^{\rm SM}\times K(\{c_n\}),
\end{equation}
where the first factor does not depend on the Wilson coefficients and
it is given by the same FK-tables that one computes for the standard
{\tt NNPDF4.0} fits~\cite{nnpdf40}, which implicitly assume the SM to be
valid at all scales at which observables are measured.
The possibility of factoring the whole dependence upon the SMEFT parameters in a multiplicative
factor simplifies the procedure highlighted in Eq.~\eqref{eq:expansion} and it thus simplifies 
the way in which the dependence upon the parameters $\{c_n\}$ is fitted within {\tt SIMUnet}, as it will be
outlined in Sect.~3.2.

%

\section{Methodology \label{sec:methodology}}

In this section we discuss the details of the new {\tt SIMUnet} methodology which
allows us to extend the deep-learning model of the {\tt n3fit}
approach~\cite{Carrazza:2019mzf,nnpdf40,nnpdfcode} to optimize a 
set of parameters of the theory and determine their best fit values to the Monte Carlo
replica representation of the experimental data.  We also highlight various
points of the {\tt n3fit} methodology that remain pertinent to our study, in
particular the hyperparameter selection as well as the cross-validation
techniques employed to avoid overfitting.

\subsection{Neural network design\label{sec:gen_fitting}}

The recent {\tt NNPDF4.0} fit~\cite{nnpdf40} that our methodology is
built upon, shares the features of the previous {\tt NNPDF} releases~\cite{Ball:2017nwa,Ball:2012cx},
specifically the use of a Monte Carlo representation
of PDF uncertainties and correlations, and the use of neural networks
as basic interpolating functions. However all the details of the
fitting methodology, such as the choice of neural network architecture and the minimization
algorithm, are now selected through an automated hyperoptimization
procedure~\cite{Carrazza:2019mzf}. 
As such, our methodology employs state-of-the-art deep-learning techniques through
publicly available and highly optimised Machine Learning libraries such
as \texttt{TensorFlow}~\cite{abadi2016tensorflow} and
\texttt{Keras}~\cite{2015CS&D....8a4008B}.
As a result, it boasts both performance and improved fit quality using the cutting edge in optimiser
technology. Moreover, through the use of \texttt{TensorFlow} graph based
execution, the code enjoys the readability \texttt{Python} is famed for,
whilst still maintaining the performance of more traditional, statically typed,
compiled languages. Additionally, we still retain parallel based execution
capabilities, allowing for the ability to fit many replicas in a scalable way,
whether on a local machine (using central or graphical processing units) or on a
cluster.

\begin{figure}[t!]
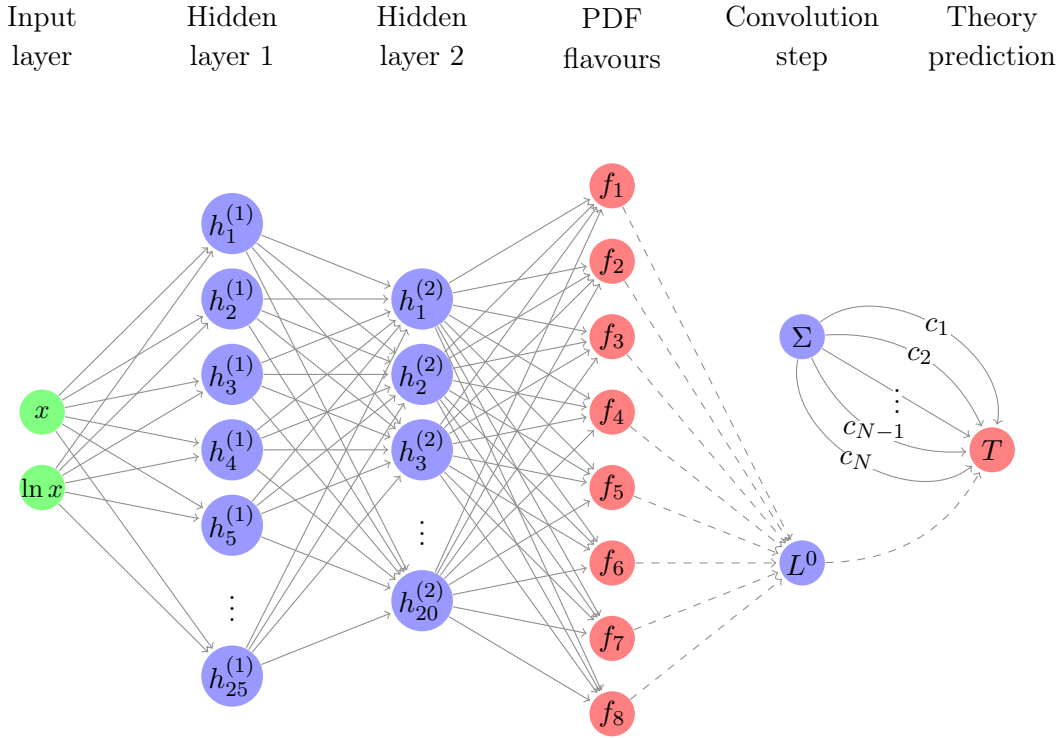

  \centering
  \begin{neuralnetwork}[height=10]
    \newcommand{\x}[2]{\IfEqCase{#2}{{1}{\small $x$}{2}{\small $\ln{x}$}}}
    \newcommand{\y}[2]{$f_#2$}
    \newcommand{\basis}[2]{\ifnum #2=4 \(L^0\) \else $\Sigma$ \fi}
    \newcommand{\hfirst}[2]{\ifnum #2=7 $h^{(1)}_{25}$ \else $h^{(1)}_#2$ \fi}
    \newcommand{\hsecond}[2]{\ifnum #2=5 $h^{(2)}_{20}$ \else $h^{(2)}_#2$ \fi}
    \newcommand{\standardmodel}[2] {$T$}
    \newcommand{\co}[4]{$c_1$}
    \newcommand{\ct}[4]{$c_2$}
    \newcommand{\cth}[4]{$c_3$}
    \newcommand{\vd}[4]{$\vdots$}
    \newcommand{\cnn}[4]{$c_{N-1}$}
    \newcommand{\cn}[4]{$c_{N}$}
    \newcommand{\convolve}[2]{$\otimes$}
    \inputlayer[count=2, bias=false, title=Input\\layer, text=\x]
    \hiddenlayer[count=7, bias=false, title=Hidden\\layer 1, text=\hfirst, exclude={6}]\linklayers[not to={6}]
    \hiddenlayer[count=5, bias=false, title=Hidden\\layer 2, text=\hsecond, exclude={4}]\linklayers[not from={6}, not to={4}]
    \outputlayer[count=8, title=PDF\\flavours, text=\y] \linklayers[not from={4}]
    \hiddenlayer[count=4, bias=false, text=\basis, title=Convolution\\step, exclude={2,3}]\linklayers[not to={1,2,3}, style={dashed}]
    \outputlayer[count=1, bias=false, text=\standardmodel, title=Theory\\prediction]
    \link[from layer = 4, to layer = 5, from node = 4, to node = 1, style={dashed, bend right=30}]
    \link[from layer = 4, to layer = 5, from node = 1, to node = 1, label=\co, style={bend left=70}]
    \link[from layer = 4, to layer = 5, from node = 1, to node = 1, label=\ct, style={bend left=35}]
    \link[from layer = 4, to layer = 5, from node = 1, to node = 1, label=\vd]
    \link[from layer = 4, to layer = 5, from node = 1, to node = 1, label=\cnn, style={bend right=35}]
    \link[from layer = 4, to layer = 5, from node = 1, to node = 1, label=\cn, style={bend right=70}]

    \path (L1-5) -- node{$\vdots$} (L1-7);
    \path (L2-3) -- node{$\vdots$} (L2-5);
  \end{neuralnetwork}
  \caption{Schematic depiction of the \texttt{SIMUnet} methodology.
    The input nodes (shown in green) are Bjorken-\(x\) and its logarithm.
    The forward pass through the deep hidden layers (blue) are performed as in the
    {\tt NNPDF4.0} methodology~\cite{nnpdf40} to yield the output
    PDFs at the initial scale (red). The initial scale PDFs are then combined in the
    initial scale luminosity $L^0$, defined in Eq.~\eqref{eq:genlumi}.
    The initial scale luminosity is then convoluted with the pre-computed FK-tables $\Sigma$ (shown in blue) to obtain
    the theoretical prediction $T$ (shown in red), which enters the figure of merit \eqref{eq:figmerit},
    which is minimised in the fit. 
    The $\Sigma$ dependence on the parameters $\{c_n\}$ is fed into theoretical prediction $T$ via the trainable
    edges of the combination layer. All trainable edges are shown by solid edges and are thus
    learned parameters determined through gradient descent, while dashed edges are
    non-trainable.\label{fig: fk_neuralnet}}
\end{figure}

The key feature of the {\tt SIMUnet} methodology is the use of a custom
\textit{combination layer},
which captures the dependence of the theoretical predictions upon the external
parameters $\{c_n\}$, with $n=1,\ldots,N$, that we fit alongside the PDFs.
The edges of the combination layer are fitted simultaneously with the
weights and biases associated with the parametrization of the PDFs at the initial scale $Q_0$.

The approach is represented schematically in Fig.~\ref{fig: fk_neuralnet}, whereby
the theory prediction, $T$, for each experimental observable included in the fit
depends on a dynamical choice of $\{c_n\}$. 
The values of $\{c_n\}$ are associated with the
weights of the trainable edges which determine the FK table, \(\Sigma\),
as in Eq.~\eqref{eq:expansion}. Such dependence enters the theoretical prediction $T$
via the bilinear produce between $\Sigma(\{c_n\})$ and the initial scale PDFs, which
in Eq.~\eqref{eq:genlumi} we refer to as $L^0$, where  $L^0$ indicates either the parametrization
of one independent PDF at the initial scale or the product of two of them.

Letting \(\theta\) denote the set of trainable neural network parameters (the
weights and biases) that parameterize the PDFs and $\{c_n\}$ the parameters that
we fit alongside the PDFs,
\texttt{SIMUnet} fits the joint \(\hat\theta=\theta\cup\{c_n\}\) parameter set, 
by letting gradient descent determine their optimum value in
order to minimize the figure of merit used in the fit, which is defined as
\begin{equation}
  \label{eq:figmerit}
  \chi^2(\hat\theta) = \frac{1}{N_\text{dat}}
  (\mathbf{D}-\mathbf{T}(\hat\theta))^T({\rm \bf cov})^{-1}(\mathbf{D}-\mathbf{T}(\hat\theta)),
\end{equation}
with \(\mathbf{D}\) being the vector of experimental central
values, \(\mathbf{T}\) the vector of theoretical
predictions and \( \rm \bf cov\) the covariance matrix
encapsulating the experimental uncertainties and the correlations
therein.

Note that the covariance matrix that we use here is the one
obtained by using the so-called $t_0$ prescription~\cite{Ball:2009qv},
where multiplicative uncertainties are multiplied by theoretical
predictions to avoid the D'Agostini bias~\cite{DAgostini:1993arp}. 
If we wanted to include also correlated sources of
theoretical uncertainties, such as those associated with missing
higher order uncertainties in the theory predictions,  we could include them using the method outlined in
Refs.~\cite{NNPDF:2019ubu,NNPDF:2019vjt,Ball:2020xqw}. We leave this
endeavour to a future analysis, once the theory covariance matrix for
missing higher orders will be available at NNLO. 

\subsection{Parameter fitting using linearisation \label{sec:fitting}}
In the case of dim-6 operators, discussed in Sect.~\ref{sec: example_wc}, including only the interference
of the SMEFT corrections with the SM diagrams 
is trivial, as we only add a linear dependence upon the Wilson coefficients.
%
Indeed, the identity of Eq.~\eqref{eq:theory_k_fac_app3} allows us to
write the theoretical predictions by \texttt{SIMUnet} at a particular
configuration \(\hat\theta\) as
\begin{equation}
  T(\hat\theta) =
  \Sigma(\{c_n\})\cdot\,L^0(\theta)=T^{\rm SM}(\theta) \cdot \left(  1+\sum_{n=1}^{N} c_n R_{\rm SMEFT}^{(n)}\right),
  \label{eq: modified_theory}
\end{equation}
where $T^{\rm SM}(\theta)=\Sigma^{\rm SM}\cdot\,L^0(\theta)$ is the SM theoretical 
prediction for each observable, that corresponds to $\{c_n=0\}$.

\begin{figure}[t]
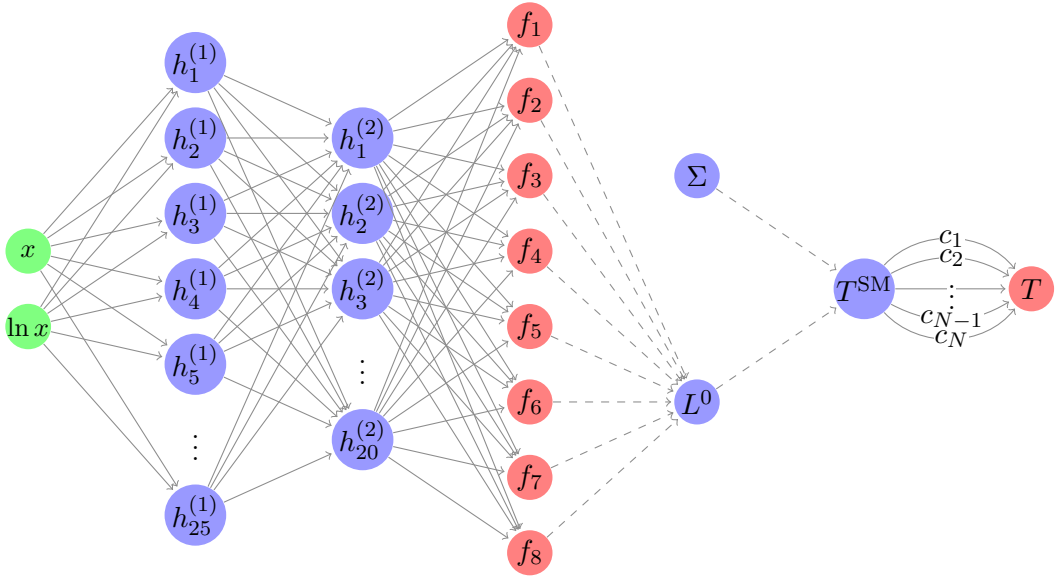

  \centering
  \begin{neuralnetwork}[height=10, layerspacing=22mm]
    \newcommand{\x}[2]{\IfEqCase{#2}{{1}{\small $x$}{2}{\small $\ln{x}$}}}
    \newcommand{\y}[2]{$f_#2$}
    \newcommand{\basis}[2]{\ifnum #2=4 \(L^0\) \else $\Sigma$ \fi}
    \newcommand{\hfirst}[2]{\ifnum #2=7 $h^{(1)}_{25}$ \else $h^{(1)}_#2$ \fi}
    \newcommand{\hsecond}[2]{\ifnum #2=5 $h^{(2)}_{20}$ \else $h^{(2)}_#2$ \fi}
    \newcommand{\standardmodel}[2] {$T^\text{SM}$}
    \newcommand{\eft}[2] {$T$}
    \newcommand{\co}[4]{$c_1$}
    \newcommand{\ct}[4]{$c_2$}
    \newcommand{\cnn}[4]{$c_{N-1}$}
    \newcommand{\cn}[4]{$c_{N}$}
    \newcommand{\vd}[4]{$\vdots$}
    \newcommand{\FK}[2]{$\sigma$}
    \newcommand{\convolve}[4]{$\otimes$}
    \inputlayer[count=2, bias=false, title=Input\\layer, text=\x]
    \hiddenlayer[count=7, bias=false, title=Hidden\\layer 1, text=\hfirst, exclude={6}]\linklayers[not to={6}]
    \hiddenlayer[count=5, bias=false, title=Hidden\\layer 2, text=\hsecond, exclude={4}]\linklayers[not from={6}, not to={4}]
    \outputlayer[count=8, title=PDF\\flavours, text=\y] \linklayers[not from={4}]
    \hiddenlayer[count=4, bias=false, text=\basis, title=Convolution\\step, exclude={2,3}]\linklayers[not to={1,2,3}, style={dashed}]
    \hiddenlayer[count=1, bias=false, title=SM\\Observable, text=\standardmodel]\linklayers[not from={2,3}, style={dashed}]
    \outputlayer[count=1, bias=false, text=\eft, title=SMEFT \\ Observable]
    \link[from layer = 5, to layer = 6, from node = 1, to node = 1, style={bend left=50}, label={\co}]
    \link[from layer = 5, to layer = 6, from node = 1, to node = 1, style={bend left}, label={\ct}]
    \link[from layer = 5, to layer = 6, from node = 1, to node = 1, label={\vd}]
    \link[from layer = 5, to layer = 6, from node = 1, to node = 1, style={bend right}, label={\cnn}]
    \link[from layer = 5, to layer = 6, from node = 1, to node = 1, style={bend right=50}, label={\cn}]

    \path (L1-5) -- node{$\vdots$} (L1-7);
    \path (L2-3) -- node{$\vdots$} (L2-5);
  \end{neuralnetwork}
  \caption{Schematic representation of the architecture used by \texttt{SIMUnet}
    in the case of the fit of SMEFT coefficients, in which the dependence of the theoretical prediction $T$
    upon the parameters $\{c_n\}$ can be factored into a multiplicative K-factor, as in Eq.~\eqref{eq:factored}.
    The scheme is the same as the one of Fig.~\ref{fig: fk_neuralnet}, however 
    the initial scale PDFs $L^0$ are first convoluted with the relevant SM FK Tables
    to obtain the Standard Model theory prediction (\(T^\text{SM}\)). The SM predictions
    are then incremented by the addition of the linear SMEFT corrections via a final 
    linear {\it combination layer}.  All solid edges are trainable and thus
    modified during gradient descent.  The full structure of the hidden
    layers are suppressed for clarity. The precise nature of the manipulation
    performed by the final layer is outlined in Eq.~\eqref{eq:
      modified_theory}\label{fig: Architecture}.}
\end{figure}

Given that in this case 
the dependence upon the parameters $\{c_n\}$ is factored out of the FK-tables into
a multiplicative factor, we can visualize the dependence upon the parameter in
the simplified schematic representation given in Fig.~\ref{fig: Architecture}. 
Thanks to linearisation, the bracketed term is implemented
using a \texttt{Keras} custom layer, 
which takes the usual SM observable predicted by the network at some
interim configuration and maps it to a SMEFT modified observable
with the strength of the new physics interaction being determined by
the weights of the combination layer, which in this case is simply
an extra sequential layer that maps $T^{\rm SM}$ into the theoretical prediction
$T$ which enters the figure of merit defined in Eq.~\eqref{eq:figmerit}.

The SMEFT-modified theory prediction, \(T\), is then split into disjoint
training and validation splits. For each split we compute a training and
validation \(\chi^2\). 
Gradient descent attempts to minimize the training \(\chi^2\) by
descending the loss surface in \(\hat{\theta}\)-space while the validation
\(\chi^2\) is monitored to assess the network's out of sample performance. The
validation \(\chi^2\) decreases initially (since the network is learning the shared
underlying laws that are common to both sets), but upon the onset of
overfitting, the out of sample performance begins to deteriorate as the network
fits the noise of the training data.  This particular point in the training
process corresponds to the validation \(\chi^2\) ceasing to decrease and instead
beginning to increase. Upon reaching this regime, training is halted and various
checks, such as positivity and integrability of the resulting PDF are assessed.
It is at this point that the best fit values of the Wilson coefficients \(\{c_n\}\) are obtained, since
gradient descent has modified the combination layer's weights such that they
best fit the input data.  Using this cross-validation procedure
\cite{bishop1995neural,DelDebbio:2007ee,Ball:2008by} we ensure the fitting of
statistical fluctuations are avoided as much as possible and it is the shared
underlying physical laws that are being fitted instead.  At the same
time, in order to ensure
that the \(\{c_n\}\) are not overfitted, we set the datasets that will be
modified by these parameters to have a good representation both within the
training set and validation set. Thus in this study we split such datasets to have
their training and validation fractions equal: \(f_\text{tr} = f_\text{val} =
0.5\), while keeping all other training and validations fractions as
in the fits of Ref.~\cite{Greljo:2021kvv}.

As with all deep learning studies, the user has freedom to choose the various
hyperparameters of their model at their own discretion. Such parameters include
the architecture of the network, the particular choice of initializer that sets the
initial values of network parameters \(\{\theta\}\), or the choice of
minimizer (and the specific settings therein) that tunes these parameters to
their optimal values such that the performance metric is minimized.
Techniques exist to automate this process such as the \texttt{hyperopt} library
\cite{bergstra2013making} employing Bayesian methods to perform this optimization.
Indeed this is employed by {\tt NNPDF4.0} and we choose to use the same settings that
were found to be optimal there~\cite{nnpdf40,nnpdfcode}.
In this study, the hyperoptimization procedure has not been performed to reassess the
optimality of the various neural network settings. Indeed, this is justified,
since we expect, \textit{a priori} that the PDF modifications due the presence
of the aforementioned SMEFT operators will be moderate, and thus the
hyperparameters selected assuming a SM-only scenario will remain
adequate. However, this assumption does not hold anymore in the
event that a more marked PDF modification is expected or obtained \textit{a
posteriori}.  This point is also worth considering if one is to introduce a
large number (relative to NNPDF4.0) of new measurements or any measurements
which introduce tensions with other datasets. Should this be the case,
hyperoptimization should be performed again on the layers preceding
the combination layer, and this can be done
straightforwardly using the standard procedure described in Ref.~\cite{nnpdf40}.

The connections to the combination layer (one for each of the \(c_n\)) are
initialized to zero, since we assume \textit{a priori} the Wilson coefficients
will be small; although we observe that if they are initialized according to a
normal distribution, then virtually identical results are obtained. These
connections are then modified during back propagation.
It is worth mentioning that for small values of Wilson coefficients (such as
those in this study), it is highly desirable to scale the units such that the
Wilson coefficients are \(\mathcal{O}(1)\) and thus comparable to the learning
rate. This will assist the optimiser during gradient descent to converge upon
the minimum in a timely fashion, since the step size will be more suited to the
characteristic scale of the Wilson coefficients. In practice, the appropriate
scaling is usually determined \textit{a posteriori}, where one can analyse the
typical values for the Wilson coefficients and refit with the normalization set
accordingly.

\subsection{Incorporating non-linear effects \label{sec: dim8}}
The effect of including the SMEFT self-interference diagrams can often introduce
a marked effect on the bounds obtained for a given SMEFT scenario
\cite{Boughezal:2021tih,Greljo:2021kvv}. Moreover, with the impetus to produce
high precision theoretical predictions for the LHC era, the inclusion of higher
order corrections in the SMEFT scattering graphs are becoming particularly pertinent
\cite{Grojean:2013kd,Baglio:2019uty,Degrande:2020evl,Dawson:2021ofa}.
Such considerations introduce a non-linear dependence on
the Wilson coefficients in the space of observables: quadratic in the former and
quadratic in the highest order in the QCD expansion in the latter. This point serves
as a major advantage of our methodology which can accommodate these effects
during the fit by the simple addition of non-trainable edges.

When computing the amplitude of a Feynman diagram related to some process one
has the schematic form \(\mathcal{A} = \mathcal{A}^\text{SM} +
\sum\mathcal{A}_i\) where \(\mathcal{A}_i\) is the amplitude
corresponding to the operator \(\mathcal{O}^{(i)}\) computed to some order in
perturbation theory. We assume here that it is LO in the Wilson coefficients,
but need not be in general and the extension to higher orders in the Wilson coefficient
expansion is discussed at the end of the present section. When computing
the observable, the matrix element, \(|\mathcal{A}|^2\), introduces terms of the
form \(\mathcal{A}_i\mathcal{A}_j\). Since these amplitudes are computed
to LO in perturbation theory, we can rewrite Eq.~\eqref{eq:
  modified_theory} as
\begin{equation}
  \label{eq: quadratic_modified_theory}
  T(\hat\theta) =
T^{\rm SM}(\theta) \cdot
    \left(
1+\sum_{n=1}^{N} c_n R_{\rm SMEFT}^{(n)}
  +\sum_{1\leq n\leq m \leq N} c_n c_m R_{\rm SMEFT}^{(n,m)} \, ,
    \right)
\end{equation}
with \(R_{\rm SMEFT}^{(n,m)}\) being defined as in Eq.~\eqref{eq:mult_k_fac_app2}. Including this
contribution in the simultaneous fit is a straightforward task and simply
requires that the manipulation performed by the combination layer correspond to
that of Eq.~\eqref{eq: quadratic_modified_theory} instead of Eq.~\eqref{eq: modified_theory}.

\begin{figure}[t]
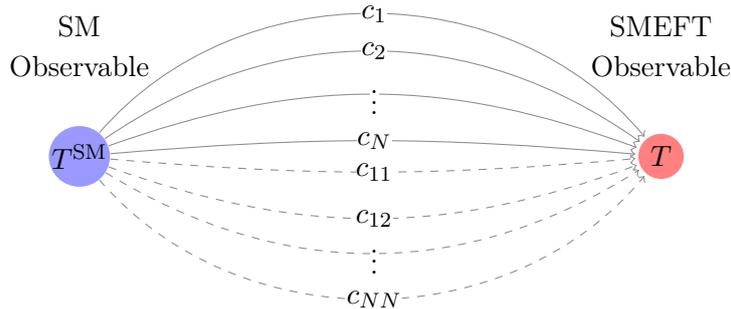

  \centering
  \begin{neuralnetwork}[height=2, layerspacing=20em]
    \newcommand{\standardmodel}[2] {$T^\text{SM}$}
    \newcommand{\eft}[2] {$T$}
    \newcommand{\vd}[4]{$\vdots$}
    \newcommand{\ci}[4]{$c_1$}
    \newcommand{\cj}[4]{$c_2$}
    \newcommand{\ck}[4]{$c_N$}
    \newcommand{\calpha}[4]{$c_{11}$}
    \newcommand{\cbeta}[4]{$c_{12}$}
    \newcommand{\cgamma}[4]{$c_{NN}$}

    \hiddenlayer[count=1, bias=false, text=\standardmodel, title=SM\\ Observable]
    \outputlayer[count=1, bias=false, text=\eft, title=SMEFT \\ Observable]
    \link[from layer = 0, to layer = 1, from node = 1, to node = 1, style={bend left=50}, label={\ci}]
    \link[from layer = 0, to layer = 1, from node = 1, to node = 1, style={bend left=35}, label={\cj}]
    \link[from layer = 0, to layer = 1, from node = 1, to node = 1, style={bend left=20}, label={\vd}]
    \link[from layer = 0, to layer = 1, from node = 1, to node = 1, style={bend left=5}, label={\ck}]
    \link[from layer = 0, to layer = 1, from node = 1, to node = 1, style={dashed, bend right=5}, label={\calpha}]
    \link[from layer = 0, to layer = 1, from node = 1, to node = 1, style={dashed, bend right=20}, label={\cbeta}]
    \link[from layer = 0, to layer = 1, from node = 1, to node = 1, style={dashed, bend right=32}, label={\vd}]
    \link[from layer = 0, to layer = 1, from node = 1, to node = 1, style={dashed, bend right=50}, label={\cgamma}]
  \end{neuralnetwork}
  \caption{Schematic representation of how \texttt{SIMUnet} allows for the
  effects of the SMEFT self-interaction diagrams (dim-6)$^2$ to be included. We show here
  how the SM observable can be transformed to a SMEFT observable which includes
  these \(\mathcal{O}(1/v^4)\) terms. The preceding PDF layers are omitted for
  clarity. The linear contributions are included in the usual way,
  with the strength of the SMEFT couplings being determined by the trainable edges,
  \(c_n\): shown by the solid lines. The SMEFT-SMEFT interference contributions
  are instead non-trainable edges with their value being fixed by the strength
  of the corresponding pair of trainable edges. These are shown by the dashed lines.
  There will in general be \(N\) trainable edges, and \(N(N+1)/2\)
  non-trainable edges. \label{fig: nontrainable}}
\end{figure}
In practice, however, it is more convenient to rewrite the terms in the right
most sum of Eq.~\eqref{eq: quadratic_modified_theory} by defining \(c_{nm} = c_nc_m\):
\begin{equation}
 T(\hat\theta) =
T^{\rm SM}(\theta) \cdot
    \left(
      1+\sum_{n=1}^{N} c_n R_{\rm SMEFT}^{(n)}
  +\sum_{1\leq n\leq m \leq N} c_{nm} R_{\rm SMEFT}^{(n,m)} \, ,
    \right)
\end{equation}
we see that both summations are of the same form and so the manipulation
required to incorporate the quadratic effects of the squared SMEFT amplitude
is reduced to introducing \(N(N+1)/2\) additional edges to the
combination layer, one for each \(c_{nm}\). These additional connections,
however, are not trainable, that is to say their value is not determined by
gradient descent during learning; since \(c_{nm}\) is completely determined
by the product of trainable edges \(n\) and \(m\). This is shown schematically in
Fig. \ref{fig: nontrainable}, whereby the trainable edges determine the
non-trainable edges that perform the manipulation corresponding to the right
most term in the brackets of Eq. \ref{eq: quadratic_modified_theory}.
Since the observable is always polynomial in the Wilson coefficients, it is
possible to include the effects of higher dimensional operators in the SMEFT
expansion (such as adding dimension-8 operators \cite{Li:2020gnx,
  Murphy:2020rsh}) in this way.

Moreover, to incorporate quantum corrections
arising from Next-to-Leading Order (NLO) terms in the QCD perturbative
expansion of the SMEFT corrections is
a similarly straightforward task, simply requiring the computation of the
corresponding K-factor and including an additional edge in the combination layer whose
value is pinned to be the value of the corresponding trainable edge raised to
some power. 
If one wanted to include also the scale dependence of the Wilson coefficients
\cite{Gonzalez-Alonso:2017iyc,Jenkins:2013zja,Jenkins:2013wua}, one would have to fit the scale dependent
Wilson coefficients at some fixed scale, \(Q_0\), and use the relevant
\(\beta\)-functions to evolve the operator to the relevant scale using some
 pre-computed evolution kernel, which would be factored into the
pre-computed FK-tables.

Finally, it is well-known that one can obtain prior knowledge on the Wilson
coefficients by assuming a UV completion of the EFT exists that is local,
unitary and causal.  Such matching conditions can often
impose positivity bounds \cite{Davighi:2021osh} on functions, \(f_i\), of the
Wilson coefficients by leveraging these standard conditions on the UV theory.
To incorporate such a prior within
our framework is analogous to the way positivity on physical cross sections is
achieved within the NNPDF methodology \cite{Ball:2010de,NNPDF:2014otw}. The
minimizer is informed of the prior by modifying the training loss function to
penalize negative values of Wilson coefficients, with one such choice of penalty
term being
\begin{equation}
  \text{loss} = \text{loss} + \lambda\sum_{i}\Theta\big(-f_i(\{c\})\big),
\end{equation}
with \(\Theta\) the usual Heaviside step function which takes unit value for
positive arguments and vanishes otherwise. The parameter \(\lambda\) is a
non-negative scalar which can be treated as a hyperparameter and encapsulates the
degree of belief in the prior: larger values impose the positivity more strictly
while smaller values allow for slight violations of the positivity constraints.
In a similar way to NNPDF it is also possible to filter the replicas \textit{a posteriori}
by discarding those that are deemed to violate positivity constraints too strongly.

\subsection{Fixed PDF analysis \label{sec: fixedpdf}}

A desirable feature of any simultaneous fitting methodology is to be able to
benchmark the simultaneous extraction with the analogous fixed PDF analysis.
Indeed, this is precisely what is done in Ref.~\cite{Greljo:2021kvv} and the
broadening of the Wilson coefficients bounds there being the key
message of the paper.

This too is achievable with our methodology and proceeds as follows.  We begin
by first freezing the combination layer parameters by making them non-trainable.
Mathematically, the combination layer thus performs an identity transformation,
effectively removing itself from the process. At this stage we are in effect
performing a regular {\tt NNPDF4.0} PDF fit.
Upon successfully achieving the stopping criterion for the PDF only portion of
the fit, we then freeze the PDF sector of the neural network (again by making
the weights and biases non-trainable) and reinstate the combination layer,
allowing for gradient descent to optimize the external parameters, \(c_i\),
only.
In this way we eliminate the cross-talk between the PDFs and the external
parameters and so the external parameters fitted in this way are equivalent to
an analogous study keeping the PDFs fixed to the baseline. 

\section{Results \label{sec: results}}
In this section we present the results we obtain by applying the 
\texttt{SIMUnet} methodology to fit the Wilson coefficients alongside the PDFs in
the two benchmark scenarios. 
We begin by discussing the specific scenarios explored in this
work by motivating our particular selection of operators.
We then outline our dataset selection, which is chosen to provide strong constraints
not only on the PDF, but also to the Wilson coefficients under
consideration in each of the two scenarios.
We then turn to describe the results we obtain in the two
scenarios, both in terms of the resulting PDFs and bounds on the Wilson
coefficients. We compare our findings to those obtained when PDFs are
kept fixed to the SM baseline, rather than fitted simultaneously alongside
the Wilson coefficients. Finally, we summarize our results, by
exploring the correlations between PDFs and the Wilson coefficients that we
consider in this study and quantifying the agreement with
respect to the previous findings of Ref.~\cite{Greljo:2021kvv}.

\subsection{Benchmark scenarios}

The two SMEFT scenarios we explore in this work are the same as those introduced in
Ref.~\cite{Greljo:2021kvv}. In this way, the \texttt{SIMUnet} methodology can be
benchmarked against the rather less efficient methodology based on
Hessian minimization and on fitting the PDFs at fixed values of the
Wilson coefficients that we used in our previous work. Also, the two scenarios highlight the interplay between the PDFs and the
EFT dynamics, illustrating in particular how the former changes and how constraints to
the latter are modified. Moreover, they allow us to test the inclusion
of purely linear SMEFT corrections (as in Benchmark Scenario I) and
of both linear and quadratic SMEFT corrections (as in Benchmark
Scenario II).

In particular, the first Benchmark Scenario belongs to the class of electroweak
precision tests and is sensitive to a broad range of UV-complete theories
proposed in the literature.
The oblique corrections, originally proposed in
\cite{Peskin:1991sw,Altarelli:1991fk}, play a key role in testing BSM theories. They parametrise the self-energy of the
electroweak gauge bosons. Out of the four operators which can be
identified with dim-6 operators in the SMEFT, two, namely $T$ and $S$, are well constrained by precision LEP
measurements~\cite{Barbieri:2004qk}  and grow slowly with the energy, while $W$ and $Y$ scale
 as \(\mathcal{O}(q^4)\) and are particularly sensitive to the tails of the high mass
Drell-Yan distributions at the LHC~\cite{Farina:2016rws,Panico:2021vav}.
Their inclusions implies the addition of the following terms to the SM Lagrangian
\begin{equation}
  {\rm \uuline{Benchmark\,\,Scenario\,\,I}:} \qquad
  \mathcal{L}_\text{SMEFT} =
  \mathcal{L}_\text{SM}
  - \frac{W}{4m_W^2}(D_\rho W^a_{\mu\nu})^2
  - \frac{Y}{4m_W^2}(\partial_\rho B_{\mu\nu})^2
\end{equation}
with \(m_W\) the W boson mass and \(D\) the usual covariant
derivative.  In this scenario the quadratic (dim-6)$^2$ effects are
much less relevant than the linear effects coming from the
interference between the SM and the SMEFT amplitudes, thus we can
define the SMEFT-modified theoretical predictions as in Eq.~\eqref{eq: modified_theory}.

The second benchmark represents a consistency check of the
existing hints of lepton universality violation in rare $B$-meson decays reported by the
LHCb collaboration~\cite{LHCb:2014vgu,LHCb:2017avl,LHCb:2019hip}. In
particular, we look at the muon-philic operator, in which the 
Lagrangian is defined as:
\begin{equation}
    {\rm \uuline{Benchmark\,\,Scenario\,\,II}:} \qquad
  \mathcal{L}_\text{SMEFT} =
  \mathcal{L}_\text{SM}
  + \frac{\mathbf{C}_{33}^{D\mu}}{v^2}
  (\bar{d}_L^3\gamma_\mu d_L^3)
  (\bar\mu_L\gamma^\mu\mu_L),
\end{equation}
where \(v\) is the Higgs vacuum expectation value, \(\mathbf{C}_{33}^{D\mu}\)
is a scalar Wilson coefficient for this scenario, and \(d^3\) (\(\mu\)) is the bottom
quark (muon) Dirac spinor. Noting that the inclusion of this operator modifies only
the muon channel whereas the electron channel is still described by
the SM: when studying this particular SMEFT scenario, we only modify the muon channel
in the high-mass DY datasets. 
Moreover, in this scenario the quadratic corrections are important, so
we do include the presence of quadratic Wilson coefficients
in the matrix element to constrain \(\mathbf{C}_{33}^{D\mu}\) and
define the SMEFT-modified theoretical predictions 
according to Eq.~\eqref{eq: quadratic_modified_theory}.

\subsection{Experimental data \label{sec: experimental_data}}

\begin{figure}[t]
  \centering
  \includegraphics[width=\textwidth]{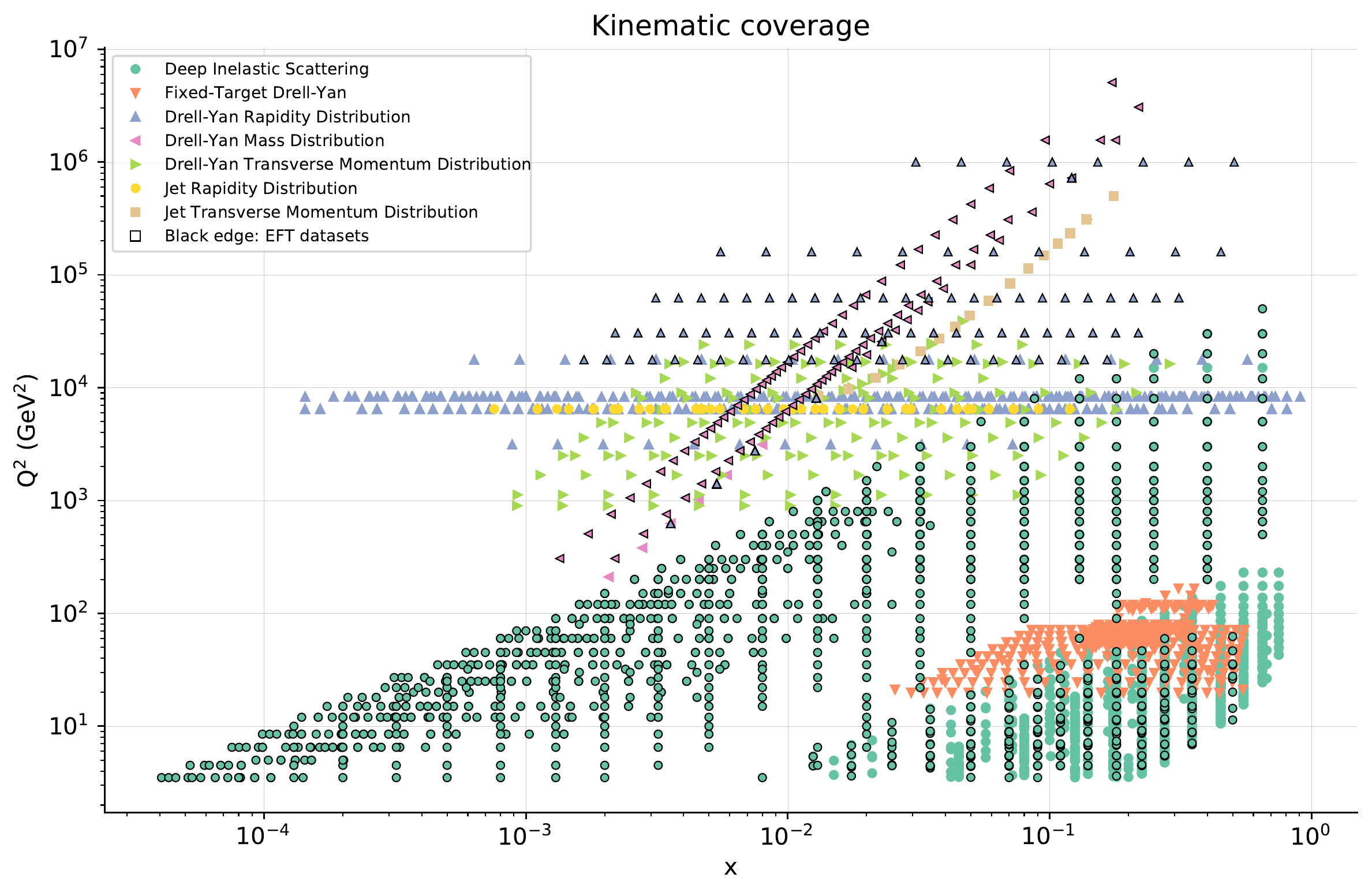}
  \caption{The kinematic coverage of the data points used in this study
  grouped according to their process. Points marked with a black edge
  are also used to constrain the SMEFT operators. Note that the HL-LHC
  projections used in this study are omitted from this figure, as only
  actual data are displayed.
  \label{fig: kinematic_coverage}}
\end{figure}

The dataset chosen for this study comprises \(4022\) data points,
spanning a broad range of processes including neutral and charged current DIS data (the vast majority of which is composed of the
HERA combined dataset) to high mass Drell-Yan measurements from the LHC.  The
dataset selection is identical to that of Ref.~\cite{Greljo:2021kvv} which is in itself an
extension of the strangeness study of \cite{Faura:2020oom} and NNPDF3.1
\cite{Ball:2017nwa}. In particular, the neutral current Drell-Yan measurements
from ATLAS at \(7\), and \(8\) TeV \cite{ATLAS:2013xny, ATLAS:2016gic} and
CMS at \(7\), \(8\), and \(13\) TeV \cite{CMS:2013zfg, CMS:2014jea, CMS:2018mdl}
are LHC measurements that are used to constrain, not only the PDF, but are also
sufficiently sensitive to the BSM scenario considered in this work to also
constrain the SMEFT operators. The kinematic coverage of the data points used in
this study are shown in Fig. \ref{fig: kinematic_coverage}. The points are shown
in \((x,Q^2)\) space with the data points that are modified by the EFT operators
highlighted with a border, such points thus also constrain the Wilson
coefficients as well as the PDFs. We note that, although DIS theory predictions are
modified by the operators we consider in the two benchmark scenarios,
the change in the HERA DIS cross sections upon the variation of the Wilson
coefficients upon consideration is minimal, as is explicitly assessed in Ref.~\cite{Greljo:2021kvv}.

\subsection{Results for Benchmark Scenario I \label{sec: WY_fit}}

We first present the results obtained within the first
Benchmark Scenario outlined in Sect.~4.1, in which the linear effects are dominant as compared to the
quadratic effect and, as a result, SMEFT-modified theoretical
predictions are defined as in Eq.~\eqref{eq: modified_theory}.
We start by employing \texttt{SIMUnet} to individually constrain the
\(W\) and \(Y\) operators with the ATLAS and CMS neutral-current (NC) high mass DY data from Run I and Run II. In the
next subsection we will be able to constrain them simultaneously 
thanks to the inclusion of charged-current (CC) HL-LHC projections. 

All details about the fit quality, once PDFs are allowed to vary
alongside the $W$ or the $Y$ parameters are given in
App.~\ref{app:fit}. We observe that the added
degrees of freedom result in a slightly better fit quality when
compared to the {\tt NNPDF4.0}-like SM baseline fit based on the
reduced dataset described in Sect.~4.2 and on SM theoretical
predictions. In particular the high-mass DY datasets that are sensitive
to SMEFT corrections display a more significant improvement in fit quality, largely driven
by the CMS measurements at \(7\) TeV,  owing to the large weight
carried by this dataset, which forms just under half the entire
high-mass DY data points hereby considered.

\begin{figure}[t]
        \centering
        \includegraphics[width=0.49\textwidth]{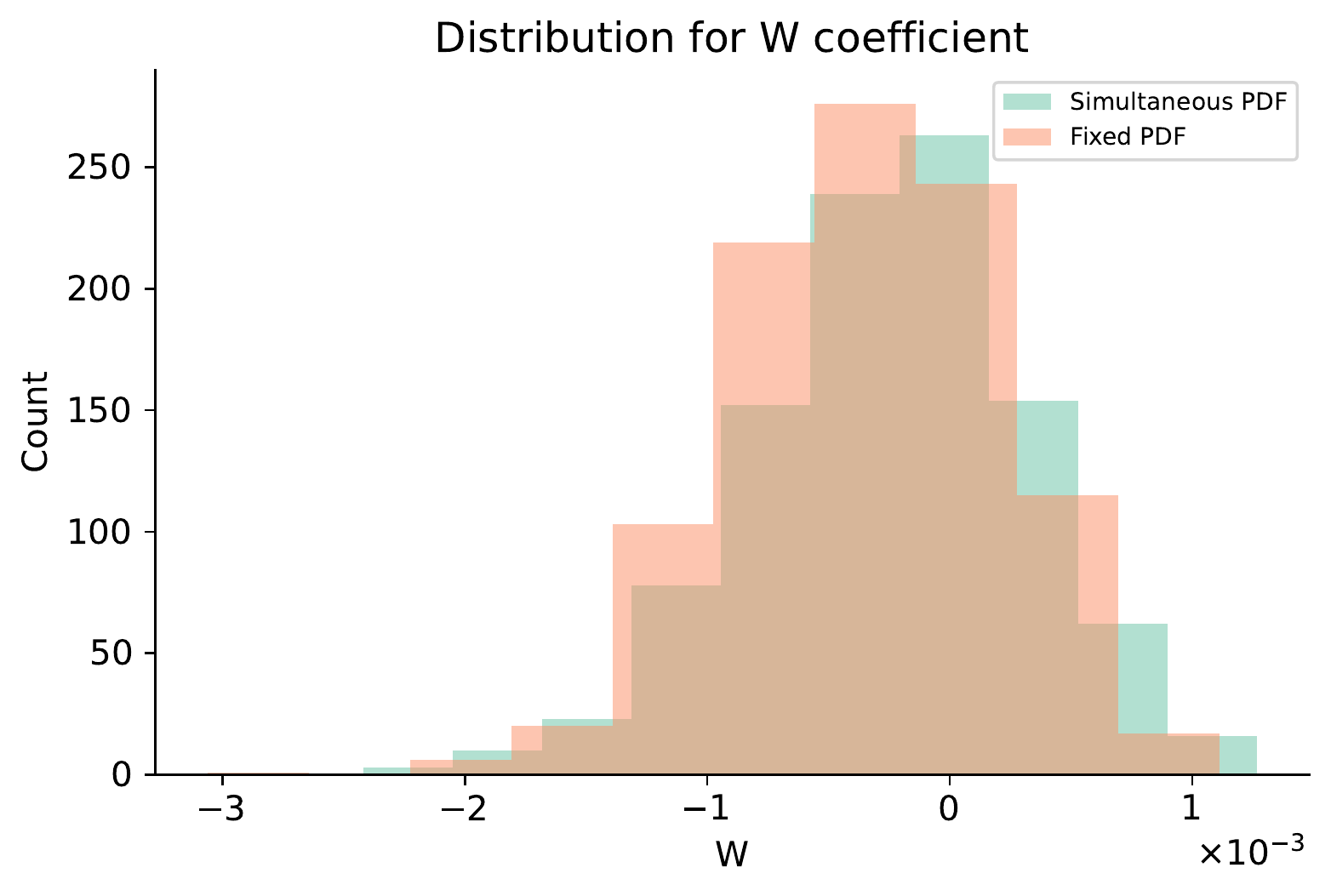}
     \includegraphics[width=0.49\textwidth]{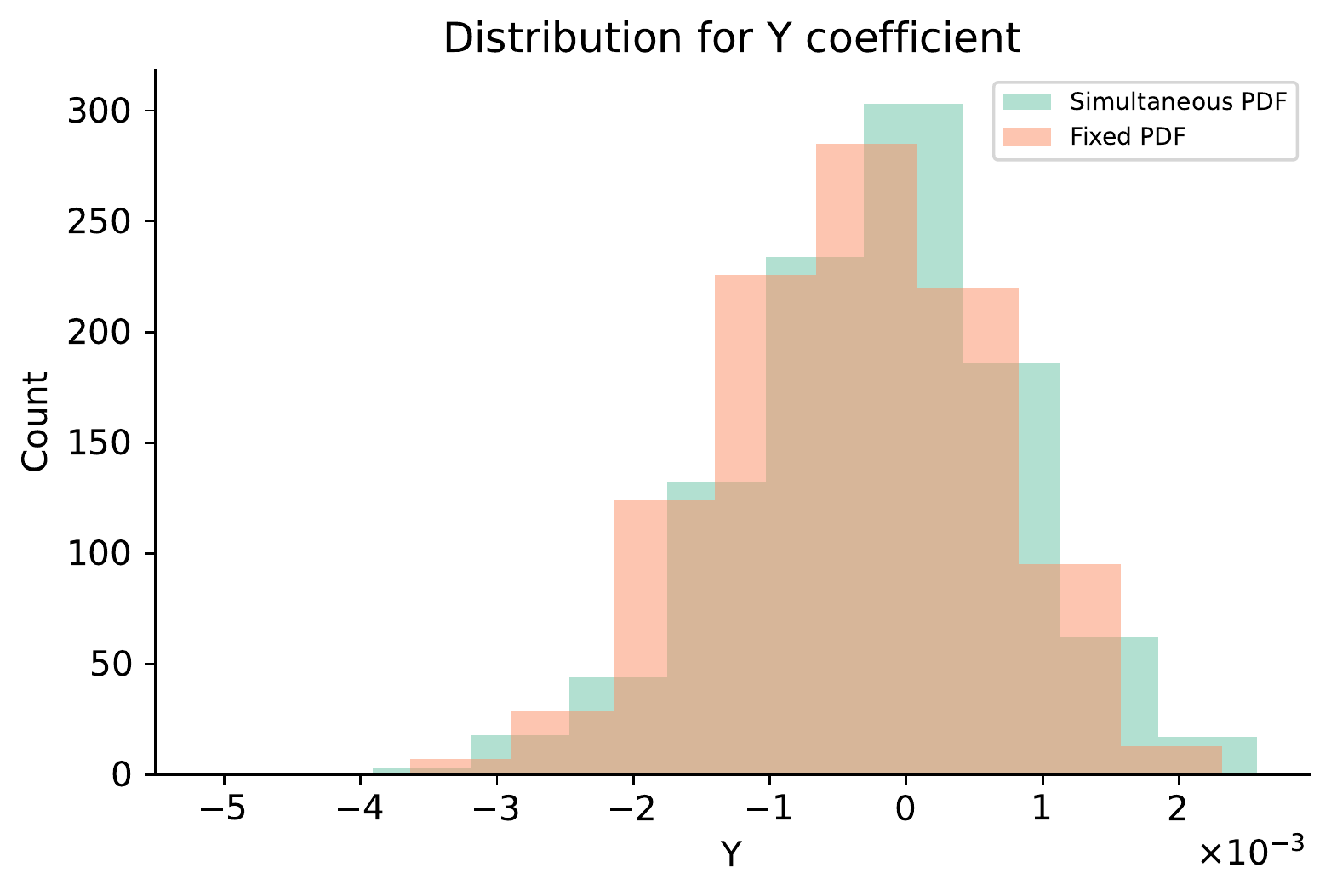}
        \caption{Distribution of best fit \(W\) (left panel)
          and \(Y\) values (right panel) obtained across 1000 replicas
          by fitting each of the Wilson coefficients alongside PDFs
          with the {\tt SIMUnet} methodology (green) compared to the
          distribution that one would obtain by keeping the PDFs
          fixed to the SM baseline (orange).
        \label{fig:WYresults}}
\end{figure}
In Figs.~\ref{fig:WYresults} we show the distribution of the optimal values of \(W\) and
\(Y\) determined for each of the 1000 replicas of the Monte Carlo
representation of the experimental data that we obtain  
during gradient descent in the simultaneous fit of either \(W\) or
\(Y\) and the PDFs. The best-fit values are normally distributed in keeping with the Monte-Carlo pseudodata
replica generation that the neural network replicas are fit to. We
compare the distribution (in green) with the one 
that we obtain by keeping the PDFs fixed to the SM baseline (in
orange), the latter obtained by using the methodology outlined in
Sect.~\ref{sec: fixedpdf}.
Both distributions are centred around zero, illustrating that the high-mass DY datasets are compatible with the SM-only
hypothesis, but admit non-zero values, with $Y$ being less constrained
by the data than $W$. The distribution of best fits obtained in a
simultaneous fit looks similar to the distribution of best fits
obtained by keeping the PDFs fixed to the SM baseline. 

\begin{table}[t]
  \centering
  \begin{tabular}{l|cccc}
    & $\quad$ SM PDFs $\quad$  & SMEFT PDFs & best-fit shift  & broadening  \\
    \toprule
    $W\times 10^4$ (68\% CL) & $[-9.0,2.1] $& $[-8.2,3.2] $  &
                                                                      $+1.0$  & +2.6\% \\
    $W\times 10^4$ (95\% CL) & $[-14.5,7.6] $& $[-13.9,8.9] $  &
                                                                      $+1.0$  & +3.2\% \\
    \midrule
    $Y\times 10^4$ (68\% CL) & $[-13.8, 5.7] $ &  $[-12.2,7.9]$ & $+1.9$ & +3.1\% \\
    $Y\times 10^4$ (95\% CL) & $[-23.5, 15.5] $ &  $[-22.2,18.0]$ & $+1.9$ &
                                                                     +3.1\% \\
    \bottomrule
  \end{tabular}
  \caption{\label{tab:w-y} The 68\% and 95\% CL bounds on the $W$ and $Y$
    parameters obtained either for a fit in which PDFs are kept fixed (SM
    PDFs) or in a fit in which PDFs are fitted simultaneously with
    either $W$ or $Y$ (SMEFT PDFs).
    The fourth and fifth column indicate the absolute shift in best-fit values,
 Eq.~(\ref{eq:shift}),
    and the percentage broadening of the SMEFT bounds, Eq.~(\ref{eq:broadening}),
    when the PDFs are allowed to change alongside the Wilson coefficients.}
\end{table}
For a more quantitative comparison, in Table~\ref{tab:w-y} we compare the bounds for the individual $W$ and $Y$
fits obtained in the simultaneous fits to those obtained by keeping 
the PDFs fixed to the SM baseline.  We indicate the shift and the broadening of the bounds
that are obtained once the Wilson coefficients are fitted alongside
PDFs, by defining them as
\begin{equation}
\label{eq:shift}
{\rm best~fit~shift}\equiv \left( \langle W\rangle \Big|_{\rm
    SMEFT\,PDFs}-\langle W\rangle \Big|_{\rm SM\,PDFs}\right) \, ,
\end{equation}
\begin{equation}
\label{eq:broadening}
{\rm broadening}\equiv \left( \Delta_W\Big|_{\rm SMEFT\,PDFs}-\Delta_W\Big|_{\rm SM\,PDFs}\right)\bigg/\Delta_W\Big|_{\rm SM\,PDFs} \, ,
\end{equation}
where $\Delta$ is the size of either the 68\% C.L. or the 95\% C.L.,
depending on the bounds we consider. The same definitions apply for
the $Y$ parameter.
%
As in ~\cite{Greljo:2021kvv}, the effect of fitting the
Wilson coefficients simultaneously with the effects that non-zero
coefficients have on PDFs changes the interpretation of high-mass DY
constraints in a moderate way, with the bounds of the simultaneous fit
being within the uncertainty of the current analysis and only slightly looser (by a factor
around 3\%) than those obtained by keeping the PDFs fixed. 

\begin{figure}[t]
  \centering
          \includegraphics[width=0.49\textwidth]{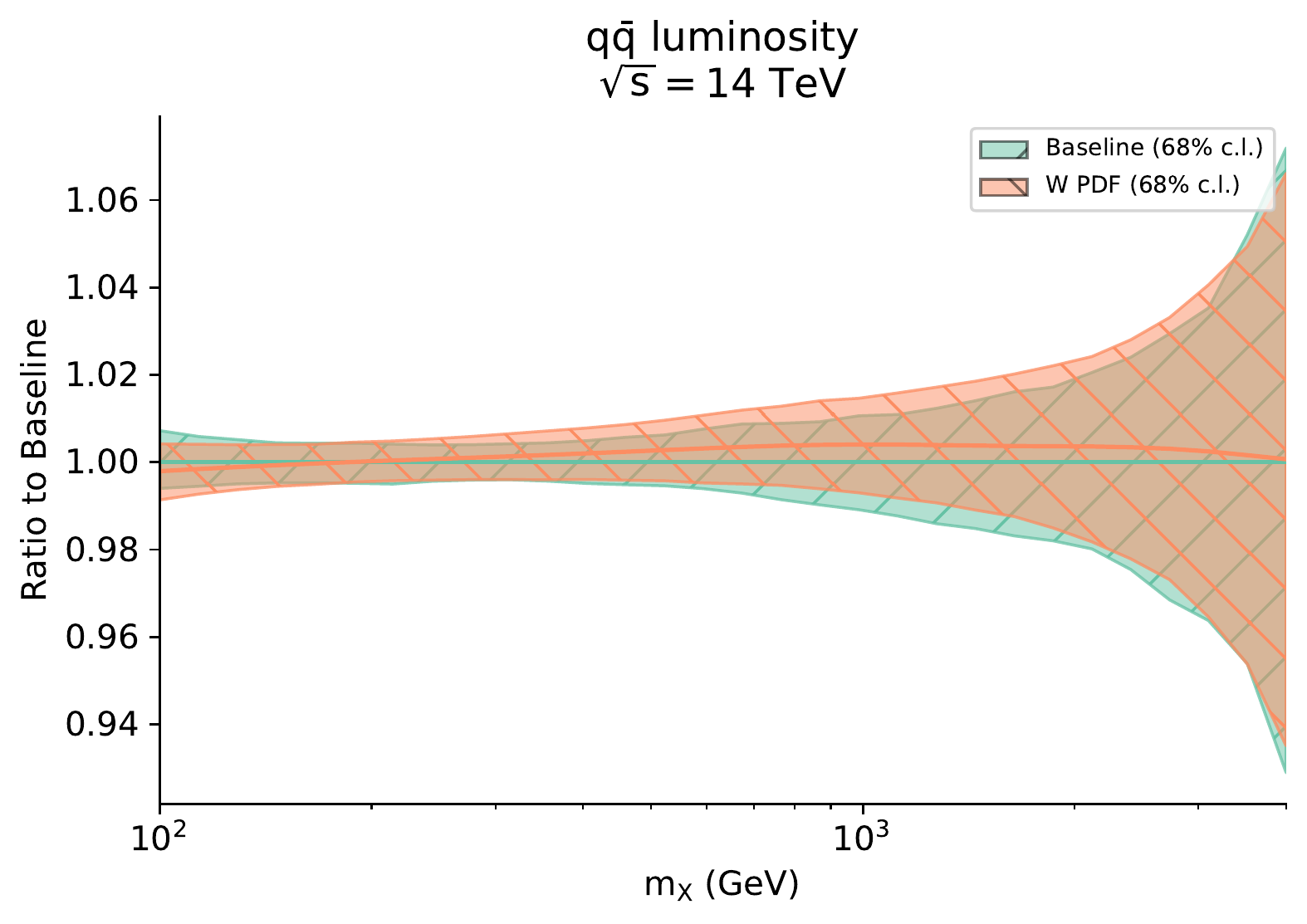}
        \includegraphics[width=0.49\textwidth]{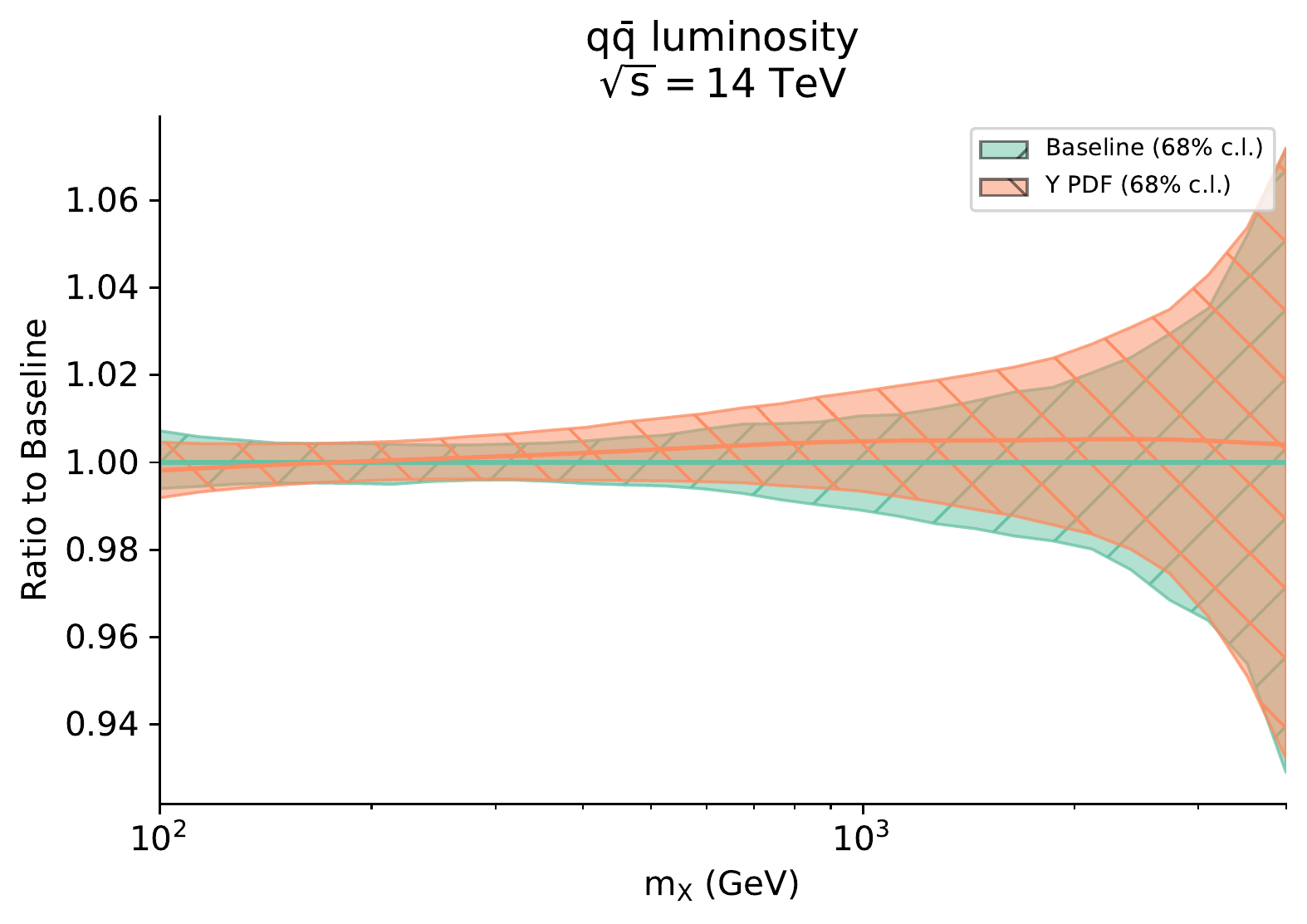}
        \caption{The one dimensional luminosity in the \(q\bar{q}\) channel for
        a PDF fitted in the presence of a non-zero \(W\) operator
        (left panel) or
        of a non-zero \(Y\) operator (right panel) shown in
        orange, normalized to the SM baseline PDF shown in green.
        \label{fig: WYlumiresults}}
\end{figure}

In Figs.~\ref{fig: WYlumiresults} we show the quark-antiquark channel
luminosity plots defined in Eq.~\eqref{eq:lumidef} with
the error bands showing \(68\%\) C.~L., normalised to the
baseline SM PDFs. We notice that, while in the previous analysis of
Ref.~\cite{Greljo:2021kvv}, we could only produce sets of PDFs obtained
at fixed values of $W$ and $Y$ (corresponding to the Benchmark Points that were
taken under consideration), here we do really produce a set of PDFs
obtained out of a simultaneous fit alongside the Wilson coefficients.
We see the
luminosity modification due to the simultaneous fit remains moderate,
with a slightly larger deviation found at
higher values of the produced lepton invariant mass, where the PDFs
exhibit slightly larger uncertainties. This is indeed a result
consistent with the indications given in Ref.~\cite{Greljo:2021kvv}.
Indeed, the dominant luminosity channel for NC DY is \(u\bar{u}\) and \(d\bar{d}\) with the
valence quark distribution being strongly constrained by DIS and the
SMEFT modification being small relative to the experimental
uncertainties of the highest invariant mass bins probed by current
experimental data. 
Again, this finding shows that the interplay between EFT effects and PDFs remains moderate
when one performs a truly simultaneous determination of both.

%
%
\begin{figure}[h]
    \centering
    \includegraphics[width=\textwidth]{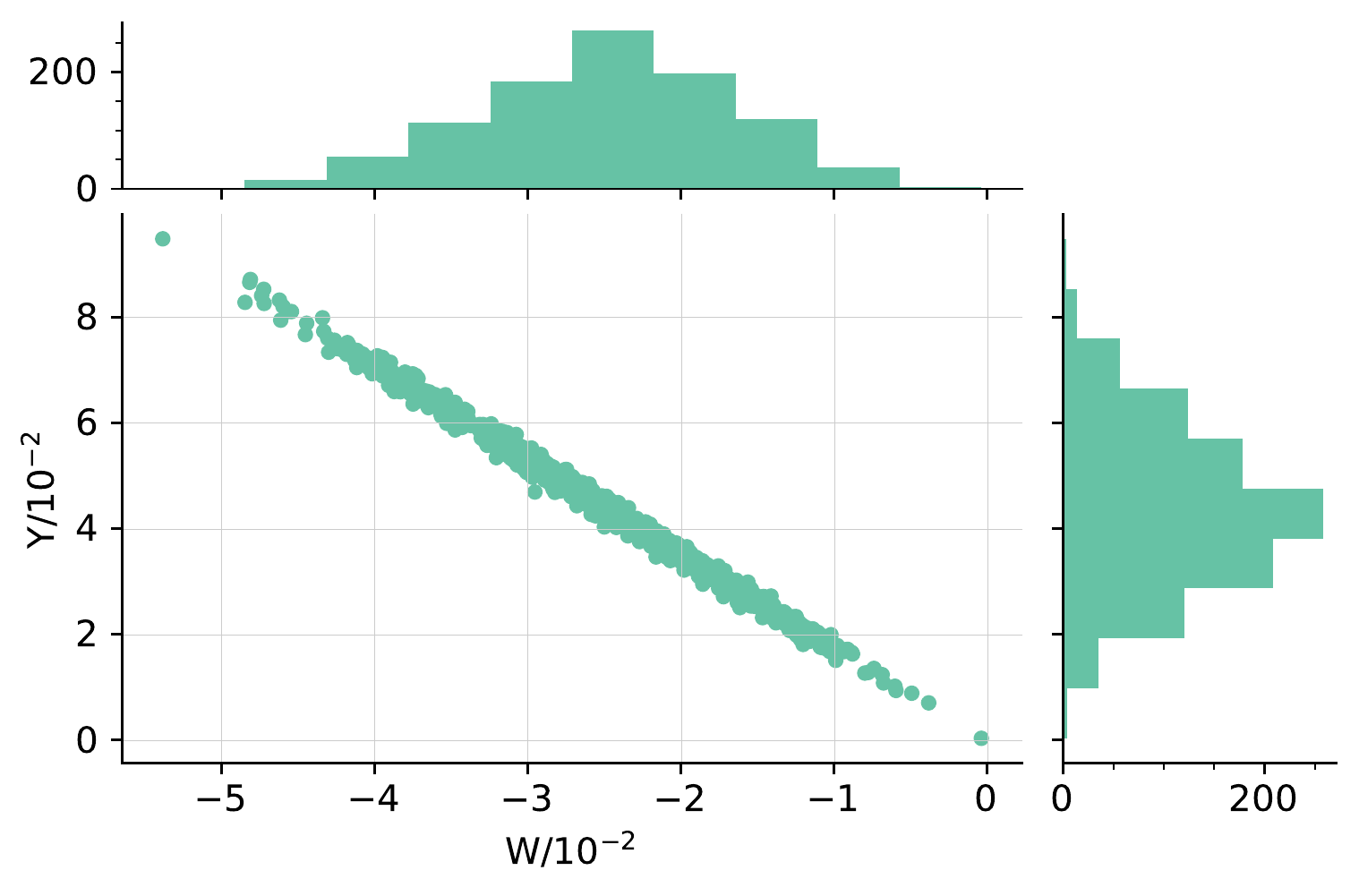}
    \caption{Scatter plot for the best fit values of \((W, Y)\) per replica using ATLAS and CMS high mass
    Drell-Yan data, which are exclusively NC observables. The upper and rightmost panels are histograms in their respective directions.
    A clear flat direction has been detected along with a strong anti-correlation.
    \label{fig: flatdirection}}
\end{figure}
It is interesting to observe that, if we try to fit $W$ and $Y$ at the same time using only the current
NC DY data, we identify both the flat direction and the strong
anti-correlation between $W$ and $Y$, which are known to exist in this
case~\cite{Farina:2016rws}.
Results are shown in Fig.~\ref{fig: flatdirection}. Both the flat direction and the anti-correlation can
be retrieved within the framework of the {\tt SIMUnet} methodology, without the user having
to be aware of the existence or particular nature of any flat directions which
may exist. Indeed, the optimizer cannot preferentially differentiate one point from
another within this landscape valley and so the points are distributed tightly
along the flat direction.
The preference for the upper left quadrant in the \(W\)-\(Y\) plane
is consistent with the findings of Ref.~\cite{Farina:2016rws} with the failure
to capture the origin possibly due to the lack of inclusion of 
\(\mathcal{O}(1/\Lambda^4)\) terms in the partonic cross section which is known
to ease such tensions when only \(\mathcal{O}(1/\Lambda^2)\)
terms are present, as it was pointed out in
Ref.~\cite{Boughezal:2021tih}, where the effect of dim-8 operator is
carefully assessed.

\subsection{Inclusion of the HL-LHC projections \label{sec: hllhc_fit}}

The flat direction illustrated in Fig.~\ref{fig: flatdirection}
can be eliminated with the inclusion of Charged Current (CC) DY 
data~\cite{Greljo:2021kvv}. 
%
No unfolded measurements of the high-mass transverse mass $m_T$
distribution have been yet released at 13 TeV, thus we will base our
analysis on the High Luminosity LHC (HL-LHC)
high-mass Drell-Yan projections that we produced in
Ref.~\cite{Greljo:2021kvv}, inspired by the HL-LHC projections studied
in Ref.~\cite{AbdulKhalek:2018rok}. The invariant mass distribution
projections are generated at \(\sqrt{s} = 14\) TeV, assuming an
integrated luminosity of \(\mathcal{L} = 6 \text{ ab}^{-1}\) (
\(3 \text{ ab}^{-1}\) collected by ATLAS and  \(3
\text{ ab}^{-1}\) by CMS).
Both in the case of NC and CC Drell-Yan cross sections, the pseudodata
were generated using the {\tt MadGraph5\_aMCatNLO} NLO Monte Carlo event
generator~\cite{Frederix:2018nkq} with additional $K$-factors to include the NNLO QCD and
NLO EW corrections. The pseudodata consist of four datasets
(associated with NC/CC distributions with muons/electrons in the final
state), each comprising 16 bins in the $m_{ll}$
invariant mass distribution or transverse mass $m_T$ distributions
with both $m_{ll}$ and $m_T$ greater than 500 GeV , with the highest energy bins reaching $m_{ll}=4$
TeV ($m_T=3.5$ TeV) for NC (CC) data.
The rationale behind the choice of number of bins and of the width of each bin was outlined 
in Ref.~\cite{Greljo:2021kvv}, and stemmed from the requirement that the expected number of events
per bin was big enough to ensure the applicability of Gaussian
statistics. The choice of binning for the $m_{ll}$ ($m_T$ ) distribution at the
HL-LHC is displayed in Fig.~5.1 of Ref.~\cite{Greljo:2021kvv}.

\begin{figure}[h]
        \centering
        \includegraphics[width=0.75\textwidth]{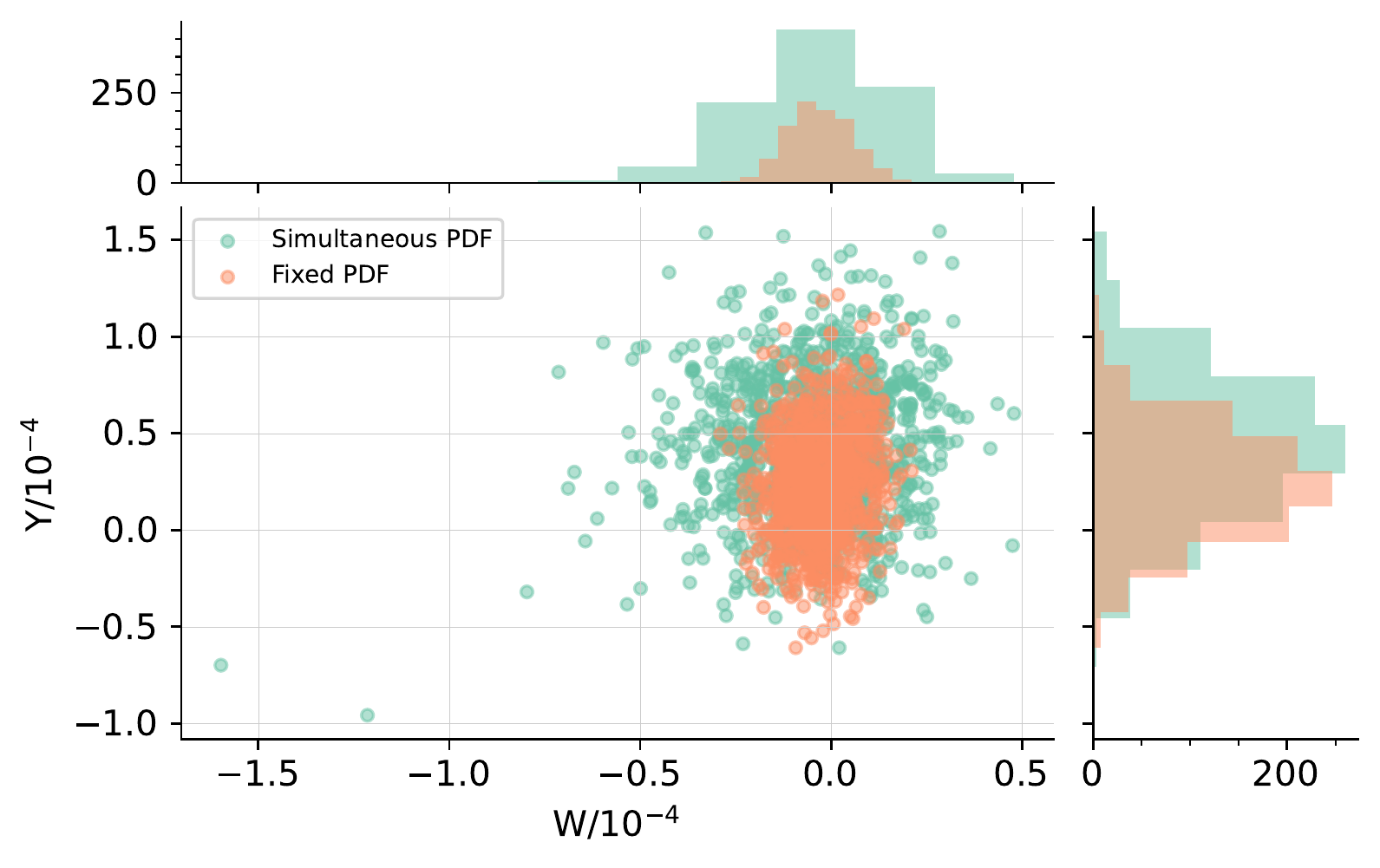}
        \caption{Scatter plot for best fit tuples of
        \((W, Y)\) for each replica obtained in the simultaneous fit
        (green) compared to those obtained when PDFs are kept fixed to
        the SM baseline (orange). The upper and rightmost panels are
        histograms in their respective directions. \label{fig: hllhc_scatter}}
\end{figure}
We performed a simultaneous fit of the PDFs and the two \((W, Y)\)
parameters by adding two trainable edges in the combination layer
displayed in Fig.~3.2, and appending the aforementioned
HL-LHC projected data to the data discussed in Sect.~\ref{sec: experimental_data}.
The best fit values of \((W, Y)\) obtained for each of the 1000
replicas of the Monte Carlo ensemble are plotted in Fig.~\ref{fig:
  hllhc_scatter}.
We see that not only is the flat direction of Fig.~\ref{fig: flatdirection} broken, but the
HL-LHC projections heavily favour vanishing Wilson coefficients. Indeed,
the origin is now covered in both the \(W\) and \(Y\) axes and much
more heavily constrained, enjoying roughly two orders
of magnitude tighter constraints in both directions. We notice a slightly favorable pull towards the upper-left quadrant
in the \(W\)-\(Y\) plane, that the ATLAS and CMS datasets seem to 
prefer. Comparing
the best-fit distribution that we get in a simultaneous fit (displayed
in green) to the one
we get in a fit in which PDFs are kept fixed to the SM baseline
(displayed in orange) we can see that the bounds are visibly tighter
once PDFs are kept fixed to the SM baseline and are not allowed to
consistently vary alongside the Wilson coefficients.

\begin{table}[t]
  \centering
  \begin{tabular}{l|cccc}
    & $\quad$ SM PDFs $\quad$  & SMEFT PDFs  & best-fit shift  & broadening  \\
    \toprule
    $W\times 10^5$ (68\% CL) & $[-1.1,0.5] $& $[-2.4,1.5] $  &
                                                                      $-0.2$  & +144\% \\
    $W\times 10^5$ (95\% CL) & $[-2.0,1.4] $& $[-4.3,3.4] $  &
                                                                      $-0.2$  & +126\% \\
    \midrule
    $Y\times 10^5$ (68\% CL) & $[-0.4, 5.2] $ &  $[0.6,8.0]$ & $+1.9$ & +32\% \\
    $Y\times 10^5$ (95\% CL) & $[-3.2,8.1] $ &  $[-3.1,11.7]$ & $+1.9$ &
                                                                     +31\% \\
    \bottomrule
  \end{tabular}
  \caption{\label{tab:hl-wy} Same as Table~\ref{tab:w-y} for the 68\%
    C.L. and 95\% C.L. marginalised bounds on the $W$ and $Y$ parameters
    obtained from the two-dimensional $(W,Y)$ fits that include the HL-LHC pseudo-data
for NC and CC Drell-Yan distributions. }
\end{table}
In Table~\ref{tab:hl-wy} we compare the bounds for the individual $W$ and $Y$
fits obtained in the simultaneous fits to those obtained by keeping 
the PDFs fixed to the SM baseline. We indicate the shift and the broadening of the bounds
according to the definitions given in Eqs.~\eqref{eq:shift}-\eqref{eq:broadening}. Consistently to what was found in
our previous study of Ref.~\cite{Greljo:2021kvv} we observe that including high-mass data at the LHC both in
a fit of PDFs and in a fit of SMEFT coefficients and neglecting the interplay between
them could result in a significant underestimate of the uncertainties associated to the SMEFT
parameters. Indeed, the marginalised bounds on the $W$($Y$)
parameter increase by about 150\% (30\%) once a simultaneous fit of the PDFs
and the $(W,Y)$ parameters is performed. The broadening is smaller
than observed in Ref.~\cite{Greljo:2021kvv} but it is still extremely
significant. A detailed comparison is given in Sect.~\ref{sec: overview}.

\begin{figure}[h]
        \centering
        \includegraphics[width=0.75\textwidth]{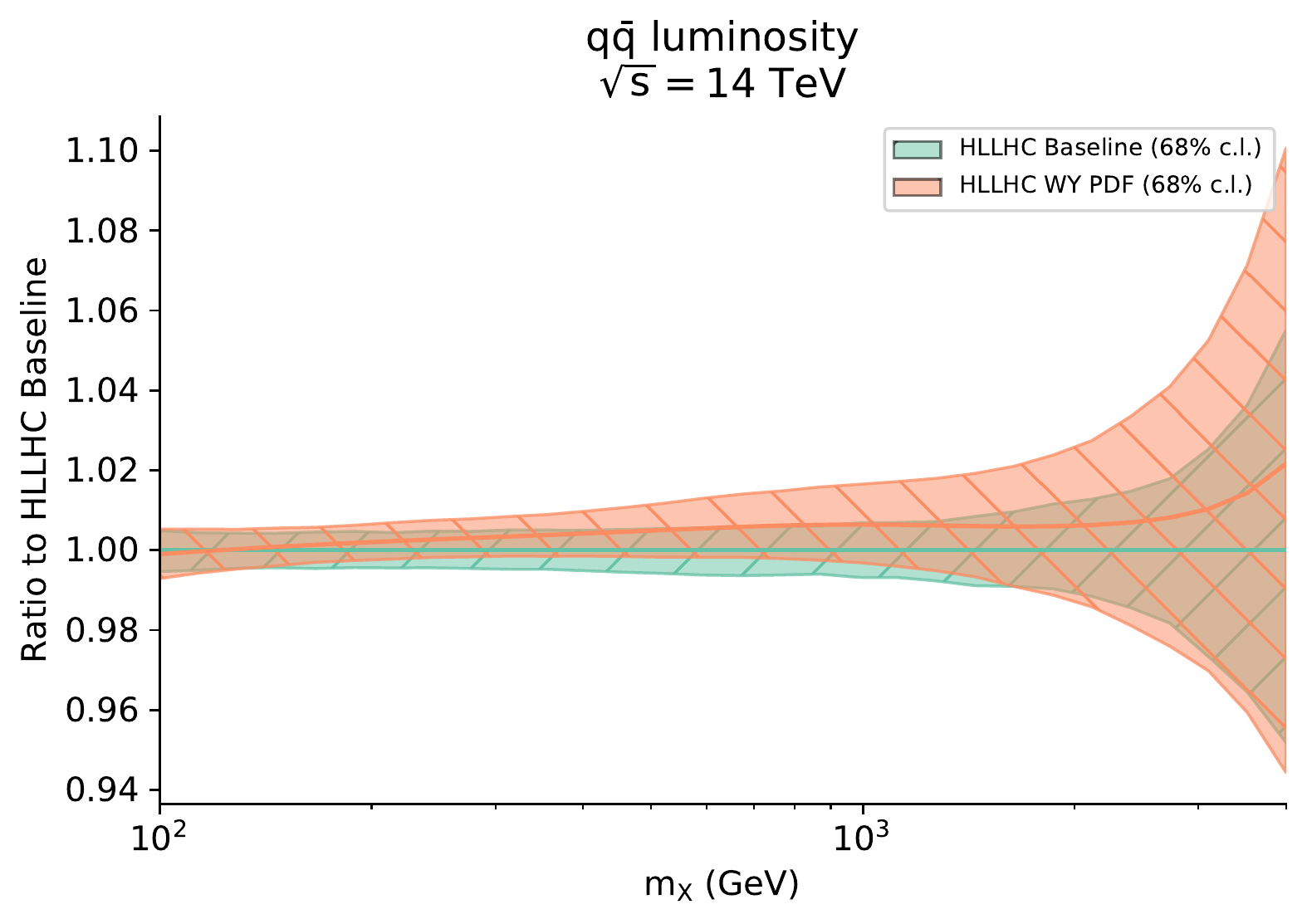}
        \caption{\(q\bar{q}\) luminosity channel of the PDF fitted in the
        presence of \(W\) and \(Y\) parameters fitted to the ATLAS and CMS high
        mass Drell-Yan data and the NC and CC DY HL-LHC projections, normalized
        to the appropriate baseline SM PDF. \label{fig: hllhc_lumi}}
\end{figure}
As far ar the quark-antiquark luminosity is concerned, we can see in Fig.~\ref{fig: hllhc_lumi} that once the PDFs are fitted
simultaneously with the $(W,Y)$ parameters two things happen. First of
all the central values of the luminosity shift upwards for large values of the invariant mass
($M_X\gtrsim$ 1 TeV) towards the edge of the 68\% C.L. error
band. Second the error band significantly increases. The shift in the central value is compatible to what we observed before, namely that the luminosity
plots, once PDFs are fitted at some representative values of the $W$
and $Y$ parameters, do change significantly, well outside the
$1\sigma$ error band of the SM PDFs, while the PDF uncertainties themselves are unchanged. However,
in this case the PDF uncertainty does increase because here we are actually
performing a simultaneous fit and, as a result, the PDF error band 
increases proportionally to the width of the range of $W$ and $Y$ that the data
allow. This is a very interesting result and shows that PDF error
bands at large-$x$ inherently have an extra source of theory
uncertainty related with possible BSM effects that the data do not
exclude. 

\subsection{Results for Benchmark Scenario II\label{sec: bs2}}

In this section we employ the left-handed muon-philic operator \(\mathbf{C}_{33}^{D\mu}\) to showcase
our methodology's ability to constrain Wilson coefficients whilst accounting for
the effect of quadratic dim-6 effects using the approach discussed in Sect.~\ref{sec: dim8}.
The fact that this second Benchmark SMEFT scenario is effectively unconstrained
\cite{Greljo:2021kvv} at the linear level serves to act as the ideal setting to
assess the ability to fit EFT operators whilst simultaneously accounting for
their quadratic contributions, as in Eq.~\eqref{eq:
  quadratic_modified_theory}. 
Furthermore, since the \(\mathbf{C}_{33}^{D\mu }\) operator affects only muon final state
observables, while electron final states are described by the SM,
the combination layer uses only those datasets that have a muon in the final
state to constrain \(\mathbf{C}_{33}^{D\mu}\). In particular, the CMS high-mass
measurements at 13 TeV, which up until now have been for the combined decay
channel, is now separated into the electron and muon channels. As a result of
splitting this particular dataset into separate channels, we have accordingly
generated a new baseline PDF used in the comparisons.

\begin{figure}[t]
        \centering
        \includegraphics[width=0.49\textwidth]{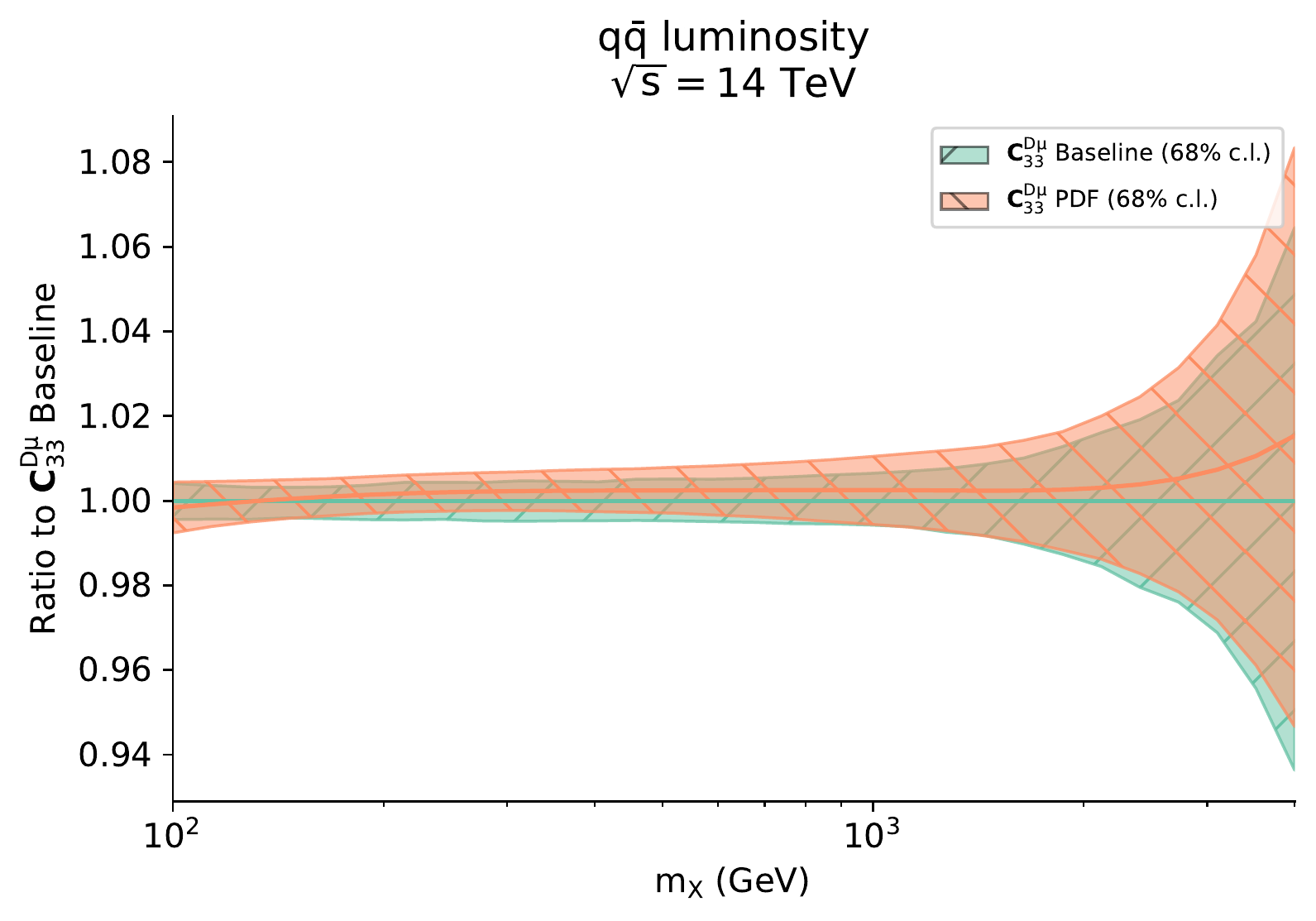}
        \includegraphics[width=0.49\textwidth]{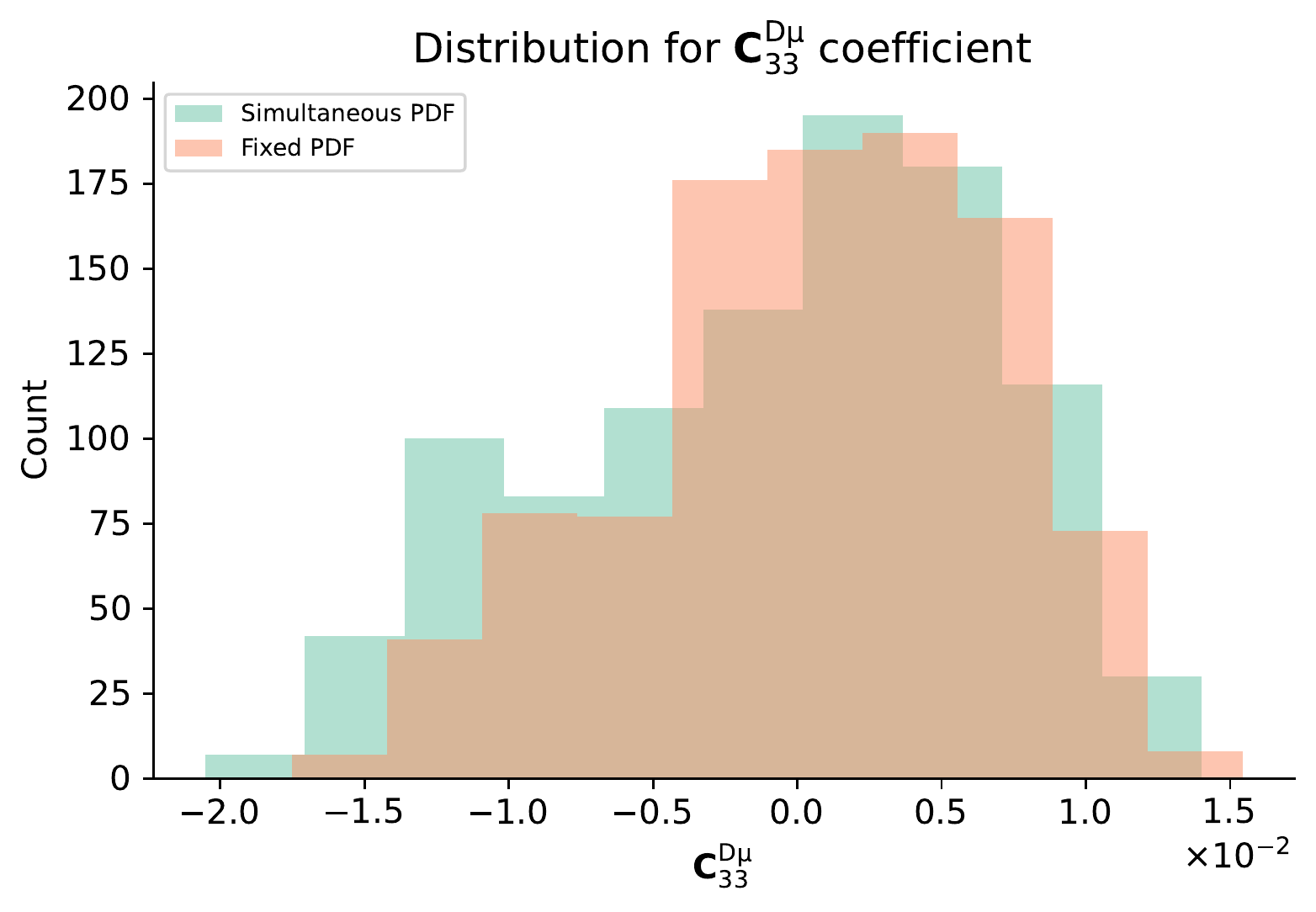}
        \caption{Left: \(q\bar{q}\) luminosity channel of the PDF fitted in the
        presence of the \(\mathbf{C}_{33}^{D\mu}\) parameter normalized to the
        baseline SM PDF.  Right: histogram plot for best fit values for
        \(\mathbf{C}_{33}^{D\mu}\) for each replica. The best-fit
        distribution over 1000 replicas obtained out of a simultaneous fit (green) is
        compared to the one obtained by keeping PDFs fixed to the SM
        baseline (orange). \label{fig: Cb_hist}}
\end{figure}
The best-fit values of  \(\mathbf{C}_{33}^{D\mu}\)  across the 1000
replicas in the fit are shown on the right-hand panel of 
Fig.~\ref{fig: Cb_hist}. We compare the distribution obtained
out of a simultaneous fit (in green) with the one obtained once PDFs
are kept fixed to the SM baseline (in orange). We
see that the distribution of best fit values is centred at the origin,
although it is skewed towards \(\mathbf{C}_{33}^{D\mu}\approx 5\cdot
10^{-3}\). The shape of the distribution is what we expect once
quadratic terms are allowed in the fit of the Wilson coefficients by
keeping PDF fixed~\cite{Ethier:2021bye}, and it is interesting to see
this feature not only holds but it is actually enhanced once the Wilson coefficient is fitted alongside
PDFs (green histogram). 

\begin{table}[t]
  \centering
  \begin{tabular}{l|cccc}
    & $\quad$ SM PDFs $\quad$  & SMEFT PDFs  & best-fit shift  & broadening  \\
    \toprule
    $\mathbf{C}_{33}^{D\mu}\times 10^3$ (68\% CL) & $[-5.6,6.9] $& $[-8.0,6.7] $  &
                                                                      $-0.9$  & +18\% \\
    $\mathbf{C}_{33}^{D\mu}\times 10^3$ (95\% CL) & $[-11.9,13.1] $& $[-15.3,14.0] $  &
                                                                      $-0.9$  & +17\% \\
 
    \bottomrule
  \end{tabular}
  \caption{\label{tab:cb} Same as Table~\ref{tab:w-y} for the 68\%
    CL and 95\% CL bounds on the \(\mathbf{C}_{33}^{D\mu}\) Wilson
    Coefficient, including both linear and quadratic terms in the SMEFT expansion.}
\end{table}
For a quantitative comparison of the bounds obtained in the
simultaneous fit to those obtained in a Wilson coefficient-only fit,
in Table \ref{tab:cb} we show the bounds that we obtain in the two
cases. The interplay between PDFs and SMEFT coefficients is quite moderate in this particular scenario. 
In contrast with the marked effects in Benchmark Scenario I, in
Benchmark Scenario II the obtained bounds on this Wilson coefficient would loosen by around 20\%. The
origin of this rather different behaviour can be traced back to the
fact that in this scenario the electron channel data do not receive
EFT corrections, and hence all the information that they provide makes it possible to exclusively constrain the PDFs. The muon channel
distributions then determine the allowed range for the \(\mathbf{C}_{33}^{D\mu}\) , restricted by the well-constrained
large-$x$ quarks and antiquark PDFs from the electron data. This claim
is backed by the luminosity comparison, displayed on the left-hand
panel of Fig.~\ref{fig: Cb_hist}: the shift and the increase in the
PDF uncertainty of the $q\bar{q}$ luminosity are visible, but less
enhanced than in the first Benchmark Scenario. 

\subsection{Results overview\label{sec: overview}}

To conclude this section, we present an overview of our results. We
have seen in Sect.~\ref{sec: WY_fit} that the current high-mass Drell-Yan data from
LHC Run I and Run II do not allow to simultaneously fit $W$ and $Y$ in
Benchmark Scenario I and loosely constrain \(\mathbf{C}_{33}^{D\mu}\)
in Benchmark Scenario II. This is mostly due to the lack of unfolded
CC Drell-Yan data, which would remove the flat direction that our algorithm is able to
detect (see Fig.~\ref{fig: flatdirection}). Moreover, our analysis
confirms what was outlined in Ref.~\cite{Greljo:2021kvv},
namely that the interplay between the individual fits of $W$ and $Y$
and the fit of the large-$x$ quark distributions is mild at the level
of current DY data. However, once the high-statistics data projections from
the HL-LHC are included, the flat direction disappears and one is able
to obtain strong constraints both on the $(W,Y)$ plane~\cite{Farina:2016rws, Greljo:2021kvv} and on the
individual \(\mathbf{C}_{33}^{D\mu}\)
coefficient~\cite{Greljo:2017vvb, Greljo:2021kvv}.  From the point of
view of showcasing our methodology, the two scenarios are interesting,
as in the Benchmark Scenario I, once the HL-LHC projections are
included, we can simultaneously fit both $W$ and $Y$ alongside the
PDFs, including only the linear SMEFT corrections while in the Benchmark
Scenario II we can fit individually \(\mathbf{C}_{33}^{D\mu}\) alongside the
PDFs, including both the linear and the quadratic SMEFT corrections. In the
first scenario, we observe that there is a strong interplay between
SMEFT and PDF fits, as the bounds for the SMEFT coefficients
significantly broaden  once PDFs are allowed to vary alongside $W$ and $Y$ (see Fig.~\ref{fig: hllhc_scatter}) and the PDFs
themselves display a sizeable shift (see Fig.~\ref{fig: hllhc_lumi}). The
interplay is more moderate in the second scenario, given that BSM
effects only affect the data with muons in the final states, while the
data with electrons in the final states constrain the large-$x$ quark
distributions. 

In this section, we focus on the results obtained including the
HL-LHC projections. We first explore the correlation patterns between the PDFs and the SMEFT
coefficients in both Benchmark Scenarios. 
These correlation coefficients can be evaluated as in
Ref.~\cite{Carrazza:2019sec}. 
For example in the case of the gluon and the $W$ Wilson coefficient,
the correlation coefficient is defined as follows
\begin{equation}
\label{eq:correlationL2CT}
\rho\lp W, g(x,Q)\rp=\frac{\la W^{\rm (best-fit)}g(x,Q)\ra - \la W^{\rm (best-fit)}\ra \la g(x,Q)\ra
}{\sqrt{\la W^{\rm (best-fit)2}\ra - \la W^{\rm (best-fit)}\ra^2}\sqrt{\la g(x,Q)^2\ra - \la g(x,Q)\ra^2} } \, ,
\end{equation}
where  $W^{\rm (best-fit)}$ is the best-fit of the $W$ coefficient
for each replica in the simultaneous fit, in which PDFs are
allowed to vary alongside $W$ and $Y$.  
In Eq.~(\ref{eq:correlationL2CT}), averages are computed over the $N_{\rm rep}=1000$ replicas.
This correlation coefficient provides a measure of how the variations
in the PDFs translate into modifications of the best fit value
of the Wilson coefficients.
\begin{figure}[tb]
         \centering
         \includegraphics[width=0.48\textwidth]{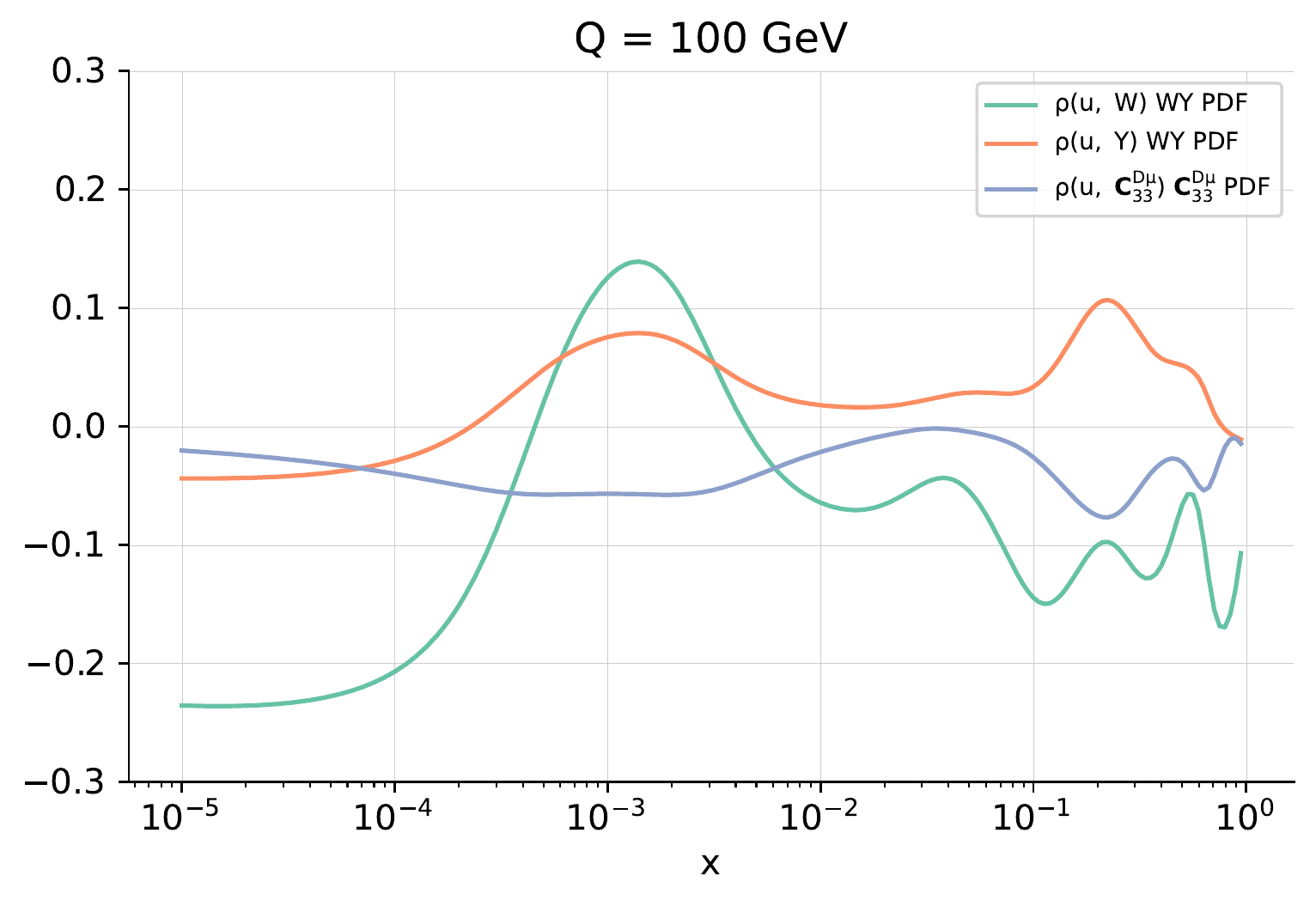}
         \includegraphics[width=0.48\textwidth]{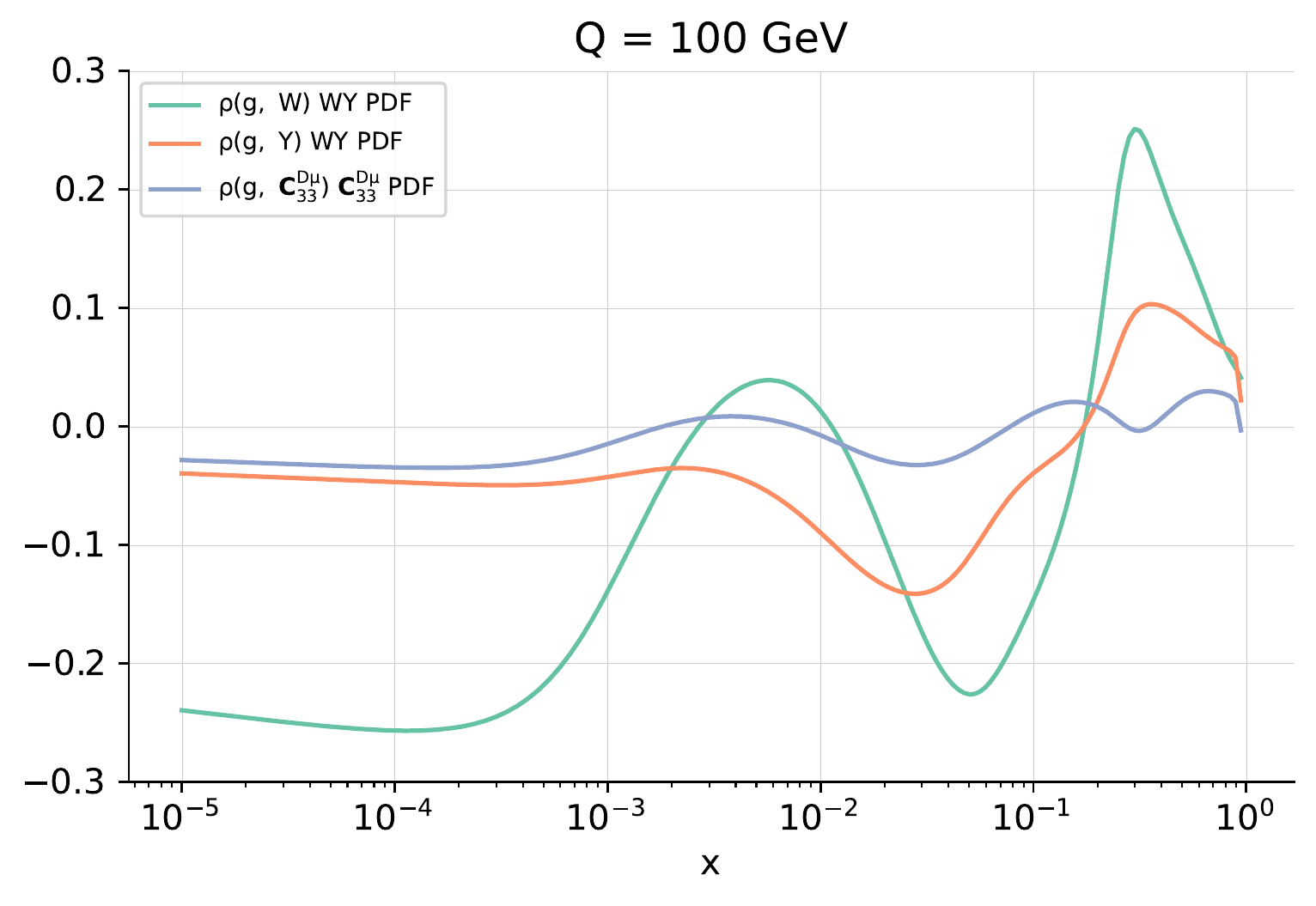}
         \caption{\label{fig:correlations} Correlation coefficients,
           defined in Eq.~\eqref{eq:correlationL2CT}, between the $W$
           (green), $Y$ (orange), $\mathbf{C}_{33}^{D\mu}$ (blue) 
           and the up quark PDF (left panel) and gluon PDF (right
           panel) computed at $Q=100$ GeV as a function of the
           momentum fraction $x$. The PDFs used to compute the
           correlation coefficients are those obtained in the
           simultaneous fit of $(W,Y)$ and the PDFs 
           described in Sect.~\ref{sec: hllhc_fit} (in the case of the green and orange
         curves) and those obtained in the simultaneous fit of
         $\mathbf{C}_{33}^{D\mu}$ and the PDFs described in Sect.~\ref{sec: bs2}
         (in the case of the blue curve).}
\end{figure}
In Fig.~\ref{fig:correlations} we show these correlation coefficients
between the up valence quark and the gluon PDFs at $Q=100$ GeV.
Each of the curves corresponds to one of Wilson coefficients
considered in the HL-LHC analysis, the $W$ and $Y$ being fitted
simultaneously alongside the PDFs (hence the ``WY PDF'' label) and the
$\mathbf{C}_{33}^{D\mu}$ being fitted individually alongside the
PDFs (hence the ``$\mathbf{C}_{33}^{D\mu}$ PDF'' label) . 
Although correlations are moderate, it is interesting to observe that the correlation/anticorrelation between $W$ and the
PDFs is much stronger than the correlation with the other Wilson
coefficients. This explains why the broadening of the bounds
in the $W$ direction are more marked than those in the $Y$
directions. Moreover the correlation is stronger in the large-$x$
region and for the gluon, which implies an anticorrelation with the
singlet in the same region, which is where we observe a broadening of
PDF uncertainties. 
%
%
%

Throughout the paper we found that the results obtained with the {\tt
  SIMUnet} methodology are in line with those presented in
Ref.~\cite{Greljo:2021kvv}. 
In Table~\ref{tab:summary} we make this comparison more quantitative,
by focussing on the results in which the effects of the interplay
between the SMEFT coefficients and the PDFs are more visible, namely
the fits obtained by using the NC and CC projections from the HL-LHC. 
The results are comparable, although the effect of fitting the
Wilson coefficients along with the PDFs is more moderate. This is not
surprising, as there are two crucial differences in the two analyses. On the one
 hand the PDF set that we use here in the fixed SM PDF case is different compared
 to the one we used in the previous analysis, being based on the same
 dataset as before but on the {\tt NNPDF4.0} methodology rather than
 the {\tt NNPDF3.1} methodology.
 Secondly, because the previous methodology was based on the use of Benchmark Points in the Wilson coefficients
 parameter space, we determined the bounds on the parameters by
 using the partial $\chi^2$  including only the data affected by the
 SMEFT corrections, rather than the global $\chi^2$. This
 approximation was forced because the statistical fluctuations of the
 global $\chi^2$ were found to
 be significantly larger than those of the partial $\chi^2$ and could only be tamed by running a very
 large batch of replicas for each Benchmark Point and by increasing the density of Benchmark
 Points in the region that is explored. This approximation is no longer necessary within
 {\tt SIMUnet}, because we no longer rely on the interpolation over Benchmark Points, rather we
 perform a truly simultaneous fit of the PDFs and the Wilson
 coefficients based on the global $\chi^2$.
 Additionally, the minimizer
 used in this study employs a momentum driven stochastic gradient descent based
 algorithm (Nesterov-accelerated adaptive moment estimation \cite{kingma2017adam,
 Nesterov1983AMF} available from the \texttt{Keras} library), while in
Ref.~\cite{Greljo:2021kvv} we employed a more traditional genetic algorithm approach
 using legacy in-house implementations. As such, one in general expects to achieve
 an improved fit quality with our methodology. Finally, the bounds quoted in our
 study are defined using Monte Carlo based statistical estimators, whereas in the latter
 approach, the geometry of the \(\chi^2\) profile is used to define bounds on the
 Wilson coefficients.
 Notwithstanding, we observe that the results and the trends are consistent with each
other, although the use of the partial $\chi^2$ over-emphasises the
broadening of the bounds.
%
\begin{table}[t]
  \centering
  \begin{tabular}{l|cccc}
    & $\quad$ SM PDFs $\quad$  & SMEFT PDFs  & best-fit shift  &
                                                                 broadening  \\
    \toprule
 $W\times 10^5$ (this work)                         & $[-2.0,1.4] $& $[-4.3,3.4] $& $-0.2$  & +126\% \\
 $W\times 10^5$ \cite{Greljo:2021kvv} & $[-1.4,1.2] $& $[-8.1,10.6] $& $-1.4$  & +620\% \\
    \hline
    $Y\times 10^5$(this work)                        & $[-3.2,8.1] $ &  $[-3.1,11.7]$ & $+1.9$ & +31\% \\
    $Y\times 10^5$\cite{Greljo:2021kvv} & $[-5.3,6.3] $ &  $[-11.1,12.6]$ & $+0.3$ & +110\% \\
    \hline
    $\mathbf{C}_{33}^{D\mu}\times 10^3$ (this work) & $[-11.9,13.1] $& $[-15.3,14.0] $  &$-0.9$  & +17\% \\
    $\mathbf{C}_{33}^{D\mu}\times 10^3$ \cite{Greljo:2021kvv} & $[-10.4,12.3] $& $[-12.5,14.6] $  &$-0.6$  & +18\% \\
    \bottomrule
  \end{tabular}
  \caption{\label{tab:summary} 95\% CL bounds on the simultaneous fit
    of the $W$ and $Y$ Wilson coefficients in Benchmark Scenario I and of the individual fit
    of the \(\mathbf{C}_{33}^{D\mu}\) Wilson
    Coefficient in Benchmark Scenario II, based on a fit inclusing the HL-LHC projections, compared to those obtained in
    the previous analysis presented in Ref.~\cite{Greljo:2021kvv}. The fourth and fifth column indicate the absolute shift in best-fit values,
 Eq.~(\ref{eq:shift})
    and the percentage broadening of the SMEFT bounds, Eq.~(\ref{eq:broadening}),
    when the PDFs are allowed to change alongside the Wilson coefficients.}
\end{table}

\begin{figure}[h]
         \centering
         \includegraphics[width=0.9\textwidth]{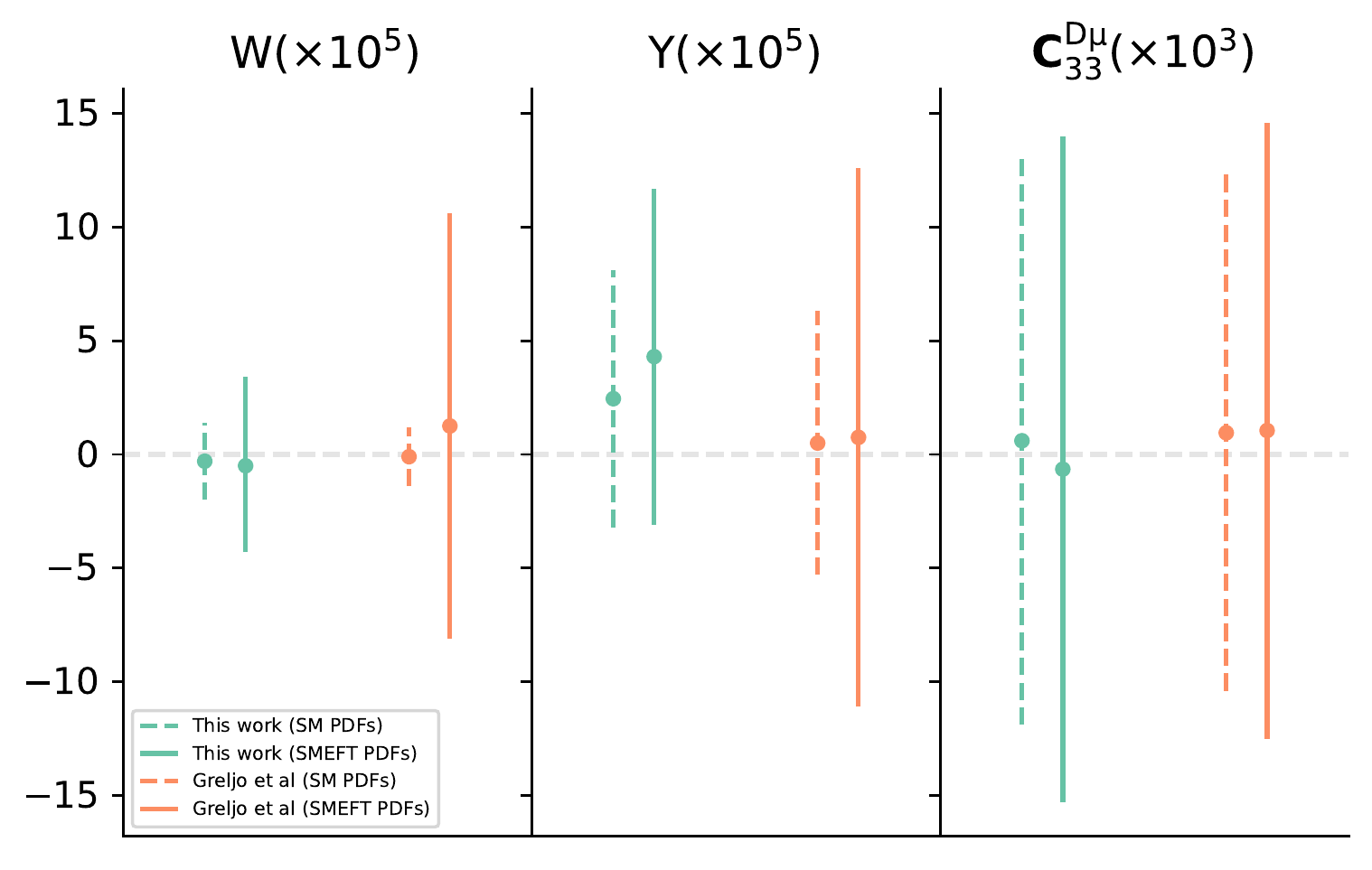}
         \caption{Comparison of \(95\%\) CL bounds between this study
           (in green) and those determined in by Greljo
         et al in Ref.~\cite{Greljo:2021kvv}.  The \(W\) and \(Y\) bounds are
         determined simultaneously in the HL-LHC scenario,
         \(\mathbf{C}_{33}^{D\mu}\) is fitted individually including
         SMEFT quadratic terms in the HL-LHC scenario. The bounds
         obtained by keeping the PDFs fixed to the SM baseline
         (dashed lines) are compared to those obtained in a
         simultaneous fit of PDFs and Wilson coefficients (solid lines). \label{fig: summary}}
\end{figure}

Results in the HL-LHC scenario are displayed in Fig.~\ref{fig: summary}.
 Shown are the results obtained including the current high-mass
 Drell-Yan data and the projected HL-LHC NC and CC Drell-Yan
 pseudo-data. In Scenario I, \(W\)
 and \(Y\) are fitted simultaneously by keeping only the linear term
 in the SMEFT expansion, while in Scenario II the
 \(\mathbf{C}_{33}^{D\mu }\) coefficients is fitted individually
 inclusing the
 \(\mathcal{O}(1/\Lambda^4)\) quadratic terms in the SMEFT expansion. 
 The results from both studies are compatible, although the bounds obtained from our study are
 slightly more conservative with differences being explained by the
 differences between the methodologies employed that are outlined at
 in this section. 
\section{Methodology validation and closure testing}
\label{sec:closure}

In this section, we assess the robustness of our approach through the closure testing
framework defined in Refs.~\cite{Hartland:2019bjb,DelDebbio:2021whr}.
Here we do not address the robustness of the PDF part of the fit,
which in a sense comes from the results presented in the {\tt
  NNPDF4.0} paper~\cite{nnpdf40} and the following dedicated study of
Ref.~\cite{DelDebbio:2021whr},
but rather focus on the robustness of the fit of the Wilson coefficients and of the PDFs in a
simultaneous fit. 

Crucially, the central values of the SMEFT-sensitive
datasets are modified by artificially contaminating them with pre-chosen,
extreme, values of Wilson coefficients. This will emulate a situation where a
dataset will favour a non-vanishing EFT operator. In the first part of
this section we show that our
methodology is sufficiently flexible to recover these chosen
values. In the second part of the section 
we consider the case where the input datasets are replaced by theory predictions
from a known underlying PDF set, rather than experimental central values. We
show that even in the case of fixing both the underlying PDF and Wilson coefficients
to some {\it a priori} values, the methodology is sufficiently robust that it can simultaneously
reproduce both.

\subsection{Closure test results on the Wilson Coefficients}
The PDF side of \texttt{SIMUnet} is of course identical to {\tt NNPDF4.0} which has
been demonstrated to be able to replicate an underlying law using the closure
testing framework~\cite{nnpdf40, DelDebbio:2021whr}.
In this section we will fix the value of the Wilson coefficients \textit{a priori} and study how effectively we can retrieve
these values using our methodology.

To perform this kind of closure tests, we artificially choose 
extreme values of the Wilson coefficients considered in Benchmark
Scenario I, where the interplay between PDF and SMEFT fits is more marked, in two separate analyses:
\begin{enumerate}
  \item  The individual fit of the \(W\) Wilson coefficient including
    the current high-mass DY data from LHC Run I and Run II along with
    all other datasets listed in Sect.~\ref{sec: experimental_data}. Specifically we set \(W=1\times
    10^{-3}\), which is outside the 95\% C.~L.~ bounds that are
    displayed in Fig.~\ref{fig:WYresults}
    \item The combined fit of the \((W, Y)\) parameters including  the
      HL-LHC projections.
      Specifically we set \((W, Y) = (1,-1)\times 10^{-4}\), which is
      also far outside the 95\% C.~L.~ contours displayed in Fig.~\ref{fig: hllhc_scatter}.
\end{enumerate}
The way we input these non-zero values of the Wilson coefficients in
the underlying law is by multiplying the Monte Carlo
pseudodata central values by the SMEFT $K$-factors obtained by setting the Wilson coefficient(s) to the aforementioned
values. The fitting methodology proceeds as before. Importantly, the entire tool chain has
no knowledge of what the value of the Wilson coefficient it is looking for are
set to.

\begin{figure}[t]
        \centering
        \includegraphics[width=0.49\textwidth]{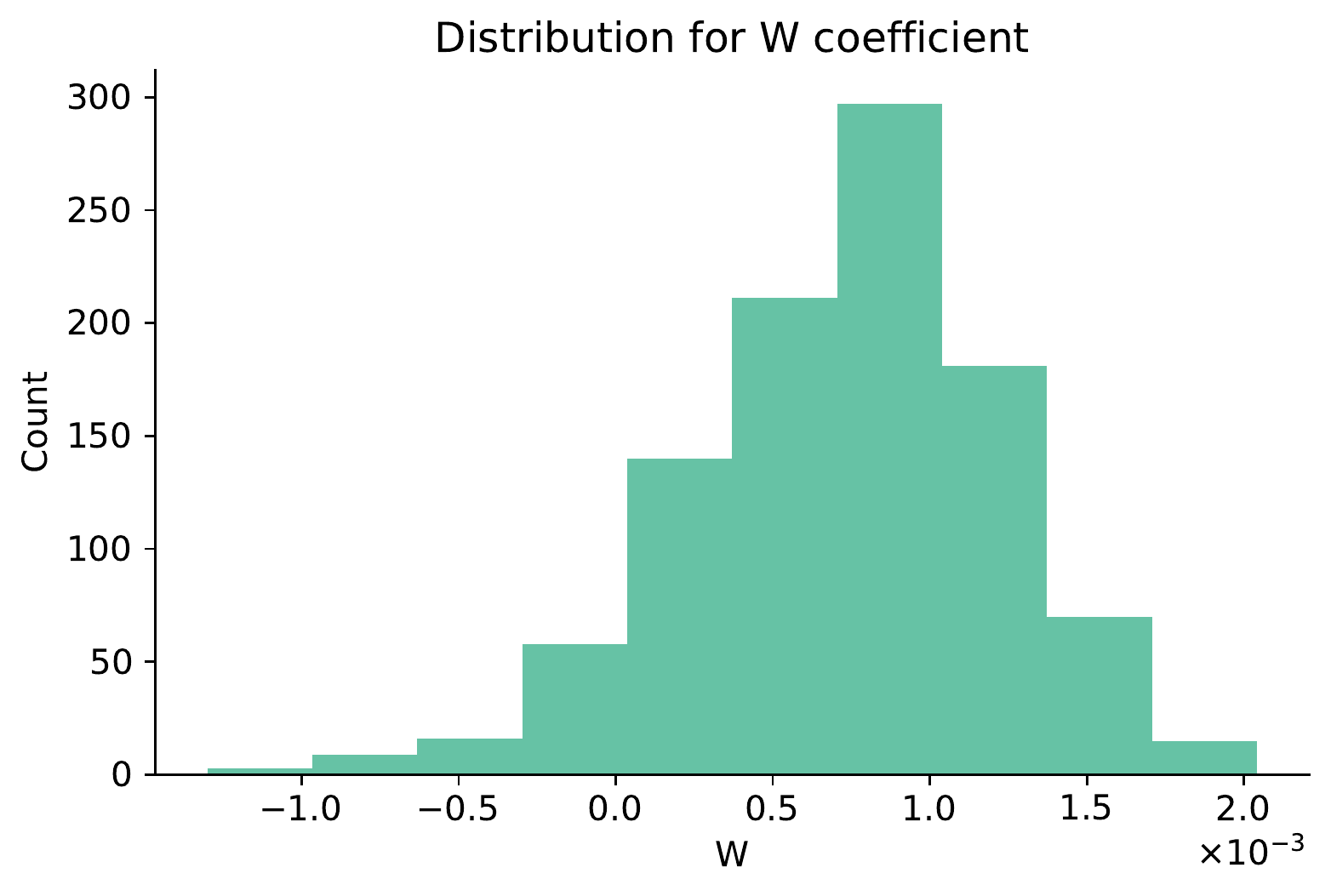}
        \includegraphics[width=0.49\textwidth]{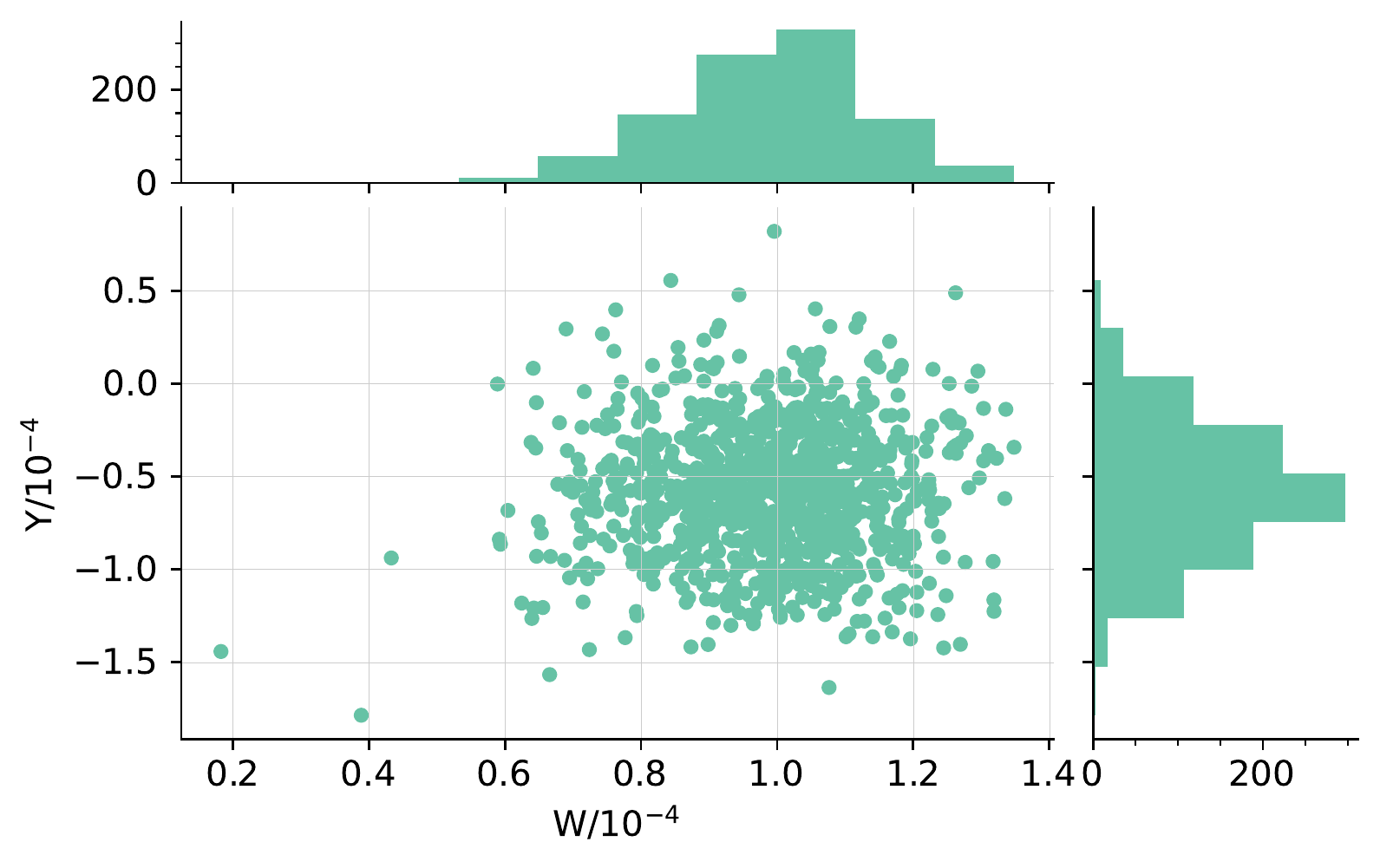}
        \caption{\label{fig: closure}Result of the closure testing framework for
        our methodology. Left: histogram for the distribution of the \(W\)
        parameter when the input data has been modified by setting \(W=1\times
        10^{-3}\). Right: distribution of \((W, Y)\) when fitting to data that
        has been modified by setting \((W, Y) = (1, -1)\times 10^{-4}\).  The
        upper and right panels show the histograms for the distribution of the
        best fit values in their respective directions.}
\end{figure}
The result of the closure test are displayed in Fig.~\ref{fig:
  closure}. On the left-hand panel we see that the distribution of
best fit values of \(W\) across 1000 replicas, that were previously centred at the origin, now moves considerably
towards the {\it a priori} value of \(W=10^{-3}\). Similarly, in the
second analysis displayed on the right-hand panel, we see that the
presence of HL-LHC data continues to eliminate the flat direction, with the distribution
of best fit Wilson coefficients resembling that of Fig.~\ref{fig:
  hllhc_scatter}, but with the best-fits values of the  \((W,Y)\) parameters consistently pushed towards the preselected{\it a priori} values of
\((W,Y) = (1, -1)\times 10^{-4}\). We see that, despite an
extremal choice of the injected values of the Wilson coefficients, the methodology is sufficiently flexible
and robust to recover them.

\begin{figure}[t]
        \centering
        \includegraphics[width=0.49\textwidth]{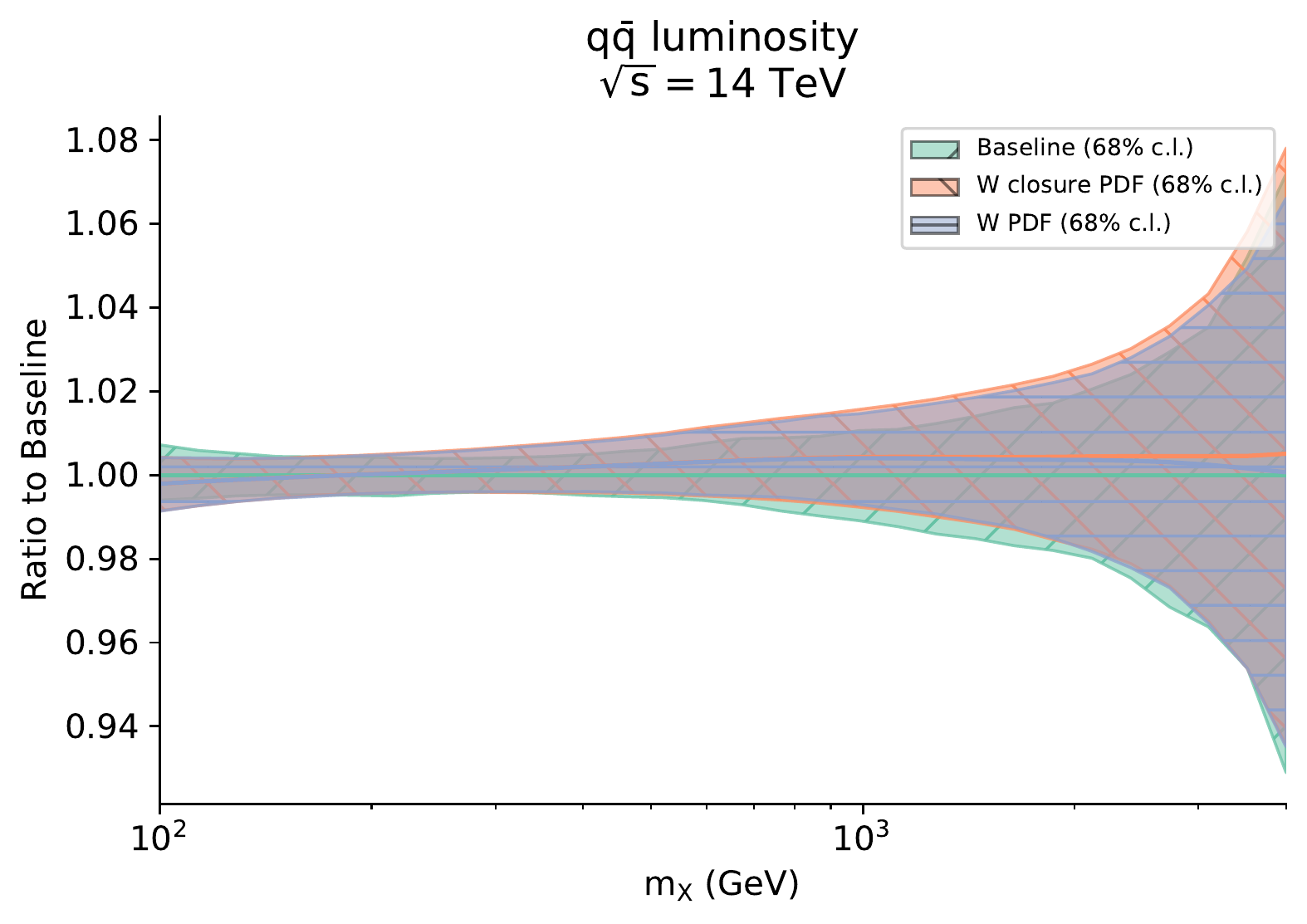}
        \includegraphics[width=0.49\textwidth]{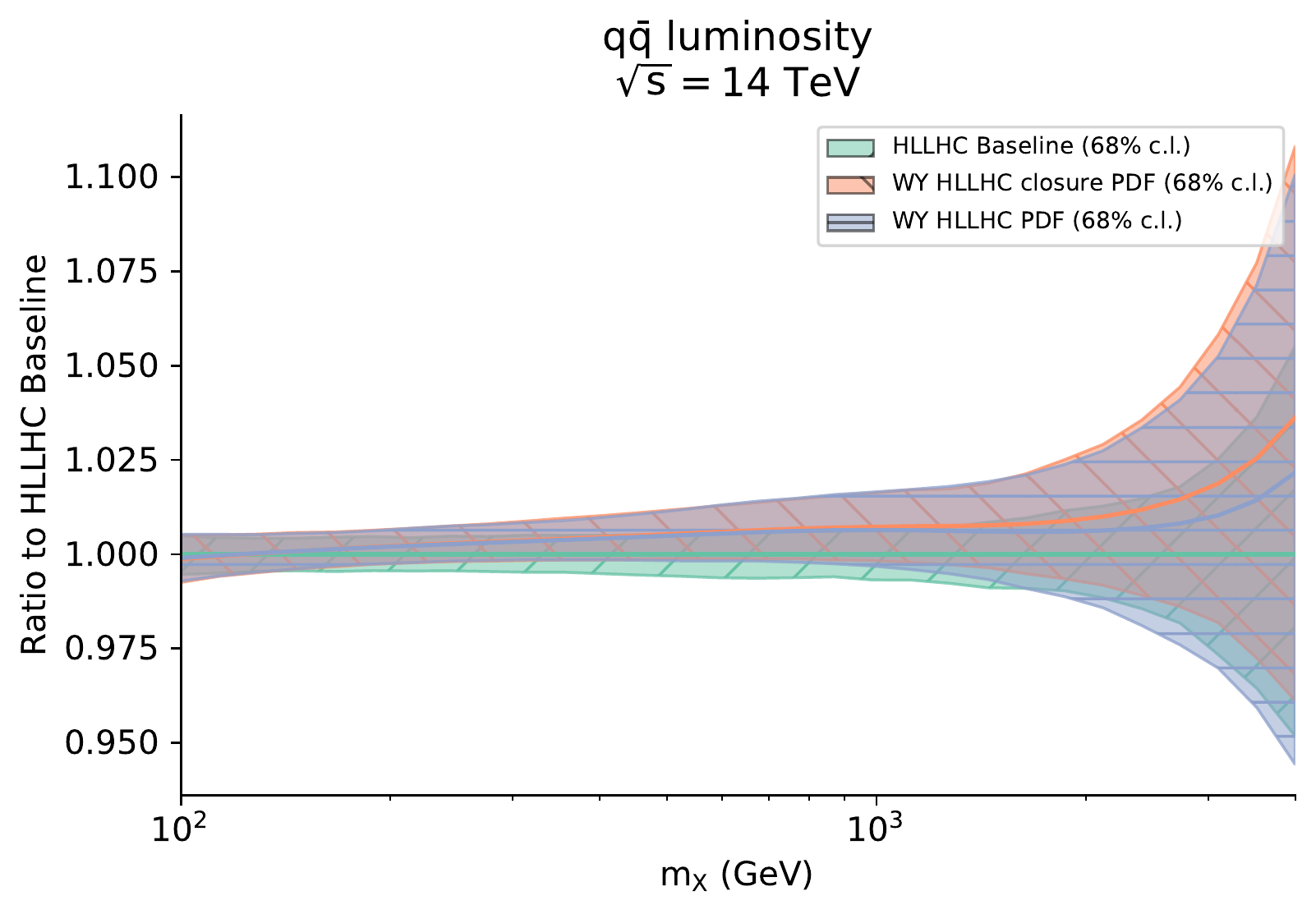}
        \caption{\label{fig: closure_lumis} The \(q\bar{q}\) luminosity obtained
        from the closure tests when the experimental central values are modified
        to encode a specific SMEFT benchmarking scenario (``closure
        PDF'' in orange), compared to the results of the simultanoues
        fits (``PDF'' in blue) normalized to the
        corresponding baseline (in green). Shown in the left panel is the \(W=1\times
        10^{-3}\) scenario while the \((W, Y) = (1, -1)\times 10^{-4}\) scenario
        is displayed in the right panel.}
\end{figure}
The PDFs generated in the context of the closure tests are displayed in
Fig.~\ref{fig: closure_lumis}. On the left panel, we plot the quark-antiquark luminosity obtained when the experimental
central values are modified by inputting $W=10^{-3}$ (called ``W
closure PDF'') and compare it to the ones that we obtain in the simultaneously fit of
the PDFs and $W$ presented in Sect.~\ref{sec: WY_fit} (called ``W PDF''), 
both normalised to the SM baseline.
We observe that the PDFs generated with our closure test for
$W=10^{-3}$ are similar to those that we obtain in the simultaneous
fit, despite the fact that the training dataset has been
heavily modified by the extreme {\it a priori} choice of $W$. This is to be
expected, since one can view the combination layer as capturing the data's
dependence on the Wilson coefficients, whilst the complementary PDF sector of
the network architecture captures the data's dependence on the underlying PDF,
which of course remains unchanged. The combination layer, in effect, subtracts
off the EFT dependence, leaving behind the pure SM contribution for the PDF
sector to parameterise. This is even more evident in the fit obtained
in the presence of the HL-LHC projections, by setting \((W, Y) = (1,
-1)\times 10^{-4}\) displayed on the right-hand panel of the figure. There we can
observe that the central value of the ``WY HLLHC closure PDF'' luminosity is pushed slightly
upwards and the uncertainty is rather smaller once the closure test is
performed as compared to the simultaneous fit ``WY HLLHC PDF''. This is once again to
be expected, given that the range of Wilson Coefficients allowed in
the simultaneous fit is larger as compared to the range that we obtain
from a simultaneous fit. 

\subsection{Closure test results on the simultaneous fit}

The natural extension of the closure test described in the previous subsection is to assess the
degree to which \texttt{SIMUnet} is able to replicate, not only fixed Wilson
coefficients, but also a known underlying PDF. For this scenario we employ the
NNPDF level 2 \cite{NNPDF:2014otw} closure test strategy. In the context of a
simultaneous fitting methodology this amounts to generating Standard Model
predictions using a known PDF set (referred to henceforth as the underlying law)
and mapping these to SMEFT observables by multiplying the SM theory predictions
with SMEFT K-factors scaled by a previously determined choice of Wilson
coefficients. These SMEFT observables, generated by the underlying law,
replace the usual MC pseudodata replicas, and are used to train the neural
network in the usual way.

Through this approach we are able to assess the degree to which the
parameterisation is able to capture not only an underlying choice of Wilson
coefficients, but ensures it is sufficiently flexible to adequately replicate a
known PDF.
For such a closure test, we use the HL-LHC baseline used in this study as
the underlying law with the SMEFT scenario being again the simultaneous
\((W, Y)\) determination, with the input data being adjusted to have
\((W, Y) = (1, -1)\times 10^{-4}\) encoded within it.

\begin{figure}[t]
        \centering
        \includegraphics[width=0.49\textwidth]{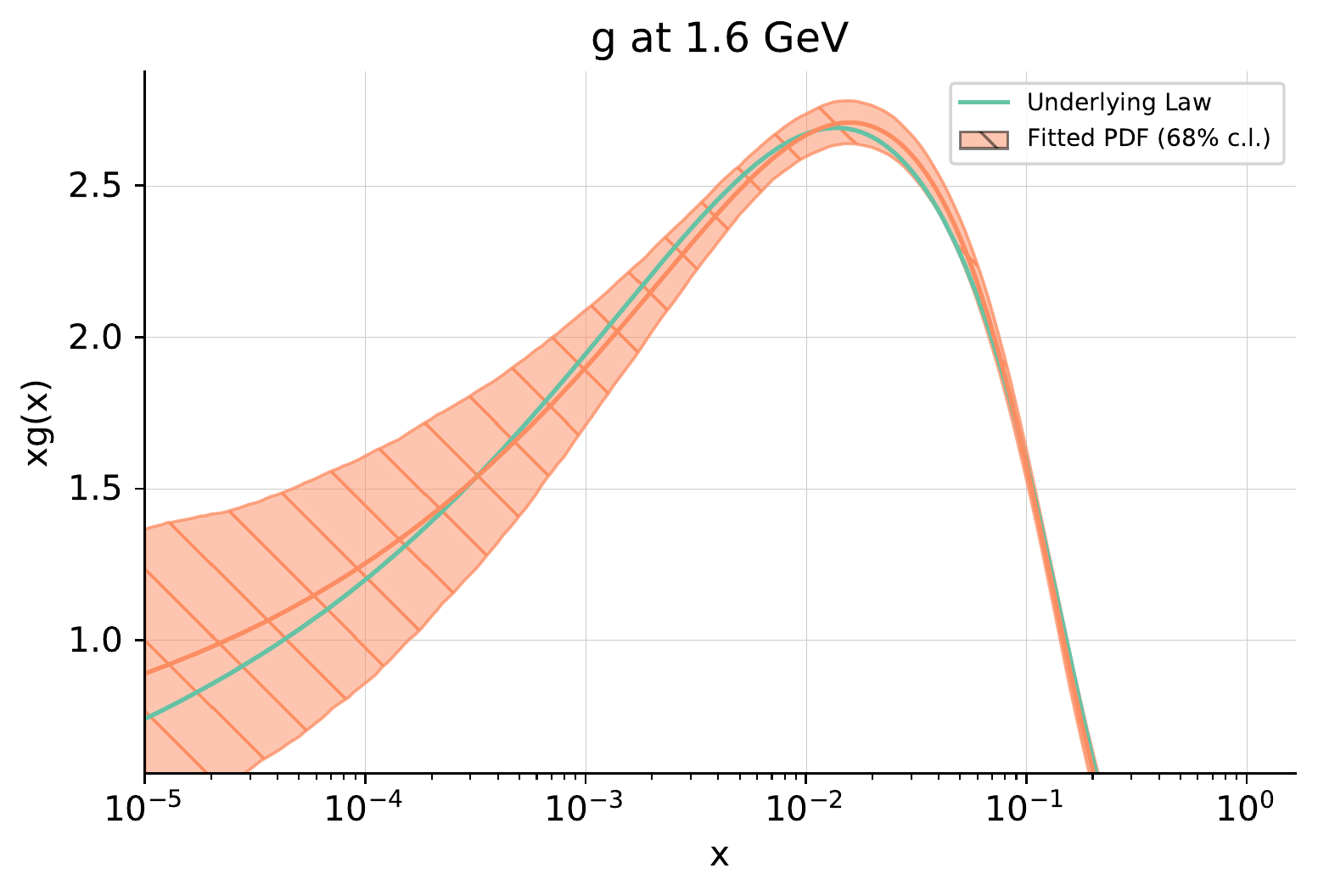}
        \includegraphics[width=0.49\textwidth]{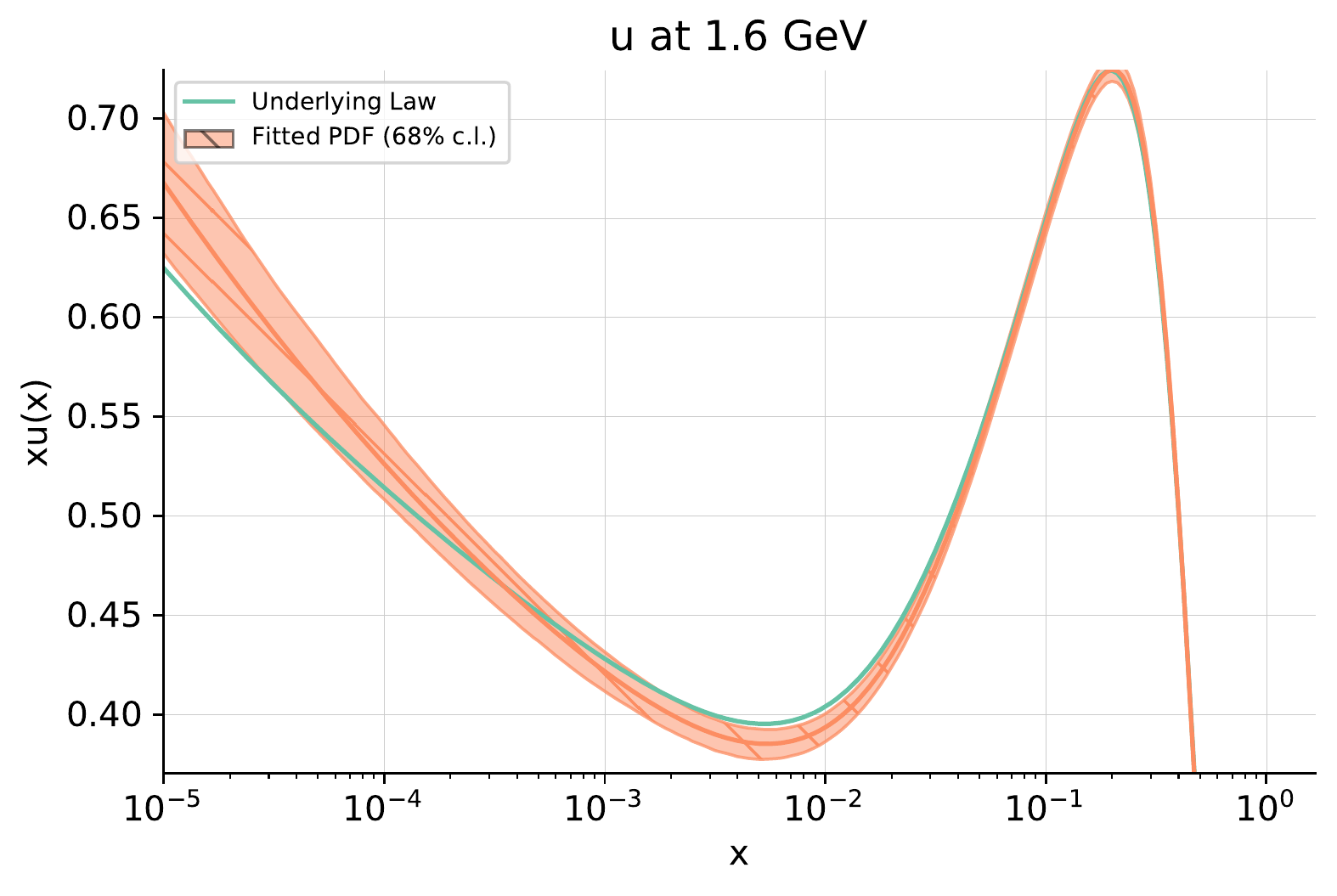}
        \caption{\label{fig: level2_pdf} The gluon (left) and up quark (right)
        PDFs obtained from the closure test framework in which both the
        underlying PDF set and Wilson coefficients are known. Shown in green is
        the PDF replica used as the underlying law which generates the fake data
        used to train our model. The resulting PDFs are shown in orange along
        with their \(68\%\) confidence level bands. The fake data generated by
        the underlying law is subsequently modified so as to encode the
        \((W,Y)=(1, -1)\times 10^{-4}\) condition.}
\end{figure}
The PDFs generated in this way are shown in Fig. \ref{fig: level2_pdf} for
representative choices of parton flavour: we display both the resulting PDFs as
well as the underlying law. We see that, despite modifying the training data
with an extreme choice of SMEFT benchmark, the methodology is sufficiently
robust so as to be able to recover the true PDF set with good precision. Indeed,
the distribution of best fit \((W, Y)\) values plotted in Fig. \ref{fig:
level2_dist} shows that not only did the methodology retrieve the underlying
law, but also managed to recover the chosen SMEFT scenario: being able to
correctly determine the \((W, Y)=(1,-1)\times 10^{-4}\) condition. Such
behaviour is reminiscent of the above closure test whereby the combination layer
is able to parameterise the BSM dynamics while the preceding layers of the model
are left to parameterise the PDFs. This conclusion thus illustrates how our
methodology is able to correctly disentangle the interplay between PDFs and BSM
dynamics.
\begin{figure}[t]
        \centering
        \includegraphics[width=0.75\textwidth]{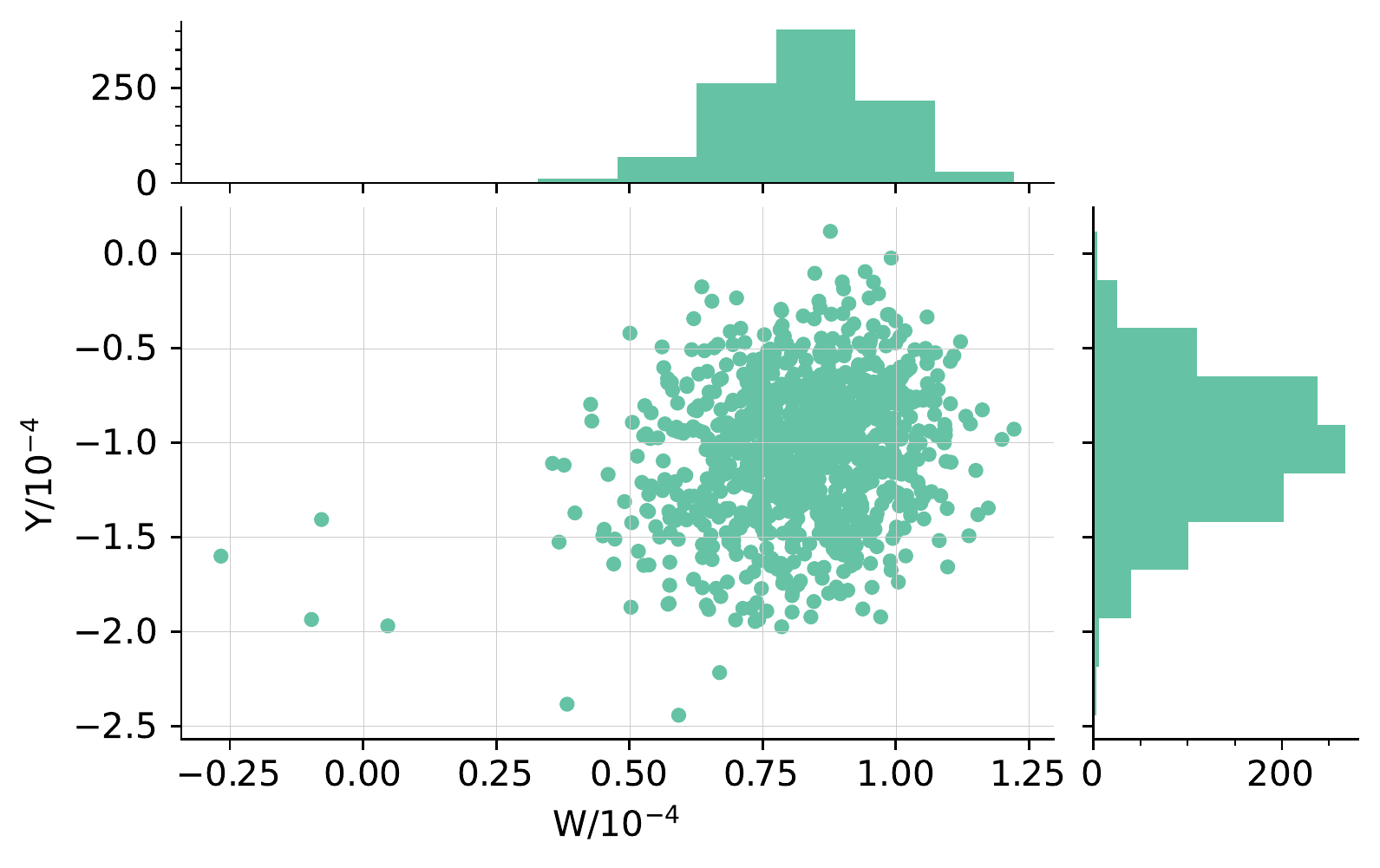}
        \caption{Distribution of best fit \((W,Y)\) parameters for each Monte
        Carlo replica. The data used for the fit was generated by a pre-selected
        PDF before the SM predictions were transformed to a SMEFT observable
        with \((W,Y) = (1, -1)\times 10^{-4}\). The upper panel is the histogram
        of best fit values in the \(W\) axis while the right panel is the
        histogram for the \(Y\) axis. \label{fig: level2_dist}}
\end{figure}

\section{Conclusions and outlook \label{sec: conclusions}}

In this work we have presented a novel methodology, dubbed {\tt
  SIMUnet}, which employs the latest in deep
learning techniques, to achieve a first truly simultaneous fit of PDFs
alongside any physical parameter that determines theoretical
predictions at the LHC. As a showcase we fit the Wilson
coefficients in the SMEFT and the PDFs in two motivated Benchmark
Scenarios.

In particular we employ our methodology to constrain each of \(W\) and \(Y\) individually
using Run I and Run II ATLAS and CMS neutral current (NC) high mass DY
data and show that the interplay between SMEFT effects and the fit of
large-$x$ PDFs remain moderate. Interestingly we show how a scenario
like this, displaying a flat direction once \(W\) and \(Y\) are fitted
simultaneously, are uncovered naturally by our approach, without the need for
any user intervention or prior knowledge, and how the framework comes to learn
that the flat direction is lifted by the inclusion of such data that eliminates
this redundancy.
Indeed, once the HL-LHC projections are added to the fit, such
projections including charged current (CC) DY data, the 
flat direction is broken and much tighter bounds are obtained.

Most
importantly, we confirm the results already shown in our previous
analysis of Ref.~\cite{Greljo:2021kvv}, namely that the inclusion of
high-mass DY data both in a fit of PDFs and in a fit of SMEFT coefficients by neglecting the interplay between
them, could result both in a significant underestimate of the uncertainties associated to the SMEFT
parameters (by a factor of 2 in the case of the $W$ coefficient) and
of the large-$x$ PDF uncertainties, that increase by about 70\% in the
high-mass luminosity regions once the SMEFT affects are accounted
for. 

We also showcase the ability of our framework to fit SMEFT parameters while including
the quadratic effect of SMEFT-SMEFT diagrams.
Such a scenario is known~\cite{Greljo:2021kvv} to be essentially unconstrained
at linear EFT level and we illustrate how our methodology replicates bounds
for \(\mathbf{C}_{33}^{D\mu}\) presented in previous analyses.

We additionally stress test the methodology by providing it artificially
contaminated data possessing data shifted by an {\it a priori} choice of Wilson
coefficients. We show that our
methodology is sufficiently flexible to recover these chosen values. Moreover,
we consider the case where the input datasets are replaced by theory predictions
from a known underlying PDF set, rather than experimental central values. We
show that even in the case of fixing both the underlying PDF and Wilson coefficients
to known values, the methodology is sufficiently robust that it can simultaneously
reproduce both.

The new methodology that we present here provides a crucial step
towards the interpretation of indirect searches under a unified
framework. We can assess the impact of new
datasets not only on the PDFs, but now on the couplings of an EFT
expansion, or on any other physical parameter.

The next steps of our work will include more SMEFT parameters to the
combination layer, which is conceptually straightforward. 
We can add the effects of dim-8 operators or
consider the effect of quadratic dim-6 operator effects. Both these
scenarios fit very naturally within our approach.  Furthermore, one could assess
the impact of other high energy datasets from the LHC, exploiting different
processes to better constrain other subsets of the Warsaw basis.  Finally, an
interesting prospect would be to explore simultaneously fitting
precision SM parameters, such as the strong coupling or electroweak
parameters.  Indeed,
our framework extends naturally, not only to BSM studies, but to any parameter
which may modify the SM prediction through the use of K-factors or
other kind of interpolation.

\section*{Acknowledgments}
We thank Richard Ball, Luigi Del Debbio, Juan Cruz-Martinez, Ben Day, Stefano Forte, James Moore,
Juan Rojo, and Zahari Kassabov for particularly useful discussion.
We thank the NNPDF collaboration for the public release of the code.
We thank the PBSP collaboration for providing the relevant SMEFT $K$-factors
for the various datasets studied in this work.
M.~U. is supported by the European Research Council under the European Union’s
Horizon 2020 research and innovation Programme (grant agreement n.950246).
M.~U. and S.~I. are supported by the Royal Society grant RGF/EA/180148.
The work of M.~U. is also funded by the Royal Society grant DH150088.
M.~U. and S.~I. are partially supported by STFC consolidated grant
ST/T000694/1.

\appendix
\section{FK table dependence on the strong coupling \label{sec: example_alphas}}

In this appendix we consider the case whereby we wish to
isolate the FK table dependence on the strong coupling constant at the
\(Z\)-pole, such that, following our notation in Sect.~\ref{sec:theory}, we have \(c=\alpha_s(M_Z)\).
We take the PDG value for the strong coupling evaluated at the \(Z\)-boson mass
\(c^*=\alpha_s(M_Z) = 0.1179(10)\) \cite{Zyla:2020zbs} to be the point about
which we perform the Taylor expansion.  The partonic cross sections are given
as a power expansion in \(\alpha_s\) using perturbation theory, which allows us
to write the exact expression%
\begin{equation}
    \label{eq: coeff_func_expansion}
    \hat\sigma(c) = \sum_p\left(c-c^*\right)^p\hat\sigma_p(c^*)
\end{equation}
where we have simply centred the order by order sum around \(c^*\). For
illustrative purposes, we restrict ourselves to the case where the process depends solely
on the non-singlet quark distribution. The evolution kernel in Mellin space to
leading order in the anomalous dimension (the Mellin transform of the splitting
function) \(\gamma_0\), is then given by \cite{Ball:2008by}:
\begin{equation}
    \Gamma(N, \alpha_s(Q^2), \alpha_S(Q_0^2)) =
    \left(\frac{\alpha_s(Q^2)}{\alpha_s(Q_0^2)}\right)^{-\gamma^0(N)/\beta_0}.
\end{equation}
If we consider the leading-log evolution of $\alpha_s$ via renormalisation group equation from $M_Z^2$ to
$Q^2$ or $Q_0^2$ and set $c=\alpha_s(M_Z^2)$, we get
\begin{equation}
    \alpha_s(Q^2) = \frac{c}{1+\beta_0c\ln\frac{Q^2}{M_Z^2}} \qquad  \alpha_s(Q_0^2) = \frac{c}{1+\beta_0c\ln\frac{Q_0^2}{M_Z^2}}
\end{equation}
where \(\beta_0=11-\frac{2}{3}N_f\). Expanding the evolution kernel $\Gamma$ according to Eq.~\eqref{eq:expansion}, taking $c^*=\alpha_s^{\rm PDG}(M_Z) = 0.1179$, it is easy to see that we get
\begin{equation}
  \label{eq:as1}
    \Gamma(c) = \Gamma_0 + (c-c^*) \Gamma_1 + \cdots,
\end{equation}
where
\begin{gather}
    \Gamma_0 = \frac{1+c^*\beta_0\ln\frac{Q_0^2}{M_Z^2}}{1+c^*\beta_0\ln\frac{Q^2}{M_Z^2}}\\
    \Gamma_1 = -\frac{\gamma_0(N)}{\beta_0}
    \Gamma_0^{-\frac{\gamma_0(N)}{\beta_0}-1}
    \left(\frac{\beta_0\ln\frac{Q_0^2}{M_Z^2}}{1+c^*\beta_0\ln\frac{Q^2}{M_Z^2}}
    - \frac{\left(1+c^*\beta_0\ln\frac{Q_0^2}{M_Z^2}\right)\beta_0\ln\frac{Q^2}{M_Z^2}}
    {\left(1+c^*\beta_0\ln\frac{Q^2}{M_Z^2}\right)^2}\right).
\end{gather}
Finally, once Eq.~\eqref{eq:as1} is combined with Eq.~\eqref{eq: coeff_func_expansion}
we obtain the order by order expansion for the FK tables:
\begin{align}
    K =& \hat\sigma_0\otimes\Gamma_0 +
    (c-c^*)\big(\hat\sigma_1\otimes\Gamma_0 + \hat\sigma_0\otimes\Gamma_1\big) + \cdots\\
    \equiv& K_0 + (c-c^*)K_1 + \cdots
\end{align}
which can, in principle, be computed and stored permanently in the usual way. In
this way we can capture the FK table dependence on the strong coupling
parameter in the neighborhood of some prior choice \(c^*\).

So far, we have limited ourselves to the leading order evolution of the PDFs and the strong coupling constant, however a similar expression can be straghtforwardly obtained at next-to-leading (NLO) and next-to-next-to-leading order (NNLO).
The only caveat is that, from NLO onwards terms of the form
\(\ln\left(1+\beta_0c\ln\frac{Q^2}{M_Z^2}\right)\) arise which spoil the
validity of the Taylor expansion. The term \(\ln\frac{Q^2}{M_Z^2}\) can in
principle be made arbitrarily large thus exiting the unit disc which is the
region of analyticity of \(\ln 1+x\). This problem can be circumvented, however,
by noting:
\begin{equation}
    \ln\left(1+\beta_0c\ln\frac{Q^2}{M_Z^2}\right) =
    \ln\left(1+\beta_0c^*\ln\frac{Q^2}{M_Z^2}\right) +
    \ln\left(1+\frac{\beta_0(c-c^*)\ln\frac{Q^2}{M_Z^2}}{1+\beta_0c^*\ln\frac{Q^2}{M_Z^2}}\right)
\end{equation}
where the rightmost term can now be Taylor expanded since \(c-c^*\) can
be made arbitrarily small so as to suppress the large logarithm. Indeed,
note that in the worst case:
\begin{gather}
    \lim_{Q^2\rightarrow\infty}
    \left((c-c^*)\frac{\beta_0\ln\frac{Q^2}{M_Z^2}}{1+\beta_0c^*\frac{Q^2}{M_Z^2}}\right)
    = \frac{c-c^*}{c^*}\leq 1\\
    \implies c\leq 2c^*
\end{gather}
which can be implemented within the optimizer to restrict it from venturing
to values greater than \(2c^*\).

A viable alternative to the Taylor expansion approach defined in Sect.~\ref{sec:theory} and detailed here for the
simultaneous extraction of $\alpha_s$ and the PDFs is to compute various FK tables using preset values of the strong
coupling and then perform an element-wise interpolation between these tensors.
In this way one can avoid having to implement the Taylor series expansion, while
still accurately replicating the linear behaviour of the FK table. Such a task
is reserved for future work~\cite{prep:as}.

Finally if one was to include electroweak corrections to the DGLAP evolution
equation, as is for example done in \texttt{APFEL} \cite{Bertone:2013vaa}, then
electroweak parameters will in general also be present in the combined QCD and
QED evolution operator.  For such a scenario, the prescription outlined in this appendix
must then be followed. However, the corrections to the
pure QCD splitting functions introduced by electroweak considerations have no
dependence on the CKM matrix elements, the weak mixing angle, \(\theta_W\), or
gauge boson masses, amongst others. Such quantities manifest solely in the
partonic cross section and in general yield a prescription much simpler than that
outlined here.

\section{Fit quality\label{app:fit}}
For each MC replica there is the pair \((\mathbf{f}_i, \mathbf{c}_i)\) which are the best
fit PDFs and Wilson coefficients respectively. Together they can be used to generate the corresponding
theory predictions for each dataset, noting that any datasets which were not modified by the SMEFT
operators will effectively have \(\mathbf{c}=\mathbf{0}\). We thus have an ensemble of \(N_\text{rep}\)
vectors of theory predictions, \(\mathbf{T}\), with the central theory prediction being given by
the average across replicas:
\begin{equation}
        \label{eq: rep0}
        \left<\mathbf{T}\right> = \frac{1}{N_\text{rep}}\sum_{i=1}^{N_\text{rep}}\mathbf{T}(\mathbf{f}_i,\mathbf{c}_i).
\end{equation}
The central \(\chi^2\) per data point is then computed in the usual way
\begin{equation}
        \chi^2 = \frac{1}{N_\text{dat}}(\mathbf{d}-\left<\mathbf{T}\right>)^TV^{-1}(\mathbf{d}-\left<\mathbf{T}\right>)
\end{equation}
with \(\mathbf{d}\) being the vector of experimental central values and \(V\) the covariance matrix
encapsulating the experimental uncertainties and the correlations therein. These values are tabulated
in Tab. \ref{tab: chi2} for each of the various SMEFT scenarios considered in this work. We also tabulate
the \(\chi^2\) for various groupings of these datasets, such as DIS only, including or excluding the high-mass DY
measurements etc. For these particular entries various correlated systematics may exist between datasets, such as
the uncertainty in beam luminosity, which introduces off-diagonal entries in the covariance matrix; as such
the grouped \(\chi^2\) is not necessarily equal to the weighted average of the individual \(\chi^2\) values
constituting the grouping. This effect is particularly marked for the HL-LHC entries.

\begin{table}[t]
        \tiny
        \centering
        \begin{tabular}{lccHcc|cc}
                \toprule
                \multirow{3}{*}{Dataset} & \multirow{3}{*}{\(n_\text{data}\)} & \multicolumn{6}{c}{\(\chi^2/n_\text{data}\)}\\
                \cmidrule{3-8}
                & & SM & \multirow{2}{*}{\((W, Y)\)} & \multirow{2}{*}{\(W\)} & \multirow{2}{*}{\(Y\)} & HL-LHC & HL-LHC\\
                & & Baseline & & & & SM Baseline & \((W, Y)\) \\
                \midrule
                SLAC & 67 & 0.866 & 0.850 & 0.855 & 0.854 & 0.837 & 0.850\\
                BCDMS & 581 & 1.285 & 1.248 & 1.265 & 1.265 & 1.294 & 1.266\\
                NMC & 325 & 1.320 & 1.308 & 1.318 & 1.318 & 1.333 & 1.320\\
                CHORUS & 832 & 1.208 & 1.198 & 1.209 & 1.210 & 1.206 & 1.208\\
                NuTeV & 76 & 0.444 & 0.492 & 0.486 & 0.487 & 0.472 & 0.497\\
                HERA inclusive & 1145 & 1.188 & 1.179 & 1.183 & 1.183 & 1.190 & 1.184\\
                HERA charm & 37 & 1.435 & 1.396 & 1.384 & 1.382 & 1.463 & 1.391\\
                HERA bottom & 29 & 1.113 & 1.114 & 1.110 & 1.110 & 1.118 & 1.110\\
                \midrule
                \textbf{Total DIS} & \textbf{3092} & \textbf{1.202} & \textbf{1.188} & \textbf{1.197} & \textbf{1.197} & \textbf{1.206} & \textbf{1.198} \\
                \midrule
                E886 \(\sigma^{d}_\text{DY}/\sigma^p_{DY}\) & 15 & 0.669 & 0.582 & 0.609 & 0.600 & 0.974 & 0.679\\
                E886 \(\sigma^p_\text{DY}\) & 89 & 1.572 & 1.507 & 1.593 & 1.604 & 1.566 & 1.631\\
                E605 \(\sigma^p_\text{DY}\) & 85 & 1.197 & 1.144 & 1.200 & 1.205 & 1.197 & 1.213\\
                \midrule
                CDF \(d\sigma_Z/dy_Z\) & 29 & 1.613 & 1.529 & 1.546 & 1.548 & 1.623 & 1.563\\
                D0  \(d\sigma_Z/dy_Z\)& 28 & 0.612 & 0.608 & 0.610 & 0.610 & 0.613 & 0.613\\
                D0 \(W\rightarrow \mu\nu\) asy. & 9 & 1.845 & 1.473 & 1.510 & 1.503 & 2.052 & 1.587\\
                \midrule
                ATLAS \(W, Z\) 2010 & 30 & 1.021 & 0.979 & 1.014 & 1.014 & 1.017 & 1.017\\
                ATLAS low-mass \(Z\rightarrow ee\) & 6 & 0.923 & 0.920 & 0.921 & 0.921 & 0.923 & 0.921\\
                ATLAS \(W, Z\) 2011 CC & 46 & 2.095 & 1.987 & 2.005 & 2.006 & 2.091 & 2.010\\
                ATLAS \(W, Z\) 2011 CF & 15 & 1.062 & 1.050 & 1.072 & 1.073 & 1.066 & 1.071\\
                ATLAS \(W + c\) rapidity & 22 & 0.453 & 0.463 & 0.463 & 0.463 & 0.447 & 0.457\\
                ATLAS \(Z_{p_T}\) & 92 & 0.959 & 0.953 & 0.938 & 0.936 & 0.939 & 0.928\\
                ATLAS \(W_{p_T}\) jets & 32 & 1.685 & 1.655 & 1.672 & 1.670 & 1.676 & 1.665\\
                \midrule
                CMS \(W\) \(e\) asy. & 11 & 0.785 & 0.782 & 0.790 & 0.789 & 0.815 & 0.804\\
                CMS \(W\) \(\mu\) asy. & 11 & 1.767 & 1.728 & 1.732 & 1.733 & 1.765 & 1.732\\
                CMS \(\sigma_{W+c}\) 7 TeV & 5 & 0.513 & 0.520 & 0.518 & 0.516 & 0.502 & 0.504\\
                CMS \(\sigma_{W^{+}+c}/\sigma_{W^{-}+c}\) 7 TeV & 5 & 1.822 & 1.804 & 1.791 & 1.796 & 1.884 & 1.848\\
                CMS \(Z_{p_T}\) & 28 & 1.303 & 1.298 & 1.312 & 1.311 & 1.287 & 1.306\\
                CMS \(W\rightarrow \mu\nu\) rapidity & 22 & 1.472 & 1.312 & 1.337 & 1.340 & 1.422 & 1.310\\
                CMS \(W+c\) rapidity 13 TeV & 5 & 0.719 & 0.722 & 0.722 & 0.721 & 0.712 & 0.711\\
                \midrule
                LHCb \(Z\rightarrow\mu\mu\) & 9 & 1.503 & 1.502 & 1.545 & 1.550 & 1.506 & 1.549\\
                LHCb \(W,Z\rightarrow\mu\) 7 TeV & 29 & 2.043 & 1.925 & 1.973 & 1.977 & 2.066 & 2.005\\
                LHCb \(W,Z\rightarrow ee\) & 17 & 1.249 & 1.231 & 1.236 & 1.236 & 1.220 & 1.231\\
                LHCb \(W,Z\rightarrow\mu\) 8 TeV & 30 & 1.621 & 1.417 & 1.497 & 1.502 & 1.615 & 1.543\\
                \midrule
                \textbf{Total DY (excl. HM)} & \textbf{670} & \textbf{1.302} & \textbf{1.243} & \textbf{1.274} & \textbf{1.276} & \textbf{1.307} & \textbf{1.286}\\
                \midrule
                ATLAS DY high-mass 7 TeV & 13 & 1.680 & 1.731 & 1.575 & 1.609 & 1.654 & 1.626\\
                ATLAS DY high-mass 8 TeV & 46 & 1.174 & 1.634 & 1.177 & 1.171 & 1.175 & 1.171\\
                CMS DY high-mass 7 TeV & 117 & 1.694 & 1.506 & 1.671 & 1.676 & 1.677 & 1.669\\
                CMS DY high-mass 8 TeV & 41 & 0.923 & 0.850 & 0.944 & 0.941 & 0.893 & 0.914\\
                CMS DY high-mass 13 TeV & 43 & 2.003 & 1.934 & 2.064 & 2.037 & 2.000 & 2.005\\
                \midrule
                \textbf{Total DY (HM only)} & \textbf{260} & \textbf{1.531} & \textbf{1.507} & \textbf{1.529} & \textbf{1.527} & \textbf{1.517} & \textbf{1.515}\\
                \midrule
                \textbf{Total (excl. HL-LHC)} & \textbf{4022} & \textbf{1.245} & \textbf{1.221} & \textbf{1.236} & \textbf{1.237} & \textbf{1.248} & \textbf{1.238}\\
                \midrule
                HL-LHC CC \(e\)& 16 & \textit{1.119} & \textit{1,264,000} & \textit{119.9} & \textit{0.922} & 0.588 & 0.544\\
                HL-LHC CC \(\mu\)& 16 & \textit{1.414} & \textit{1,246,000} & \textit{112.8} & \textit{1.162} & 0.894 & 0.803\\
                HL-LHC NC \(e\)& 12 & \textit{1.164} & \textit{3,481} & \textit{17.65} & \textit{7.495} & 1.104 & 0.961\\
                HL-LHC NC \(\mu\)& 12 & \textit{1.041} & \textit{3,444} & \textit{13.32} & \textit{5.048} & 0.964 & 1.071\\
                \midrule
                \textbf{Total HL-LHC only} & \textbf{56} & \textbf{\textit{1.298}} & \textbf{\textit{80,220}} & \textbf{\textit{72.88}} & \textbf{\textit{5.546}} & \textbf{0.894} & \textbf{0.836}\\
                \midrule
                \textbf{Total} & \textbf{4078} & \textbf{\textit{1.246}} & \textbf{\textit{11,020}} & \textbf{\textit{2.220}} & \textbf{\textit{1.296}} & \textbf{1.243} & \textbf{1.232}\\
                \bottomrule
        \end{tabular}
        \caption{Values of the \(\chi^2\) per data point across all datasets
        used in this study. We tabulate values for the baseline PDF set as well
        as those obtained in the various SMEFT scenarios.  Shown also is the
        \(\chi^2\) for the HL-LHC scenario. The rows indicating total
        \(\chi^2\) values are computed accounting for any relevant correlated
        systematic errors. Values in italics indicate the dataset was not used
        in the corresponding fit. \label{tab: chi2}}
\end{table}

In all scenarios considered, the added degrees of freedom result in the
\(\chi^2\) per data point to drop when a simultaneous determination is
performed when contrasted to the purely Standard Model fits. The SM
\(\chi^2\) in this sense serves as an upper bound, since the optimizer
is free to determine \(\mathbf{c} = 0\) and so the goodness of fit can
be no worse than the SM fit.
In particular we see the SMEFT sensitive high-mass DY datasets
experience a large overall improvement in fit quality, largely driven
by the CMS measurements at \(7\) TeV owing to the large weight
carried by this dataset: forming just under half the entire
high-mass DY data points considered in this study.
Moreover, the DIS only grouping experiences a marked improvement in fit quality
with the \(\chi^2\) per data point dropping by \(0.014\) across \(3092\) data
points in the case of the simultaneous \((W,Y)\) determination. The reader
is again reminded that the HERA combined dataset, which forms the majority
of the DIS data points, is included in the set of datasets that are modified by
the \(W\) and \(Y\) operators.
It is interesting to observe that the improvement fit quality is propagated
down to those datasets, such as the low-mass measurements, that are not used
to explicitly constrain the Wilson coefficients. Such datasets, however,
are correlated through shared sources of systematic errors, such as the
luminosity uncertainty or detector effects, thus improving the fit
to one such dataset necessarily affects others.
Also tabulated are the fit quality values for the HL-LHC projections, even for
those fits which do not incorporate these projections in their training data.
These entries illustrate the pull the HL-LHC projections have on the Wilson
coefficients. We see that not including these data points in the fit renders the
fit virtually useless in the context of the projected data.  This is indeed
reflected in the values of the K-factors used for these projections, reaching
\(K\simeq 5\) for the highest transverse mass bins.

\begin{table}[t]
    \centering
    \begin{tabular}{lccc}
      \toprule
      \multirow{2}{*}{Dataset}& \multirow{2}{*}{$n_\text{data}$} &SM& \multirow{2}{*}{\(\mathbf{C}_{33}^{D\mu}\)}\\
      && Baseline&\\
      \midrule
      ATLAS DY high-mass 7 TeV & 13 &1.631 &1.625\\
      ATLAS DY high-mass 8 TeV & 46 &1.179&1.168\\
      CMS DY high-mass 7 TeV & 117 &1.666 &1.664\\
      CMS DY high-mass 8 TeV & 41 &0.858 &0.893\\
      CMS DY high-mass 13 TeV \(ee\)* & 43 &2.579 &2.574\\
      CMS DY high-mass 13 TeV \(\mu\mu\)& 43 &0.836 &0.837\\
      \midrule
      \textbf{Total DY (HM only)} & \textbf{303} &1.501 &1.503\\
      \midrule
      HL-LHC CC \(e\)*& 16 &0.437&0.424\\
      HL-LHC CC \(\mu\)*& 16 &0.841 &0.752\\
      HL-LHC NC \(e\)*& 12 &1.108 &1.130\\
      HL-LHC NC \(\mu\)& 12 &0.943 &0.969\\
      \midrule
      \textbf{Total HL-LHC only} & \textbf{56} &0.830 &0.826\\
      \bottomrule
    \end{tabular}
      \caption{Values of the \(\chi^2\) per data point across all high mass
      Drell-Yan measurements. We tabulate values for the baseline PDF set as well
      as those obtained in the EFT scenario II.  Shown also is the
      \(\chi^2\) for the HL-LHC projections. Quadratic terms in the EFT parameter are used in the \(\chi^2\) calculation. The rows indicating total
      \(\chi^2\) values are computed accounting for any relevant correlated
      systematic errors. Datasets marked with an asterisk indicate that they
      were not used to constrain \(\mathbf{C}_{33}^{D\mu}\).}
\end{table}
\section{Stability on replica number \label{app: stability}}

When performing purely PDF fits to experimental data, it is often the case one
will have to compute \(\sim 100\) Monte Carlo PDF replicas in order to achieve a
percent level, faithful, uncertainty estimation \cite{Ball:2010de}. As this study acts as a
proof-of-concept for our methodology, we perform high statistic fits composed of
\(1000\) MC replicas. However, with each replica requiring approximately 3 hours
of compute time, one necessarily requires access to a cluster of nodes in order
to asynchronously compute the roughly \(3000\) hours of total wall clock time that is
needed for each fit. It is possible, however, to reduce this time by an order of
magnitude by instead computing \(\sim 100\) MC replicas per fit, and in this appendix
we show that doing so poses little risk in underestimating the statistics when
compared to a high replica fit.

\begin{figure}[t]
        \centering
        \includegraphics[width=0.49\textwidth]{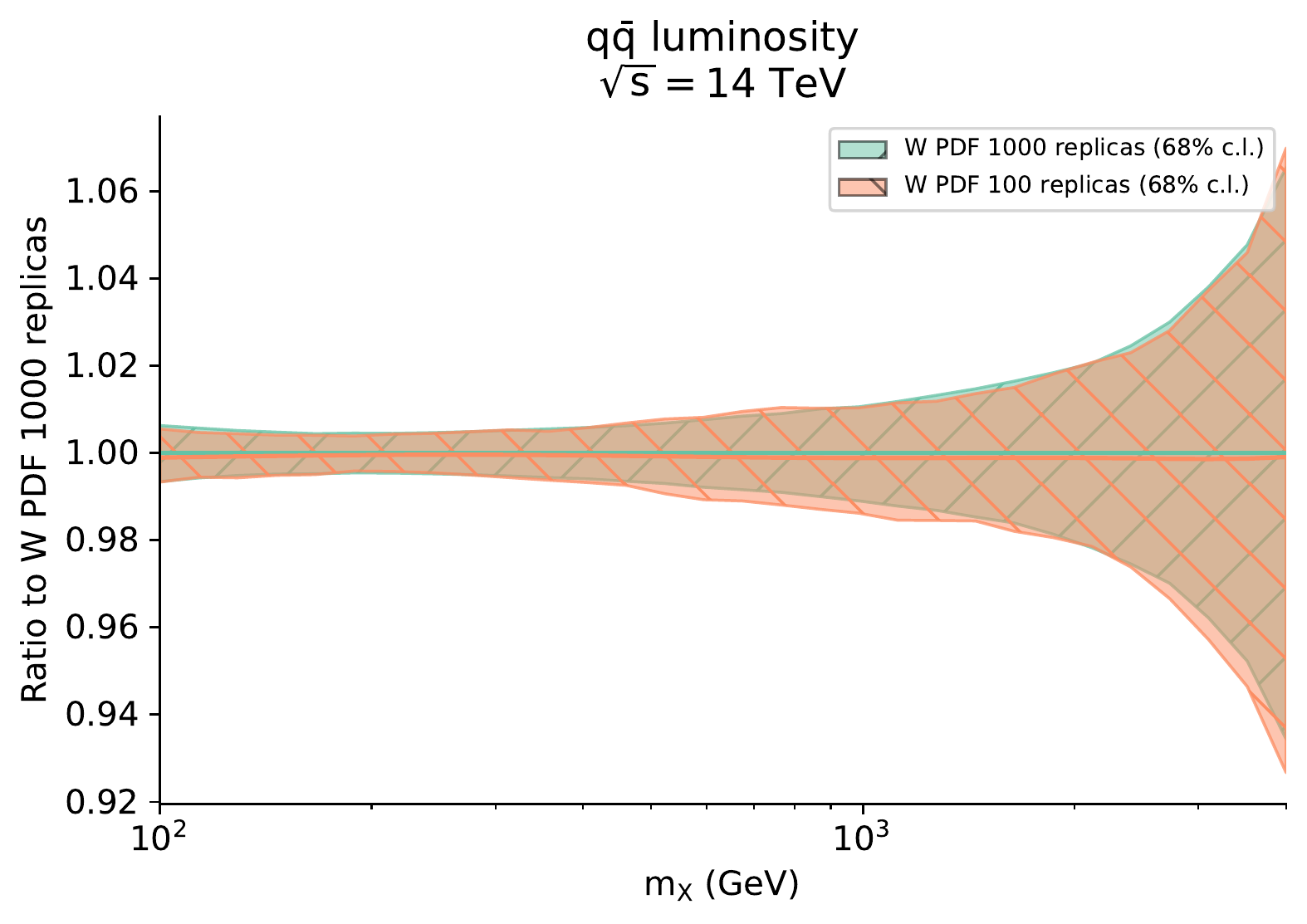}
        \includegraphics[width=0.49\textwidth]{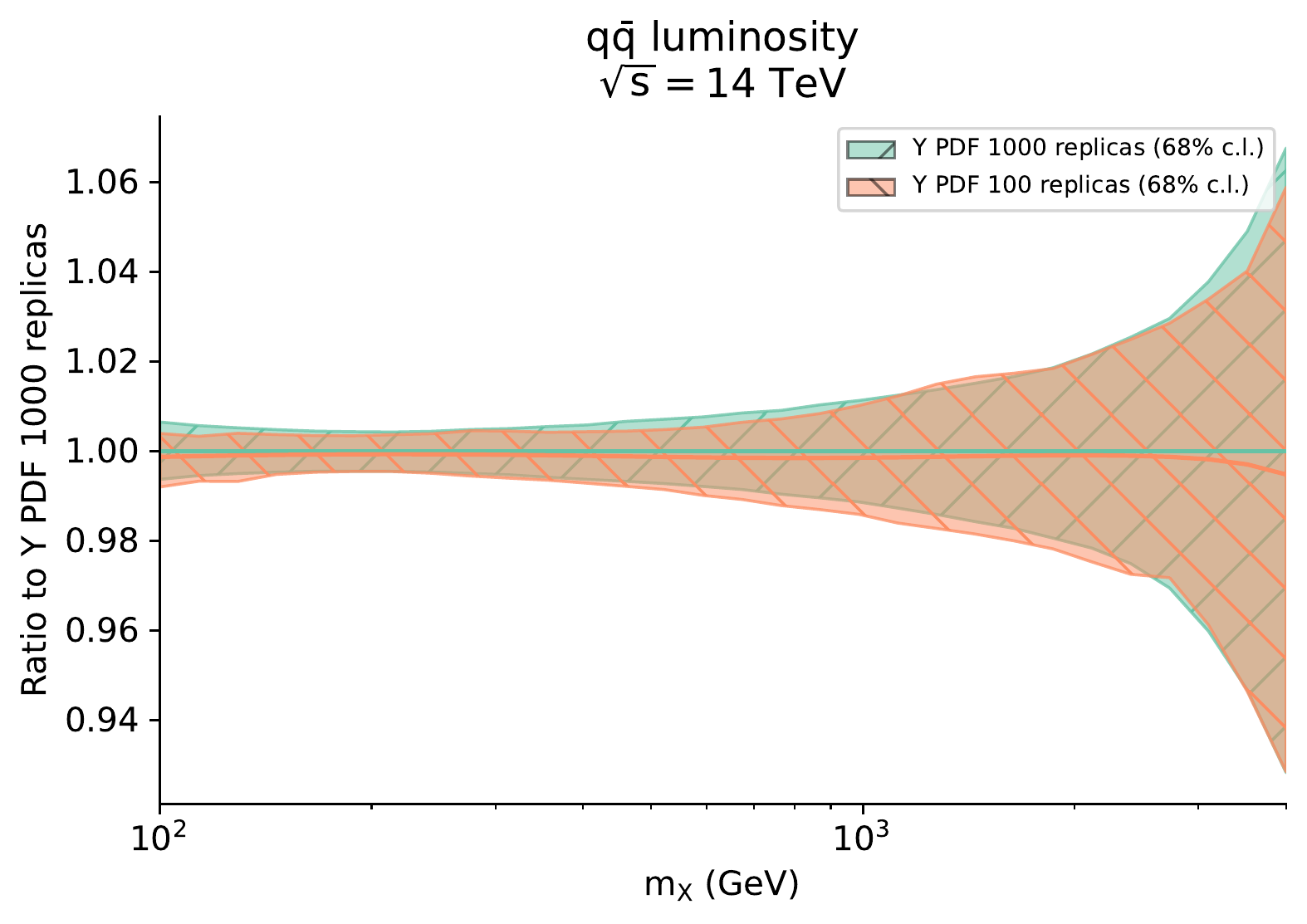}\\
        \includegraphics[width=0.49\textwidth]{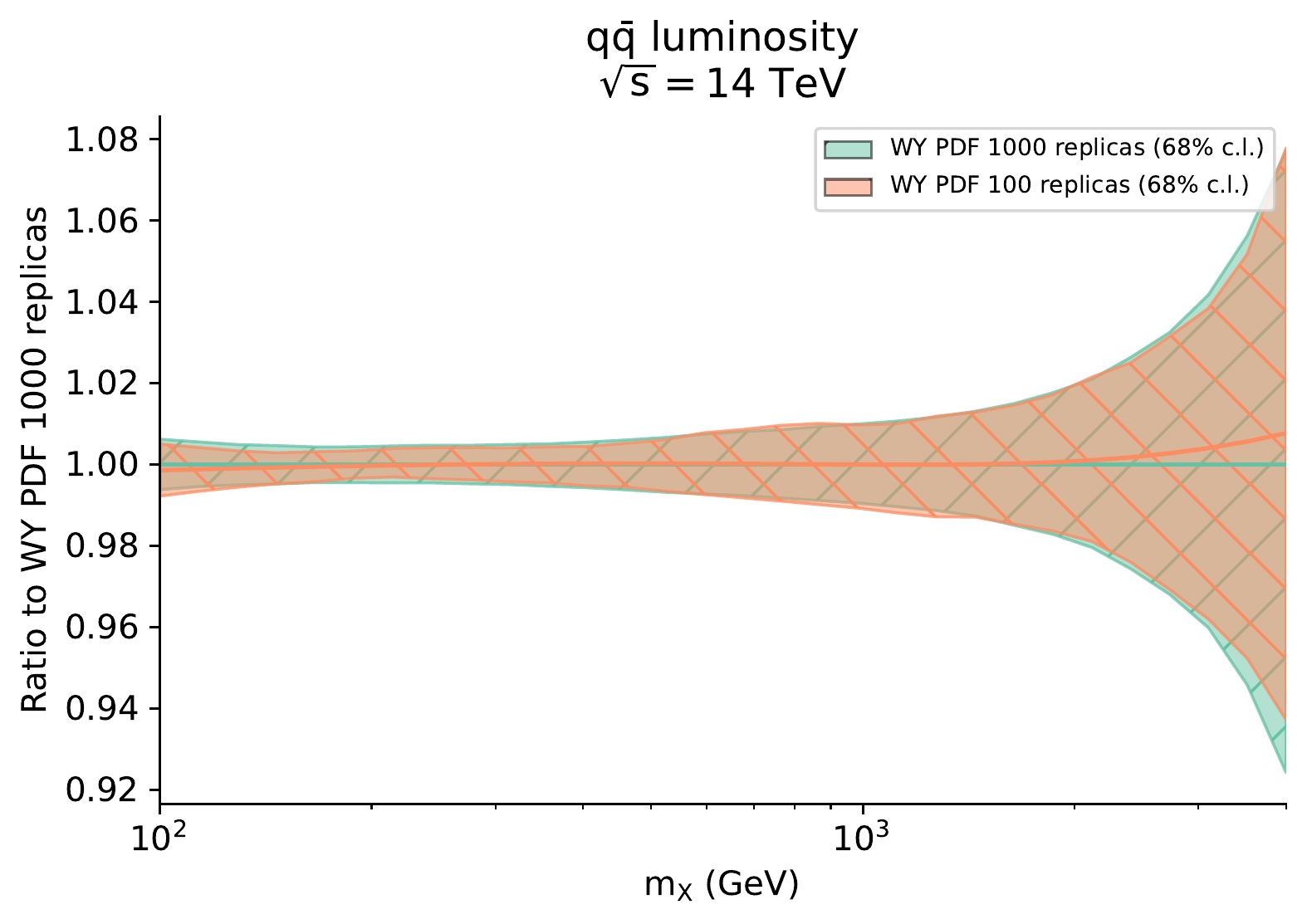}
        \caption{\label{fig: lumi_replica} The \(q\bar{q}\) luminosity of the
        100 replica (low statistics) fits normalized to the 1000 replica (high
        statistics) fits for various SMEFT scenarios considered in this work.
        Shown in the top left (top right) is the individual \(W\) (\(Y\)) scenario,
        while the lower panel is the combined \((W, Y)\) scenario with HL-LHC
        projections being used in the fit.}
\end{figure}

From each of our high replica fits we randomly sample, without replacement, a
set of \(100\) replicas thereby effectively emulating the scenario whereby the user
would have performed a low statistics fit. In Fig. \ref{fig: lumi_replica} we
plot the \(q\bar{q}\) luminosity for the various SMEFT scenarios considered in this
work. We plot the low replica luminosity normalized to the analogous high
replica set. We see that the luminosities remain virtually identical, with no
discernible difference at the ensemble level.  Such behaviour is inherited
directly from NNPDF4.0 where a typical fit will typically only possess \(100\) MC replicas
after the post-fit selection has filtered poorly performing replicas.

\begin{figure}[t]
        \centering
        \includegraphics[width=0.49\textwidth]{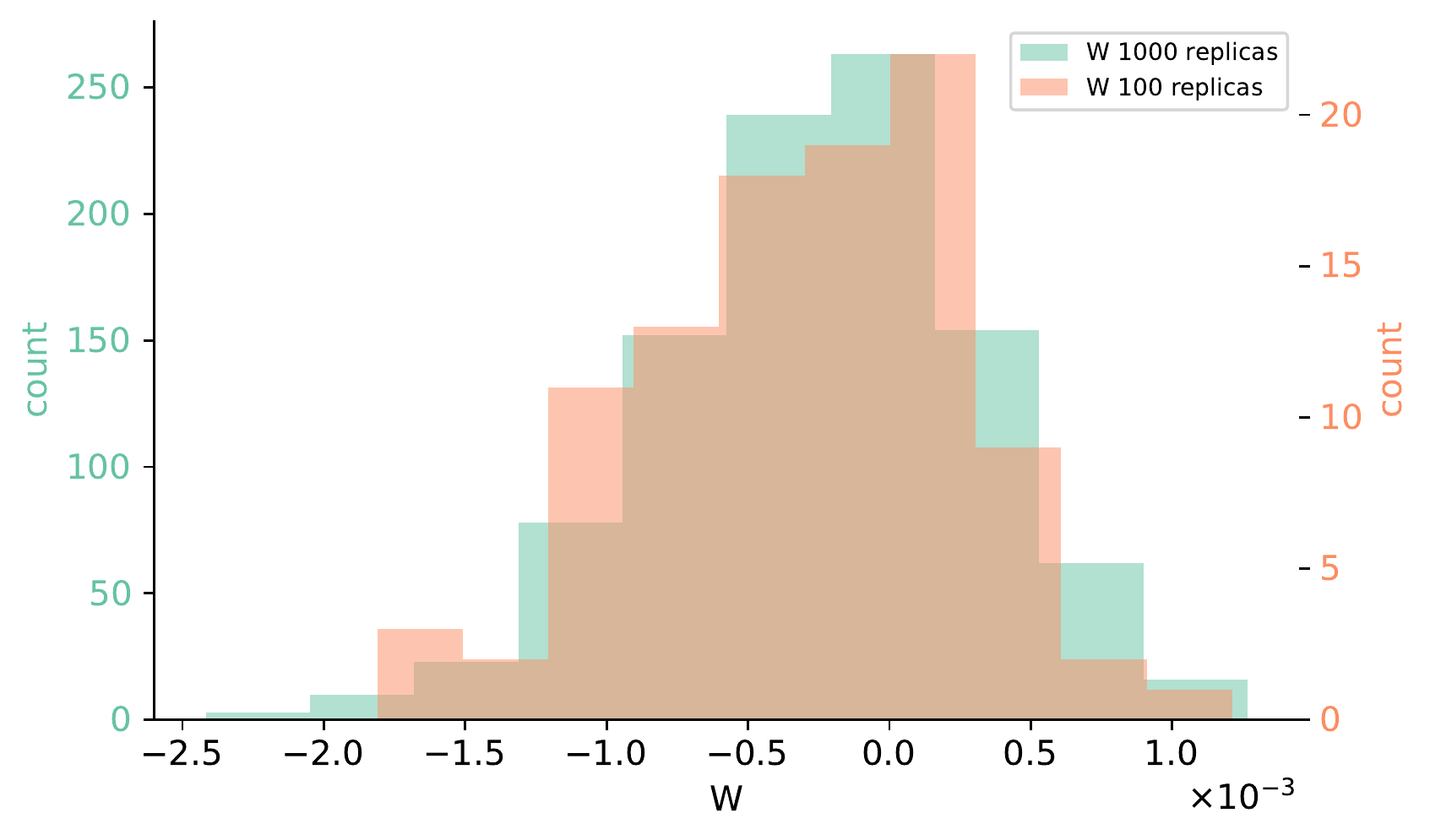}
        \includegraphics[width=0.49\textwidth]{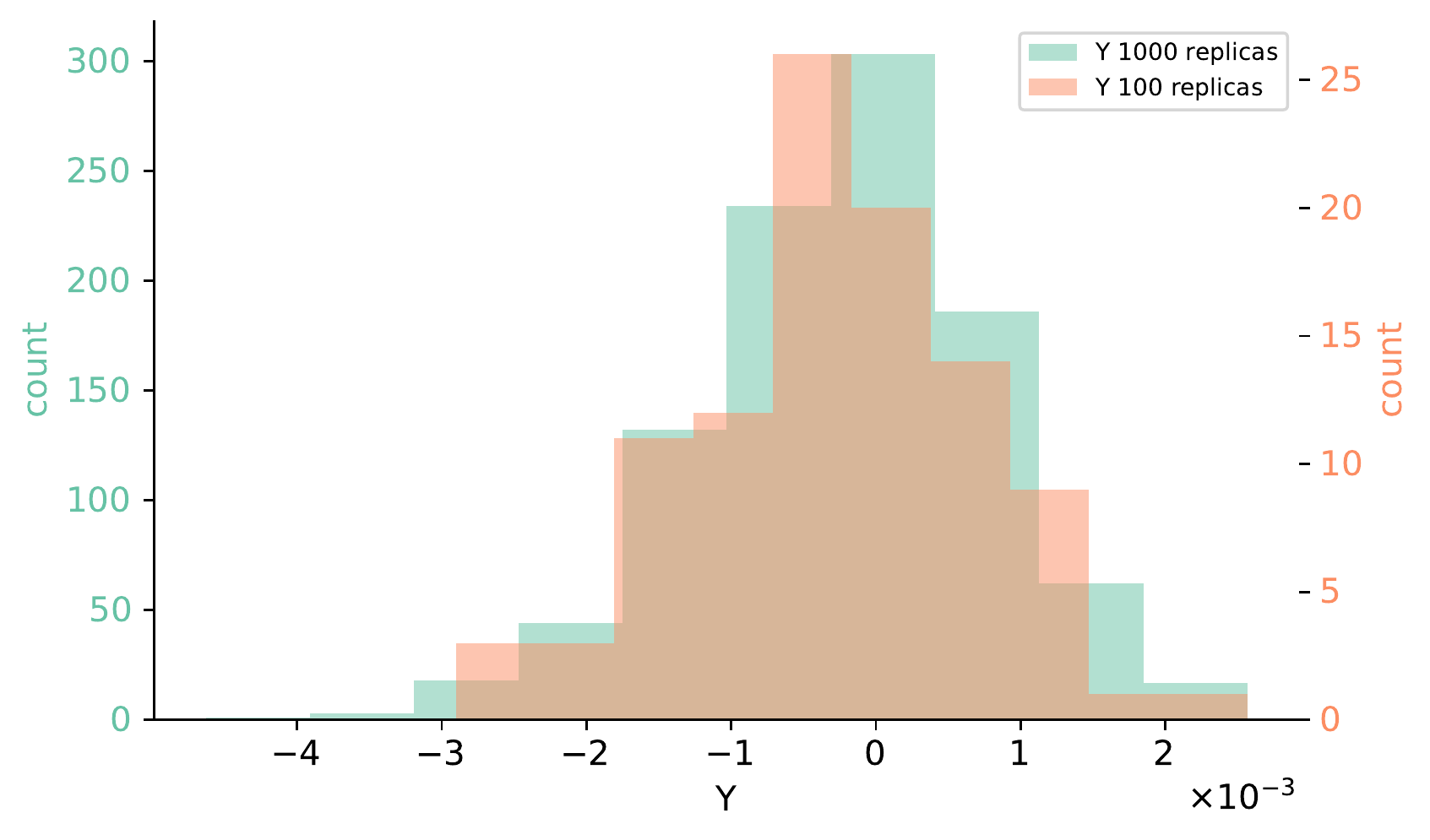}
        \caption{\label{fig: hist_replica}Distributions of the best fit \(W\)
        (left) and \(Y\) (right) parameters using both high
        (\(N_\text{rep}=1000\)) and low (\(N_\text{rep}=100\)) statistics fits.
        We bring the readers attention to the different \(y\) axes for the
        histogram overlays.}
\end{figure}

In Fig. \ref{fig: hist_replica} we present the distributions of best-fit \(W\)
and \(Y\) values using both the high replica and reduced sets. We see that the
distribution of best fit Wilson coefficients are accurately reproduced by the
low statistics set: implying that, had we stopped the fitting process with 100 MC
replicas, the additional 900 replicas would have changed the ensemble statistics
such as the mean, standard deviation, or bounds very little. This is a typical pattern
in the Monte Carlo approach to PDF fitting, whereby one quickly reaches saturation after
\(\sim 100\) MC replicas and further replicas only serve to accurately reproduce
the experimental correlations.

\clearpage

\bibliography{bib/simul.bib}

\end{document}